\documentclass[twocolumn,aps,superscriptaddress]{revtex4-1}
\usepackage{palatino}
\usepackage[latin1]{inputenc}
\usepackage{epsf}

\usepackage{graphicx}

\usepackage{float}

\usepackage{amsmath,amssymb}
\usepackage{latexsym}
\usepackage{calc}
\usepackage{color}
\usepackage{shadow}
\usepackage{epsfig}
\usepackage[usenames,dvipsnames]{xcolor}

\def\bea{\begin{eqnarray}}
\def\eea{\end{eqnarray}}
\def\ben{\begin{equation}}
\def\een{\end{equation}}
\def\benu{\begin{enumerate}}
\def\enu{\end{enumerate}}

\def\n{n}

\def\sss{\scriptscriptstyle\rm}

\def\1var{(\bx_1...\bx\N)}

\def\sss{\scriptscriptstyle\rm}
\def\x{_{\sss X}}
\def\c{_{\sss C}}
\def\s{_{\sss S}}
\def\xc{_{\sss XC}}

\def\H{_{\sss H}}

\def\Hxc{_{\sss HXC}}
\def\ee{_{\rm ee}}

\def\br{{\bf r}}

\def\bx{{\br t}}

\def\bj{{\bf j}}

\def\ext{_{\rm ext}}

\begin{document}
\title{Charge Transfer in Time-Dependent Density Functional Theory}
\author{Neepa T. Maitra}
\affiliation{Department of Physics and Astronomy, Hunter College and the Physics Program at the Graduate Center of the City University of New York, 695 Park Avenue, New York, New York 10065, USA}

\date{\today}

\begin{abstract}
  Charge transfer plays a crucial role in many processes of interest
  in physics, chemistry, and bio-chemistry. In many applications the
  size of the systems involved calls for time-dependent density
  functional theory (TDDFT) to be used in their computational
  modeling, due to its unprecedented balance between accuracy and
  efficiency. However, although exact in principle, in practise
  approximations must be made for the exchange-correlation functional
  in this theory, and the standard functional approximations perform
  poorly for excitations which have a long-range charge-transfer
  component.  Intense progress has been made in developing more
  sophisticated functionals for this problem, which we review. We
  point out an essential difference between the properties of the
  exchange-correlation kernel needed for an accurate description of
  charge-transfer between open-shell fragments and between
  closed-shell fragments. We then turn to charge-transfer dynamics,
  which, in contrast to the excitation problem, is a highly
  non-equilibrium, non-perturbative, process involving a transfer of
  one full electron in space. This turns out to be a much more
  challenging problem for TDDFT functionals.  We describe dynamical
  step and peak features in the exact functional evolving over time,
  that are missing in the functionals currently used. The latter
  underestimate the amount of charge transferred and manifest a
  spurious shift in the charge transfer resonance position. We discuss
  some explicit examples. 
  \end{abstract}

\maketitle

\section{Introduction}
Charge transfer could be argued to be one of the most important
interactions in science, underlying central processes in
photosynthesis, photovoltaic devices, photo-catalysis, molecular
switches, nanoscale conductance, reactions at interfaces and in
solvents. In some applications the excitation
energies are primarily of interest, while in other applications a full
account of the dynamics of the transferring charge in real time is
critical. Note that we use the term ``charge transfer'' in a
somewhat casual way: first, we really only mean electron transfer, and second, we mean it to include 
electronic motion accompanied  by nuclear motion (sometimes referred to as "charge separation") as well as  when there is no accompanying nuclear motion (sometimes termed "charge migration").
A quantum-mechanical treatment of the electronic system is
necessary, yet the processes above typically involve systems large
enough that approximate wavefunction methods that capture 
electron correlation accurately enough are not computationally feasible. 

\begin{figure}[h]
\begin{center}
\includegraphics[width=0.45\textwidth,height=0.3\textwidth]{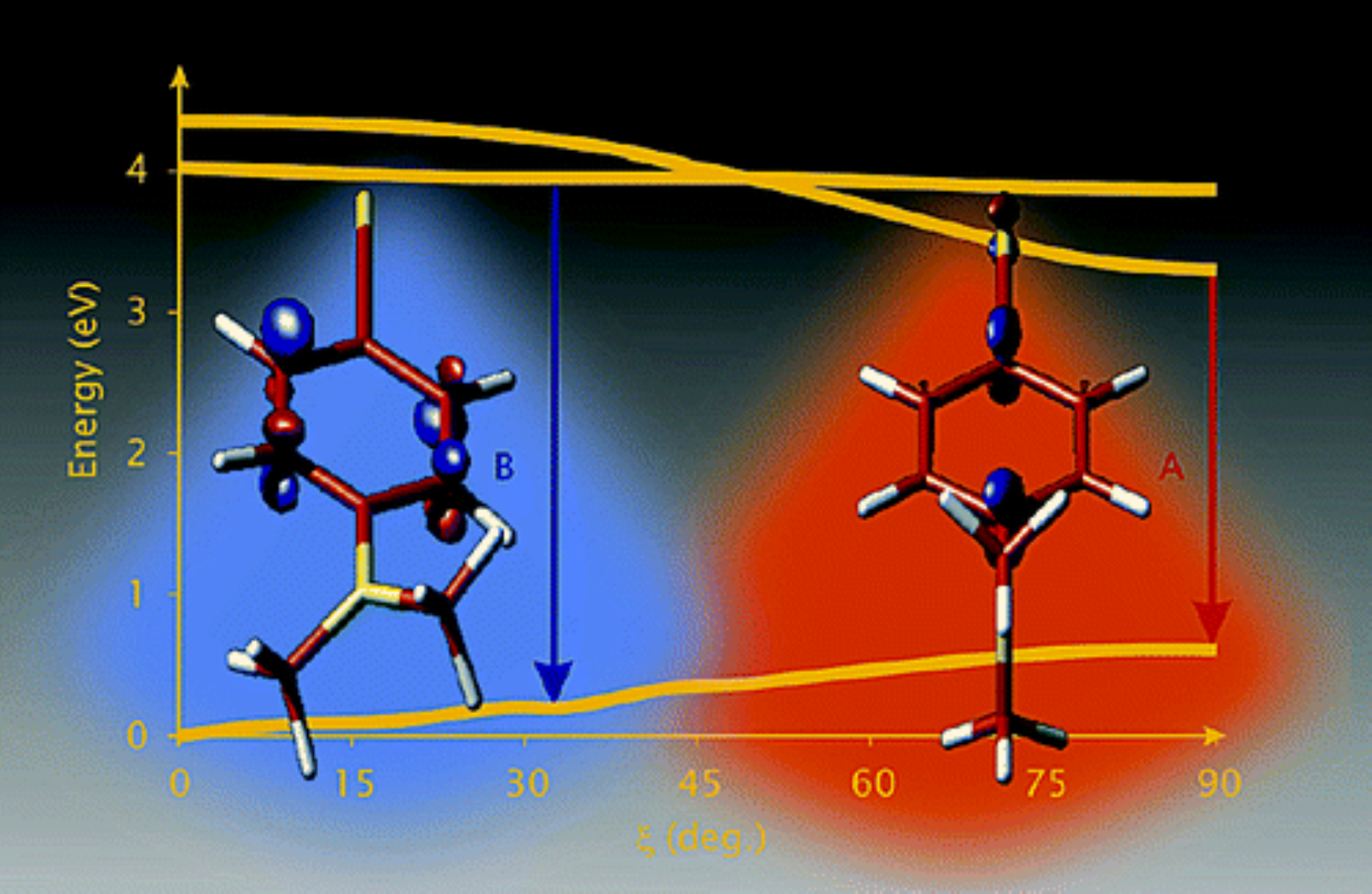}
\end{center}
\caption{Photo-induced intramolecular charge-transfer, leading to dual fluorescence in DMABN when placed in a polar solvent. The band labelled "B" is from fluorescence due to a largely local excitation, while that labelled "A" is reached via crossing of the excited energy surfaces and is due to fluorescence from a charge-transfer excitation (See text).  Reprinted with permission
from Ref.~\cite{RF04}. Copyright 2004, American Chemical Society.}
\label{fig:dmabn}
\end{figure}

Time-dependent
density functional theory (TDDFT) offers a promising
alternative, as it recasts the correlated problem in terms of a
non-interacting problem in an in-principle exact way with the use of an
exchange-correlation functional~\cite{RG84}. In practise, approximations must be made for the latter, and the success of TDDFT hinges on both the general accuracy of the approximation made and its computational efficiency.
A beautiful example of the interpretive and predictive value TDDFT has had for a charge-transfer process is the breakthrough explanation of the dual fluorescence of 4-dimethyl-aminobenzonitrile (DMABN) by Rappoport and Furche in 2003~\cite{RF04}, which finally put to rest a forty-year old debate about the mechanism. 
When this molecule is placed in a polar solvent, in addition to a "normal" emission band that is always present, it had been observed experimentally that an additional red-shifted band appears, dubbed "anomalous" (see Fig.~\ref{fig:dmabn}). It had been thought, since its experimental discovery~\cite{Lippert1959}, that this band was due to an intramolecular charge-transfer state but the nature of this state  (twisted, or planar quinoidal) had been, until Ref.~\cite{RF04}, very controversial, as it was difficult to resolve without accurate but  computationally-efficient first-principles methods for excited states. Comparing results from transient spectroscopy measurements with TDDFT predictions of emission energies, excited dipole moments and vibrational frequencies, the authors could finally without doubt identify the nature, electronic and geometric, of the state and the mechanism that led to the dual fluorescence. (We return briefly to this work in Section~\ref{sec:CT_excitations}). 

The value of the electronic excitation energies were not so relevant to the study above, and in fact, in practise the standard approximations that enter for the exchange-correlation functional of TDDFT have
performed poorly for charge-transfer excitation energies as well as in modeling the dynamics of a transferring electron in real time. 
Given the importance of 
charge transfer in physics, chemistry, and biology, the past
decade has seen enormous effort to understand and remedy this problem, with the
development of more reliable approximate functionals to capture
charge-transfer excitations in many situations, but the problem is not entirely satisfactorily solved. 
Time-resolved
dynamics of the transferring charge remains a challenge
today: the system evolves far from its ground state, and the
description must go beyond the linear response regime. When the charge
transfers out of a system initially in its ground-state, features that
are necessary for an accurate description of the dynamics develop in
the exact exchange-correlation potential that require a spatially
non-local and time-non-local (non-adiabatic) dependence on the density, and this is missing in all
approximations in use today. The prognosis for standard functionals
for charge-transfer from a photo-excited state is better, and some
success has been seen here, although open questions remain.

There have been reviews of progress in TDDFT for charge-transfer excitations~\cite{A09}, and for non-perturbative charge-transfer dynamics~\cite{F16}. Here we provide a overview of both regimes, focussing on the behavior of the exact functionals and of approximations. 
We begin in Section~\ref{sec:tddft} with a brief reminder of the theory behind TDDFT and its linear response formalism. We then turn to the description of charge-transfer excitations in Section~\ref{sec:CT_excitations} with an
explanation of why standard functional approximations have difficulty, and a review of some recent progress in
developing more suitable functionals. Section~\ref{sec:CTdynamics} turns to the description of the charge-transfer process in real time, where it is necessary to step out of the linear response regime to capture the motion of the transferring charge.

\section{TDDFT in a nutshell}
\label{sec:tddft}
TDDFT is an exact reformulation of the quantum mechanics of
interacting identical many-particle systems, whose fundamental theorem states that all
properties of the system evolving from a given initial wavefunction
can be obtained from knowledge of the time-dependent one-body
probability density~\cite{RG84,M16, GM12chap,TDDFTbook12,Carstenbook}, 
\ben
n(\br,t)
= N \sum_{i=1}^N \sum_{\sigma_i = \uparrow \downarrow}
\int d^3r_2...d^3r_N \vert
\Psi(\br\sigma_1,\br_2\sigma_2...\br_N\sigma_N, t)\vert^2\;,
\een
a far simpler
object than the complex many-particle time-dependent wavefunction $\Psi$. Attempting
to numerically solve Schro\"odinger's equation for $\Psi$ for systems
of more than a few electrons, one rapidly hits an exponential
wall. The theorem was  derived for systems with Hamiltonians of the
form
\ben
\hat H(t) = \hat{T} + \hat{V}\ee +\hat{V}\ext(t)
\een
where $\hat{T} = -\frac{1}{2}\sum_i^{N}\nabla_i^2$ is the kinetic
energy operator and $\hat{V}\ee$ is the particle-particle interaction; in particular, for electrons, $\hat{V}\ee = \frac{1}{2}\sum_{i\neq j}\frac{1}{\vert\br_i - \br_j\vert}$. Note that atomic units are used throughout this review, $e^2 =\hbar = m_e = 1$.
 The external potential
$\hat{V}\ext(t) = \sum_i^Nv\ext(\br_i,t)$ is the term that
distinguishes systems of a fixed type of particle: for electrons, it
represents the potential the electrons experience due to the nuclei
and to any field applied to the system, and the theorems state that a
given density evolution $n(\br,t)$ of a system composed of this type of
particle, beginning in a given initial state, can at most be
reproduced by one $v\ext(\br,t)$. This means that the one-body density
$n(\br,t)$ completely identifies the many-particle system, since it
points to a unique external potential, therefore to a unique $H(t)$
and therefore, to the time-dependent many-particle
wavefunction $\Psi(t)$ whose one-body density is $n(\br,t)$.  The implication is that if we could find
directly the one-body density, then, in principle, we would be able to extract from
it, together with the initial many-body wavefunction $\Psi(0)$,  all properties of the interacting many-body system, i.e. every observable is a  functional of the density and the initial state, $O[n;\Psi(0)]$.

One can formulate an exact evolution equation for the density, but the
terms in it require the kinetic and interaction stress tensors of the
many-particle wavefunction~\cite{L99}, which are challenging to
accurately approximate as functionals of the density. Instead, in
practise, TDDFT operates via the non-interacting Kohn-Sham system,  similar to that in
ground-state DFT~\cite{HK64,KS65} but now time-dependent: the density of the true interacting system
is obtained as the density of this auxiliary system of non-interacting
fermions, evolving in a one-body potential. One begins by choosing an
initial state (typically a Slater determinant of orbitals $\{\phi_i(\br,0)\}$) for the Kohn-Sham propagation, $\Phi(0)$, that must have
the same density and first time-derivative of the density as the
physical initial state $\Psi(0)$. The latter condition arises because,
due to the equation of continuity that states $\partial_t n(\br,t) =
-\nabla\cdot\bj(\br,t)$, where $\bj$ is the one-body current-density, the initial wavefunction fixes both the initial density as well as its
initial time-derivative.  Then one solves
\ben
\left(-\frac{\nabla^2}{2} + v\s(\br,t)\right)\phi_i(\br,t) = i\partial_t\phi_i(\br,t)
\label{eq:tdks}
\een
where the Kohn-Sham potential $v\s(\br,t)$ is usually written as
\ben
v\s(\br,t) = v\ext(\br,t) + v\H[n](\br,t) + v\xc[n,\Psi(0),\Phi(0)](\br,t)\;.
\label{eq:vs}
\een
Here $v\H[n](\br,t) = \int d^3\br' \frac{n(\br',t)}{\vert \br -  \br'\vert}$ is the Hartree potential, and
$v\xc[n;\Psi(0),\Phi(0)](\br,t)$ is the exchange-correlation (xc)
potential. If the exact functional was known, then solution of these equations would yield orbitals which, at each time, yield the exact density of the interacting system via $n(\br,t) = \sum_{i=1}^N\vert\phi_i(\br,t)\vert^2$.

In practice, approximations enter first when
selecting an initial Kohn-Sham state (the exact initial density and time-derivative are
generally only known approximately, e.g. via a ground-state DFT
calculation, if the calculation begins in the ground state, or, more generally, via a
correlated many-body calculation for $\Psi(0)$), and second in approximating the xc
functional. It is known that the exact xc functional has
``memory-dependence'': that is, at time $t$, $v\xc[n,\Psi(0),\Phi(0)](\br,t)$ depends on the history
of the density $n(t'<t)$ as well as the physical initial state
$\Psi(0)$ and the choice of Kohn-Sham initial state $\Phi(0)$.
Almost always however, this memory dependence is neglected (exceptions include e.g.~\cite{VK96,VUC97,KB06}), and an adiabatic approximation is used, which inserts the time-dependent density into a
ground-state (gs) approximation:
\ben
v\xc^A[n,\Psi(0),\Phi(0)](\br,t) = v\xc^{\rm gs}[n(t)](\br)
\een
For example, in the simplest approximation, adiabatic local-density approximation (ALDA), $v\xc^{\rm ALDA}(\br,t) = \frac{d (n\epsilon^{\rm unif}\xc[n])}{dn} \vert_{n = n(\br,t)}$ where $\epsilon^{\rm unif}\xc[n]$ is the xc energy per particle of a uniform electron gas of density $n$. In real-time propagation, almost always an adiabatic local or semi-local approximation is used, but not always (e.g. Ref.~\cite{FCG15,TGLC15}). Calculations that use hybrid functionals which mix in a fraction of Hartree-Fock exchange, within a generalized Kohn-Sham framework, may be thought of as containing some memory-dependence in their exchange-component, in the sense that the instantaneous orbital is a functional of the history of the density and the Kohn-Sham initial state. 
Rarely, when a ground-state functional is used that is an explicit orbital-functional (e.g. as in exact exchange, or self-interaction corrected LDA) within TDOEP or KLI~\cite{L98,UGG95,MKLR07,HKK12,WU08}, then likewise such an approximation contains some memory dependence.

In any adiabatic approximation there are then two sources of error: one from making the adiabatic approximation itself (neglecting memory), and the second from the choice of ground-state xc functional approximation. To disentangle the effects of these two errors, it is instructive to define the "adiabatically-exact" (AE) potential, where the instantaneous density is inserted in the exact ground-state xc functional. Finding this potential is only possible for small or model systems, and often it proceeds by finding the interacting $v\ext^{\rm ex. gs}$ and non-interacting  $v\s^{\rm ex. gs}$ potentials whose ground-state density is exactly equal to the instantaneous density~\cite{HMB02,TGK08,M16}, and then
 \ben
 v\xc^{\rm AE}[n](\br,t) = v\s^{\rm ex. gs}[n(t)](\br) - v\ext^{\rm ex. gs}[n(t)](\br) - v\H[n(t)](\br)\;.
 \label{eq:AE}
 \een

There are several intriguing foundational properties
($v$-representability and the existence of the Kohn-Sham system,
assumptions of time-analyticity underlying the proof, choices of the
initial Kohn-Sham state, known conditions that the exact xc functional
satisfies, for example) that have been uncovered in recent years, and
we refer the reader to the recent review~\cite{M16} for a discussion
and references to some of this work. We here put those considerations
aside, and continue with the formalism for the special case of linear
response, which is the regime that has propelled TDDFT to its
fame in achieving an unprecedented balance between accuracy and
efficiency in obtaining excitation spectra.

\subsection{Linear Response TDDFT}
\label{sec:lr_tddft}
Linear response is measured by applying a small perturbation to a
ground-state and measuring the change in an observable to first order
in the perturbation. TDDFT response theory revolves around the
response of the density, and the central equation~\cite{GK85,PGG96,GM12chap} takes the form:
\ben
\chi(\omega) = \chi\s(\omega) + \chi\s(\omega) \star f\Hxc(\omega) \star \chi(\omega)
\label{eq:dyson}
\een
where $\chi(\omega) = \chi(\br,\br',\omega)$ is the density-density response function (or susceptibility) of the physical system, defined as the time-frequency Fourier transform of  
\ben
\chi[n_0](\br,\br',t-t') = \left.\frac{\delta n(\br,t)}{\delta v\ext(\br',t')}\right\vert_{n_0(\br)}\;,
\een
 and $\chi\s(\omega)$ is the analogous quantity for the Kohn-Sham system. They are related via the Hartree-exchange-correlation kernel,
$f\Hxc(\omega) = f\H + f\xc(\omega)$  where $f\H(\br,\br') = 1/\vert\br - \br'\vert$ and 
\ben
f\xc[n_0](\br,\br',t-t') = \left.\frac{\delta v\xc[n](\br,t)}{\delta n(\br',t')} \right\vert_{n=n_0}
\label{eq:fxc}
\een
is the xc kernel, a functional of the ground-state density $n_0(\br)$. 

Equation~(\ref{eq:dyson}) is proven by equating the density-response of the physical system to an applied external perturbation $\delta v\ext(\br,t)$  to that of the Kohn-Sham system under the corresponding Kohn-Sham perturbation. The density-density response function $\chi(\omega)$ contains complete information about the excitation energies $\Omega_I$ and transition densities $\langle\Psi_0\vert\hat{n}(\br)\vert\Psi_I\rangle$, as evident from its spectral, a.k.a Lehmann, representation:
\ben
\chi(\br,\br',\omega)= \sum_I \frac{\langle\Psi_0\vert\hat{n}(\br)\vert\Psi_I\rangle\langle\Psi_I\vert\hat{n}(\br')\vert\Psi_0\rangle}{\omega - \Omega_I + i 0^+} + c.c.(\omega \to -\omega)\;,
\label{eq:chi}
\een
where $\hat{n}(\br) = \sum_i \delta(\br - \hat{\br_i})$ is the density operator,  $\Omega_I = E_I - E_0$ is the excitation frequency of the $I$th excited state, and the  sum goes over all interacting states $\Psi_I$. The notation $c.c.(\omega \to -\omega)$ means the complex conjugate of the first term evaluated at $ -\omega$. TDDFT thus provides a route to calculating excitations and transition probabilities without having to solve the many-body problem: instead, one solves Eq.~(\ref{eq:dyson}) for $\chi(\omega)$.

There are two steps, and each step requires in practise an
approximation. First, the ground-state problem is solved, yielding the
occupied and unoccupied Kohn-Sham orbitals that live in the one-body
potential $v\s[n_0](\br)$. Second, the xc kernel is applied, and this
shifts and mixes the Kohn-Sham excitations, which are simple
orbital-energy differences (single excitations), to yield the true
ones. These are often called dynamical corrections. 
If the exact ground-state xc potential and xc kernel were known
(as they are in simple cases), this process would yield the exact
response of the physical system. In practise, approximations must be
made for both, and, technically, the approximation used for the
ground-state xc potential and that used for the xc kernel should be
linked as they arise from the same perturbed Kohn-Sham potential. But
in practise, the two approximations are often treated
independently. Almost always, an adiabatic approximation is used for
the kernel, which renders the xc kernel independent of frequency. This
can be seen by the fact that in such an approximation,
$v\xc^A[n](\br,t)$ depends only on $n(t)$ at the present time, so the
right-hand-side of Eq.~(\ref{eq:fxc}) becomes proportional to
$\delta(t-t')$ and its Fourier transform frequency-independent,
$f\xc^A(\br,\br',\omega) = f\xc(\br,\br')$.

Typically, for finite systems, Eq.~(\ref{eq:dyson}) is re-cast in the form of
 matrix
 equations~\cite{C95,C96}. The eigenvalues of the matrix
\ben
R_{qq'}(\omega) = \omega_q^2\delta_{qq'} + 4\sqrt{\omega_q\omega_{q'}}f\Hxc^{qq'}(\omega)\;, {\rm where}
\label{eq:casida}
\een
$f\Hxc^{qq'}(\omega)=\int d^3r d^3r' \phi_i(\br)\phi_a(\br) f\Hxc(\br,\br',\omega)\phi_{i'}(\br')\phi_{a'}(\br')$, lie at the squares of the excitation energies, and oscillator strengths can be obtained from the eigenvectors. Here $q=(i,a)$ represents a double-index, with $i$ labelling an occupied orbital and $a$ an unoccupied one. 
Different equivalent derivations and formulations appear in Refs.~\cite{BA96a,GPG00}; note that the derivation in Ref.~\cite{BA96a} is valid only when an adiabatic approximation is assumed from the outset.

For molecules, the matrix formulation is usually used, while for
solids with their continuous spectra, Eq.~(\ref{eq:dyson}) is usually
solved.  Significant progress has been made for efficient solution of the matrix equations, e.g. Ref.~\cite{FKNK16} achieves the same scaling and similar cost per excited state as a ground-state calculation.  One
can also obtain linear response spectra in TDDFT via real-time
propagation~\cite{YNIB06} under a weak perturbation: applying a
$\delta$-kick to uniformly stimulate all excitations, and examining
peaks of the Fourier-transformed dipole moment which lie at the
resonant frequencies. In addition to the matrix formulation, this is
available in codes such as octopus~\cite{octopus,octopus2} and
NWChem~\cite{nwchem, LN11}. This can be
computationally advantageous over the matrix formulation for large
systems especially when a broad picture of the spectrum is of
interest, i.e. not just the lowest excitations.  Finally, 
another formulation to obtain response from TDDFT is the Sternheimer
approach, also known as density functional perturbation theory, or
coupled perturbed Kohn-Sham theory, that operates directly in
frequency-space but avoids the calculation of many unoccupied orbitals
by instead considering perturbations of the occupied orbitals in
frequency-space~\cite{Sternheimer54,Andrade07,Strubbechapter}.

\section{Charge-transfer excitations}
\label{sec:CT_excitations}
Consider a charge-transfer excitation, where compared to the ground state,
the excited state has a significant fraction of electron density transferred
from one part of the molecule, the donor, to another part, the
acceptor. In the large separation limit, it is straightforward to deduce that the exact lowest charge-transfer excitation frequency of a neutral molecule must approach
\ben
\Omega_{\rm CT} = I^D - A^A - 1/R
\label{eq:omega_CT}
\een
 where $I^D = E^D(N_D-1) - E^D(N_D)$ is the first ionization energy of the $N_D$-electron donor,
$A^A = E^A(N_A) - E^A(N_A+1) $ is the electron affinity of the $N_A$-electron acceptor, and $-1/R$ is the
 electrostatic correction that is lowest-order in the separation $R$ between the fragments after the
charge-transfer. We will focus only on neutral molecules throughout this review. 
The following subsections will analyse what the TDDFT linear response formalism gives for such an excitation with the standard approximations, properties of the exact functional in TDDFT responsible for yielding the correct value, and the recent development of approximate functionals that have had some success in capturing these excitations in a number of cases.

\subsection{Charge-transfer between closed-shell fragments}
\label{sec:closed}
We first consider the case where the molecule consists of two closed-shell fragments at large separation, so that, in the Kohn-Sham description of the ground-state, each fragment has paired electrons doubly-occupying the orbitals of interest, localized on one fragment or another. 
As mentioned above, Eqs.~(\ref{eq:dyson}) and (\ref{eq:casida}) show that TDDFT
excitations are obtained from shifting and mixing single-excitations
of the Kohn-Sham system via matrix elements of $f\xc$.  A long-range charge transfer excitation is composed of Kohn-Sham excitations in which the unoccupied orbitals have only
little spatial overlap with the occupied orbitals.
Then, the matrix elements of the conventional approximate xc
kernels that are relevant for this excitation vanish, and the TDDFT excitation energy reduces to the
Kohn-Sham orbital-energy difference. This typically yields a drastic underestimate of the excitation energy, up to several eV, as was first observed in the literature in Ref.~\cite{TAHR99}. 
 Several useful diagnostics of the expected accuracy  of a
given functional approximation have been used to  gauge whether the result of an approximate TDDFT calculation cannot be trusted for this reason; of particular note are the $\Lambda$-index of Ref.~\cite{PBHT08} based on considering the overlaps of molecular orbital moduli, and more recently the $\Delta\br$-index, based on the charge-centroids of the orbitals involved in the transition~\cite{GCMA13}. The particle-hole map, related to the transition-density map, gives a direct visualization of the origin and destination of electrons and holes for a given excitation, so can indicate the role of charge-transfer for that excitation~\cite{LU16,MPD15}.

To understand this  underestimation further, consider the case where the charge-transfer excitation is dominated by a single excitation from occupied Kohn-Sham orbital on the donor, $\phi_D$ ,to an unoccupied orbital on the acceptor, $\phi_A$. That is, there is negligible mixing with other Kohn-Sham excitations (which happens, for example if the charge-transfer excitation energy does not lie close to any other excitation energy). Then, it is justified to take only the diagonal element in the matrix~(\ref{eq:casida}), yielding
\ben
\Omega^{\rm TDDFT}_{\rm SMA} \approx \sqrt{\omega_q^2 + 4 \omega_qf\Hxc^{qq}(\omega)} 
\label{eq:sma}
\een
where 
\ben
f\Hxc^{qq} = \int d^3r d^3r' \phi_D(\br)\phi_A(\br)f\Hxc(\br,\br',\omega)\phi_D(\br')\phi_A(\br')\;,
\label{eq:fHxcmatrixelt}
\een
 and $\omega_q$ is the Kohn-Sham orbital-energy difference, $\epsilon_A -\epsilon_D$. Eq.~\ref{eq:sma} is often referred to as the "small matrix approximation" (SMA)~\cite{VOC99,AGB03,GPG00}, giving a diagonal correction to the Kohn-Sham excitation energy for any case (not just charge-transfer) when coupling to other excitations is assumed small, and it can be a useful diagnostic tool. 
If we assume the $f\Hxc^{qq}$ correction is small compared to the Kohn-Sham excitation energy, for example for weakly-correlated systems, then Eq.~(\ref{eq:sma}) reduces to the single-pole approximation (SPA)~\cite{VOC99,GPG00,PGG96}, 
\ben
\Omega^{\rm TDDFT}_{\rm SPA} = \omega_q + 2 f\Hxc^{qq}(\omega_q)
\label{eq:spa}
\een

Now we ask, if we were to use the {\it exact} ground-state xc functional and the {\it exact} xc kernel, how would Eq.~(\ref{eq:sma}) reproduce the exact charge-transfer excitation energy, Eq.~\ref{eq:omega_CT}? For the  lowest charge-transfer excitation, $\epsilon_A$ is the lowest unoccupied molecular orbital  (LUMO) energy of the acceptor, and $\epsilon_D$ is the highest occupied molecular orbital (HOMO) energy of the donor, so  
\ben
\omega_q =\epsilon^A_L -\epsilon^D_H = I^D - A^A + \Delta\xc^A
\label{eq:exact_omq}
\een
where the second equality uses the fact that in DFT with the exact ground-state functional, the HOMO energy is equal to minus the ionization potential while the LUMO differs from minus the affinity by the derivative-discontinuity~\cite{GKKG00,PL83,SS83,P85b,AB85,PPLB82}, i.e.
\ben
I = -\epsilon_H \;, {\rm while}\; A = -(\epsilon_L + \Delta\xc) \;.
\label{eq:IA}
\een

The derivative-discontinuity expresses the fact that the ground-state energy as a function of particle number $M$ has a piecewise linear structure, with discontinuities in the slopes at integer values $N$: $\Delta = \left.\frac{\partial E}{\partial M}\right\vert_{M = N+ \delta} -  \left.\frac{\partial E}{\partial M}\right\vert_{M = N- \delta}  = I - A$. Recognizing that the partial derivatives on the right are equal to functional derivatives with respect to the density,  $\frac{\partial E}{\partial M}= \left.\frac{\delta E[n]}{\delta n(\br)}\right\vert_{M}$, and that all ground-state energy components are continuous except for the Kohn-Sham kinetic energy $T\s[n]$ and possibly the xc energy $E\xc[n]$, one deduces  that $\Delta\xc = \Delta -\Delta\s = \epsilon_H(N+1) -\epsilon_L(N)$, and this underlies the relations of $I$ and $A$  to $\epsilon_H$ and $\epsilon_L$ , respectively, in Eq.~(\ref{eq:IA}). $\Delta\xc$ plays a critical role in obtaining accurate band-gaps and band-structure from DFT and dissociation behaviour~\cite{PPLB82,SS83,P85b,AB85,SGVML96,CMY08,MCY08,YCM12}.
A useful note for later is that since the xc potential is the density-functional derivative of $E\xc[n]$, then a derivative-discontinuity in the latter imparts a spatially-uniform discontinuity in $v\xc[n](\br)$ at integer values of the number of electrons. Using the convention that $v\xc(N) = v\xc(N-\delta)$, then if a fractional number of electrons is added to the system, the ground-state xc potential jumps up uniformly by $\Delta\xc$: $v\xc(N+\delta) - v\xc(N)$; this in turn imparts a discontinuity on the xc kernel~\cite{MK05,HG12}.

Comparing Eq.~(\ref{eq:exact_omq}) and Eq.~(\ref{eq:spa}) with Eq.~(\ref{eq:omega_CT}) we see that the exact $f\Hxc^{qq}$ matrix element must provide the derivative-discontinuity correction to the affinity as well as the $-1/R$ tail. Within the SPA, we find that the exact  kernel matrix element in Eq.~(\ref{eq:fHxcmatrixelt}) must be
\ben
2f\Hxc^{qq} = \Delta\xc^A - 1/R
\een
in the large-separation limit. 
 Because the donor and acceptor orbitals have an overlap that decays exponentially with donor-acceptor separation $R$, this means that the {\it exact kernel $f\Hxc[n_0](\br,\br',\omega_q)$ must diverge exponentially with inter-fragment separation $R$} (Note, this need not diverge with $\vert\br - \br'\vert$ but, rather, diverges with $R$, the information about which is contained in the density-functional dependence).  
 
 The question then naturally arises: which approximate kernels have the necessary divergence?
 We discuss below what happens with various often-used TDDFT approximations.
 
 \subsubsection{LDA/GGA (and corrections on top)}
 It is clear from the analysis above that with a  functional that depends only locally or semi-locally on the density, such as LDA or GGA, the $f\xc$ matrix element goes to zero exponentially in $R$, as the spatial-dependence of $f\xc(\br,\br')$ cannot compensate the exponential decays of the donor-acceptor orbital overlaps. Further, although the Hartree term has a long-range Coulombic tail, the donor-acceptor overlaps still kill off the integral in Eq.~(\ref{eq:fHxcmatrixelt}). One obtains then  $\Omega^{\rm LDA/GGA} = \epsilon_L^A -\epsilon_H^D$, simply the Kohn-Sham orbital energy difference. 
 
 Not only are the derivative-discontinuity and $-1/R$ tail missing, but moreover $\epsilon_H^D$ within LDA/GGA is a severe underestimate of the true ionization energy (and also underestimates the LDA/GGA ionization energy as computed from total energy differences). This underestimate is because the xc potential falls off exponentially away from the atom since it is proportional to the local electronic density and gradients, instead of as $-1/r$ as the exact potential does. The valence levels get pushed upward by the too-rapid fall-off, and hence the LDA/GGA HOMO eigenvalue is too small~\cite{P85b}. 
 (It is in fact possible to extract derivative discontinuities from LDA and GGA using ensemble theory, as has been pointed out in Refs.~\cite{KK13,G15}, but this is not how these functionals are usually used.)
 
 This what gives rise to the notorious underestimation of charge-transfer excitation energies in TDDFT when the traditional functionals are used~\cite{DWH03,T03,DH04,GB04c}. Tozer showed that the error in these functionals tends to the average of the derivative-discontinuities of the donor and acceptor in the limit of large separation, $\Omega^{\rm LDA/GGA}  - \omega^{\rm exact} \approx -\frac{1}{2}(\Delta\xc^D + \Delta\xc^A)$. 
 
An early correction was presented in Ref.~\cite{DWH03} (which was one of the first papers to point out the severe underestimation LDA/GGA yield). Here, configuration-interaction singles (CIS), which captures the $-1/R$ asymptotic behavior, 
is used to simply shift the local/semilocal DFT values. (CIS on its own, tends towards the HF orbital energy difference, which is generally much too large~\cite{S11}).
  Ref.~\cite{GB04c} designed a kernel to be applied for charge-transfer excitations, that switches on an asymptotic correction to ALDA when the diagonal coupling matrix of Eq.~\ref{eq:fHxcmatrixelt} vanishes.

\subsubsection{Exact Exchange}
Although the required diverging property of the xc kernel may seem tortuous, it is in fact contained in one of the fundamental approximations in TDDFT, time-dependent exact-exchange (TDEXX). This was first shown by Ipatov, He{\ss}elmann, and G\"orling~\cite{HIG09,GIHG09} by analyzing the structure of the exact-exchange kernel in terms of the occupied and unoccupied Kohn-Sham orbitals and eigenvalues~\cite{G98}.  Hellgren and Gross~\cite{HG12,HG13} later showed this from the many-body stand-point, and we will outline their argument here. TDEXX is perhaps the first approximation in TDDFT one might consider, given that it arises in first-order perturbation theory in terms of the interaction, when the perturbation is done within the Kohn-Sham framework, i.e. maintaining the same density at each order in the perturbation. This is known as G\"orling-Levy perturbation theory~\cite{GL93}, and Ref.~\cite{GS99} showed that, at least within the single-pole approximation, excitation energies obtained from TDEXX are equal to those from first-order G\"orling-Levy perturbation theory. That is~\cite{GS99},
\bea
\nonumber
2f\x^{qq}(\omega_q) &=& \langle\phi_a\vert\hat{\Sigma}_x - v\x\vert\phi_a\rangle - \langle\phi_i\vert\hat{\Sigma}_x - v\x\vert\phi_i\rangle \\ 
&-& \int d^3r d^3r' \frac{\vert\phi_a(\br)\vert^2\vert\phi_i(\br')\vert^2}{\vert\br - \br'\vert}\;.
\label{eq:tdexx_spa}
\eea
where $\hat{\Sigma}_x = \hat{v\x}^{\rm HF}$ is the non-local Fock operator. 
This gives exactly the desired exchange correction to the Kohn-Sham charge-transfer excitation energy: Taking $\phi_i = \phi_H^D$ and $\phi_a = \phi_L^A$, the first term gives $\Delta\x$, the second term vanishes~\cite{GKKG00}, and the third term is an electrostatic energy that goes like $-1/R$ in the limit of large separation. Hence, within SPA,
\bea
\Omega^{\rm TDEXX}  = \epsilon_L^A  -\epsilon_H^D  + \Delta\x^A - 1/R
\eea
in the large-separation limit, and is expected to be close to the full TDEXX result. In weakly-correlated systems, $\Delta\x$ can be a good approximation to the full derivative discontinuity $\Delta\xc$. 

It is important to realise that the frequency-dependence of the kernel is critical here. The exact-exchange kernel must be evaluated at the Kohn-Sham charge-transfer energy difference to yield this correction.  If instead $f\x^{qq}(\omega = 0)$ is used, as in adiabatic exact exchange (AEXX), the correction vanishes as $R \to \infty$. One finds that the AEXX kernel yields a correction to the Kohn-Sham excitation energy at small and intermediate fragment separations but that it falls away to zero for larger separations. 

Ref.~\cite{HG12} argued that the underlying feature of the full non-adiabatic exact-exchange kernel that allows the finite correction is the derivative-discontinuity of the kernel with respect to the particle number: Just as the ground-state xc potential has a discontinuity if one adds a fractional number of electrons to the system as discussed above, the time-dependent xc potential does too~\cite{MK05}, and so its functional-derivative $f\xc(\br,\br',\omega)$ as well has a discontinuity and this is strongly frequency-dependent.

The lack of the particle-number discontinuity in a functional has been argued to be closely related to its self-interaction error~\cite{P90}: this is the violation of the fact that for one-electron systems, the exchange-energy must exactly cancel the Hartree energy and the correlation energy must be zero.
 The self-interaction-corrected LDA (SIC-LDA), applied within a generalized TD optimized effective potential scheme~\cite{KKM08}, has been shown to produce reasonable charge-transfer energies~\cite{HKK12}.  The computational effort required has prevented this method from being more extensively explored.

\subsubsection{Global hybrids}
Global hybrid functionals combine a uniform fraction of Hartree-Fock exchange with a local or semilocal xc functional so the ground-state xc energy has the form: 
\ben
E\xc^{\rm hyb} = a\x E\x^{\rm HF} + (1-a\x) E\x^{\rm LDA/GGA} + E\c^{\rm LDA/GGA}
\label{eq:hybrid}
\een
where parameter $a\x$ is typically chosen as $\frac{1}{4}$. The B3-LYP hybrid functional~\cite{B93,LYP88}, popular for molecules, involves three parameters and can be written in the form
\bea
\nonumber
E\xc^{\rm B3-LYP}& =& a_0 E\x^{\rm HF} +  a\x E\x^{\rm GGA} + (1 -a_0 - a\x)E\x^{\rm LDA}\\
&+& a_c E\c^{\rm GGA} + (1 - a\c) E\c^{\rm LDA}
\label{eq:b3lyp}
\eea
The $E\x^{\rm HF}$ is the Hartree-Fock exchange energy expression evaluated on Kohn-Sham orbitals that are obtained via self-consistent calculations using Eq.~(\ref{eq:hybrid}) or Eq.~(\ref{eq:b3lyp}). The effective potential determining the ground-state Kohn-Sham Slater determinant, is therefore non-local and orbital-dependent, as it includes a fraction of the Hartree-Fock potential along with a one-body term.   In this sense, it steps out of the original Kohn-Sham framework, but instead fits into what is known as a "generalized Kohn-Sham" framework~\cite{SGVML96, YCM12, LSWS97,Savin96,GL97}. 
A significant advantage is that the inclusion of a fraction $a\x$ of Hartree-Fock exchange means that the LUMO eigenvalue includes this fraction of the exchange-contribution to the discontinuity~\cite{SGVML96}.
Also, with hybrids, the HOMO eigenvalue is a better approximation to the ionization energy than in LDA/GGA, because the Kohn-Sham potential seen by the hybrid falls away as $-a\x/R$ instead of exponentially, closer to the exact $-1/R$ behavior, and so it does not push up the HOMO level as much. 

It is important to appreciate that even with $a\x=1$ and $E\c$ taken to zero, i.e. Hartree-Fock, this is {\it not} equivalent to TDEXX previously discussed.  TDEXX still operates within the original Kohn-Sham framework (local one-body potential), and unoccupied eigenvalues represent excitations of the neutral system as their orbital equation sees an $N$-electron system, while in Hartree-Fock the unoccupied eigenvalues represent approximate electron addition energy levels (affinities), since their orbital equation sees an $(N+1)-$electron system. The latter point underlies why the LUMO-value in hybrid functionals includes a fraction of the derivative discontinuity $a\Delta\x$.  

So, even at the level of {\it bare} (generalized) Kohn-Sham energy eigenvalue differences, a hybrid will reduce the underestimation of the charge-transfer energies.  Further, the $f\xc$ correction is modified to include the Fock-exchange~\cite{TC03}, giving, for functional of Eq.~(\ref{eq:hybrid}), within the single-pole approximation (written for GGA to avoid clutter),
\bea
\nonumber
&2 f\xc^{qq \; {\rm hyb}}(\omega_q) = \int d^3r d^3r' \phi_i(\br)\phi_a(\br)\left(f\c^{\rm GGA} +\right.&\\
\nonumber
& \left. (1 - a\x)f\x^{\rm GGA}\right)\phi_i(\br')\phi_a(\br') + a\x\int d^3r \int d^3r'\frac{\vert \phi_i(\br)\vert^2\vert\phi_a(\br')\vert^2 }{\vert \br - \br'\vert}&
 \label{eq:fxchybrid}
\eea
This equation shows that, when applied to the orbitals in the charge-transfer excitation,  the asymptotic $R$-dependence of the charge-transfer excitation energy is partially captured as $-a\x/R$ due to the last term; as in the LDA/GGA case, the other terms vanish due to the exponentially small donor-acceptor overlap, uncompensated by the LDA/GGA kernel.  Unless one mixes in 100\% Hartree-Fock exchange, the full $-1/R$ tail will not be captured. TDDFT with a hybrid functional thus reduces the LDA/GGA underestimation of the charge-transfer excitation energy by including a fraction of the derivative discontinuity in the  orbital energy difference, and incorporating a scaled tail.

In fact the calculations of Ref.~\cite{RF04} that enlightened the dual fluorescence observed in DMABN (Figure~\ref{fig:dmabn}) used B3-LYP. The charge-transfer energies were not the focus of the study, and were likely underestimated (although not as much as if a GGA was used). Actually the calculations were performed in the gas phase while the phenomenon occurs in solvent which would lower the excitation energy, so the discrepancy in the electronic excitation energy does not appear as bad as it would if compared with experimental values in the gas phase. Excited state properties such as vibrational frequencies and force constants appear to be generally less sensitive and very close to experiment in this case, which enabled the mystery of the nature of the charge-transfer state and dual fluorescence mechanism to be solved.

Double-hybrid functionals that further mix in a perturbative second-order correlation part to the GGA for correlation have also been explored for charge-transfer excitations~\cite{GN07}. Highly-parameterized functionals such as the M06-HF meta-GGA hybrid~\cite{ZT06} that includes 100\% Hartree-Fock exchange, and MN15~\cite{YHLT15} have also been developed, using more complicated functional forms with many parameters fit to datasets.

\subsubsection{Range-separated hybrids (RSH)}
The idea of developing a DFT around a range-separated interaction dates back to the mid-eighties~\cite{Savin96,SS85,LSWS97} with the motivation to capture short-range dynamical correlation using local or semi-local DFT while using wavefunction methods for long-range effects. Such functionals sometimes have an LC preface, for "long-range corrected". One splits the Coulomb interaction into a long-range and short-range term, for example
\ben
\frac{1}{\vert \br_1 - \br_2\vert} = \frac{\rm{erf}(\gamma \vert \br_1- \br_2\vert)}{\vert \br_1 - \br_2\vert}  + \frac{1-\rm{erf}(\gamma \vert \br_1- \br_2\vert)}{\vert \br_1 - \br_2\vert} 
\label{eq:RSH}
\een
where $\gamma$ is a "range-separation parameter". 
\begin{figure}[h]
\begin{center}
\includegraphics[width=0.48\textwidth,height=0.35\textwidth]{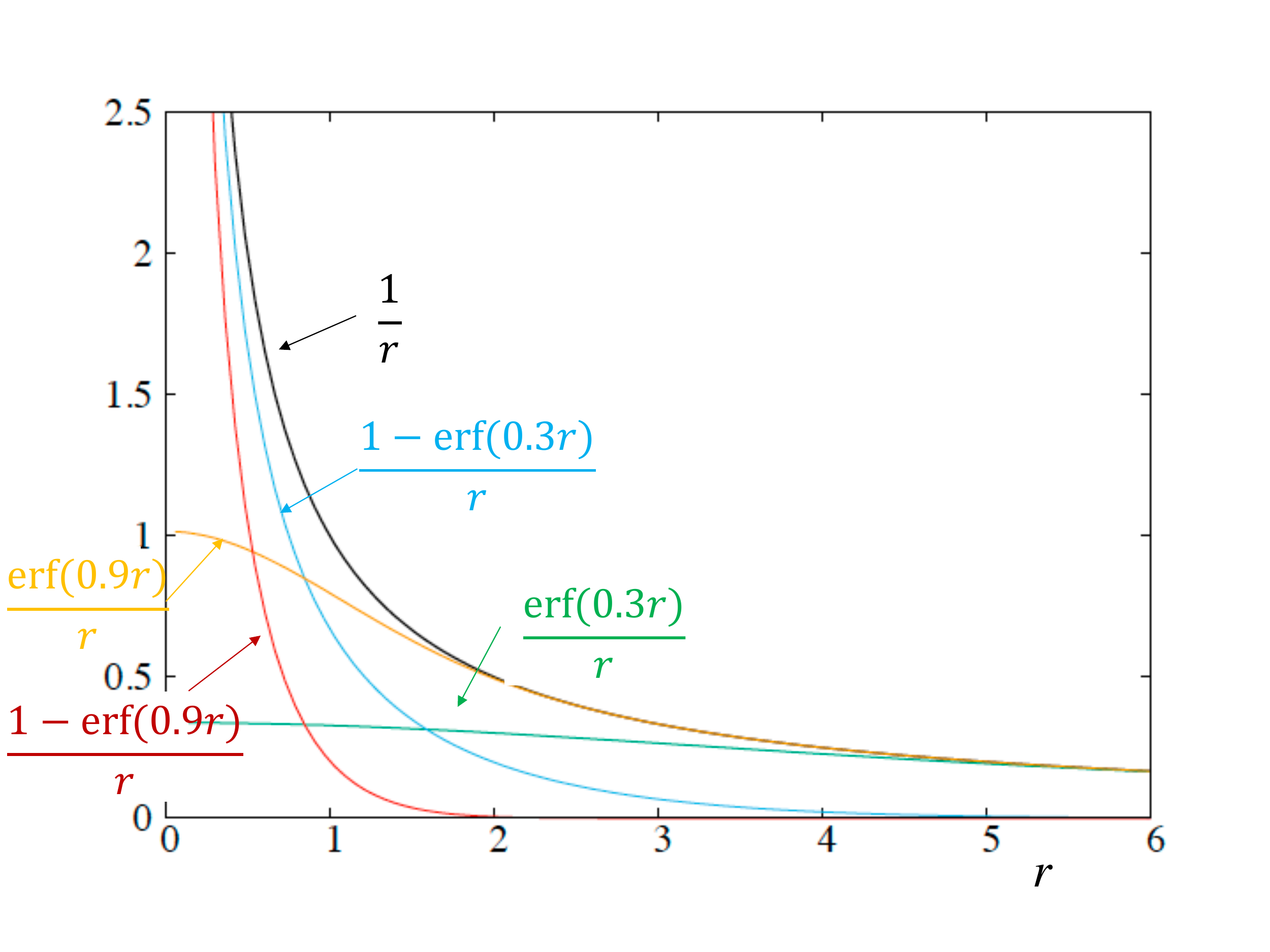}
\end{center}
\caption{Illustration of the effect of range-separation. The Coulomb repulsion $1/r$ is shown in black, while its range-separated form of Eq.~(\ref{eq:RSH}) is shown for two parameters: $\gamma = 0.3$ (blue and green) and $\gamma = 0.9$ (orange and red). }
\label{fig:RS}
\end{figure}

The first term on the right survives at long-range but is killed off for small $(\br_1 - \br_2)$ while the second term picks up at short range and dies off at long range (See Fig.~\ref{fig:RS}), and $\gamma$ determines the "range" of the short/long parts; the larger the $\gamma$ the faster the short-range part vanishes. 
The idea then is to use LDA/GGA exchange for the second term, which, at short-range, benefits from the dynamical correlation contained in DFT and error cancellation between exchange and correlation, while using Hartree-Fock for the long-range interaction that is poorly captured by local/semilocal approximations but where exchange dominates. The parameter $\gamma$ determines how much Hartree-Fock gets incorporated;  larger $\gamma$ means more Hartree-Fock is built in.   Usually correlation is treated semi-locally without range-separation. 
More general forms have also been explored, e.g. including some Hartree-Fock at both long and short-range (see shortly, and e.g. Ref.~\cite{HJS08} for a review),  range-separating the correlation also. As in the case of global hybrids, the approach formally lies within the generalized Kohn-Sham framework~\cite{KSRB12}. 

Tawada et al. in Ref.~\cite{TTYY04} realized that RSH's would yield the full correct asymptotic $-1/R$-dependence for charge-transfer excitation energies and, similar to global hybrids, would provide a discontinuity correction to the LUMO orbital energy as well as lowering the HOMO energy compared to the local/semilocal value. They also found this functional improved Rydberg excitation energies and oscillator strengths. In fact, the excitation energies in the DMABN example discussed earlier and in Fig.\ref{fig:dmabn} were revisited with a range-separated functional in Ref.~\cite{CTH07}. Yanai et al. combined the idea of the range-separated hybrid with B3LYP~\cite{ITYH01,YTH04} in order to incorporate the good performance for atomization energies of B3LYP, developing the "Coulomb-Attenuated Method"-B3LYP (CAM-B3LYP), which is based on the 3-parameter separation:
\ben
\frac{1}{\vert \br_1 - \br_2\vert} = \frac{\alpha +\beta\rm{erf}(\gamma \vert \br_1- \br_2\vert)}{\vert \br_1 - \br_2\vert}  + \frac{1-\alpha - \beta\rm{erf}(\gamma \vert \br_1- \br_2\vert)}{\vert \br_1 - \br_2\vert} 
\label{eq:RSH3}
\een
where $0\leq \alpha+\beta \leq 1, 0\leq \alpha,\beta \leq 1$. Taking $\beta = 0$ and $\alpha$ non-zero reduces to a global hybrid, while taking $\alpha=0$ and $\beta=1$  yields the original RSH, but having both non-zero allows to include a (non-uniform) fraction of Hartree-Fock exchange everywhere. Only when the parameters are such that the long-range portion is not scaled, will the correct $-1/R$ be reproduced. Sometimes, charge-transfer excitations are improved at the expense of worsening ground-state properties or local valence excitations, and the choice of the parameters can heavily sway results in different directions. Several other long-range corrected functional forms have been explored e.g.~\cite{RMH09,HJS08b} that mitigate these problems.

These methods clearly involve empirical parameters, usually fit to some training set, and then applied in a system-independent way. For example, Ref.~\cite{TTYY04} used a value of $\gamma = 0.33$, chosen from fitting to equilibrium dimer distances. 
 Typically $\gamma$ assumes a value in the range of $0.3-0.5 a_0^{-1}$, but the "best" value for $\gamma$ can have a strong dependence on the system, and so sometimes a global hybrid is more reliable. It has also been argued that in fact $\gamma$ itself is a functional of the density~\cite{BN05, BLN06} but when this is done, size-consistency is violated. What has become the most popular way to use an RSH without empiricism was pioneered by Baer, Kronik, and co-workers, and  where the value of $\gamma$ is tuned to satisfy an exact condition for the system of interest~\cite{BLS10,SKB09,KSRB12,KB14}. In what is now known as "optimally-tuned RSH", $\gamma$ is chosen to minimize 
\ben
\sum_{i=D(N_D),A(N_A+1)}\vert\epsilon_{H,i}^\gamma + E_i^\gamma(N_i-1) - E_i^\gamma(N_i)\vert\;,
\een
self-consistently fitting $\epsilon_H^D(N_D)$ to the ionization-potential $I_D$ as determined by energy-difference calculations for the same functional  {\it and} $\epsilon_L^A(N_A) = \epsilon_H(N_A+1)$ to the electron-affinity $A_A$. When $\gamma$ is tuned in this way, the RSH is able to yield not only good charge-transfer excitations but also good valence excitations~\cite{EWRS14}. To improve the accuracy of higher charge-transfer excitations, a state-specific tuning has been suggested~\cite{MLMD15}.
It has been argued that tuned RSH functionals tend to reduce the occurrence of triplet instabilities and generally improve triplet excitation energies~\cite{SKZB11,PWT11,KB14} but it has also been shown that the tuning procedure leads to erratic zig-zagging potential energy surfaces for triplets and singlets which do not appear when the range-separation parameter is fixed~\cite{KKK13}; also ground-state properties and binding energies can be significantly in error. 
This, and other problems, in particular size-consistency violation~\cite{KKK13,KB14} arising from system-dependence of the tuned range-separation parameter and hence of the
functional means caution must be applied for general usage.

\subsection{Charge-transfer between open-shell fragments}
\label{sec:open}
There is a fundamental difference between the Kohn-Sham description of a diatomic molecule consisting of open-shell fragments, and that of a molecule consisting of closed-shell fragments. In the former, the HOMO orbital is delocalized over both fragments, doubly-occupied by an up- and a down- spin electron, while in the latter the HOMO is localized on one of the fragments. This means that for the open-shell-fragment case, the Kohn-Sham wavefunction is fundamentally different than the true interacting wavefunction, which is close to a Heitler-London form. The Kohn-Sham ground-state is a Slater determinant, while in the open-shell-fragment case, the  interacting wavefunction requires minimally two Slater determinant states to describe it; this is referred to as static, or strong, correlation, and small fractions of other determinants also appear to account for dynamical correlation.
For example, for two electrons in a (fictitious) diatomic molecule consisting of open-shell atoms, the spatial part of the  exact wavefunction, in the limit of large separation, has the form $\Psi_{gs}(\br_1,\br_2) =\left( \phi_a(\br_1)\phi_b(\br_2) +\phi_b(\br_1)\phi_a(\br_2)  \right)/\sqrt{2}$, with $\phi_{a(b)}$ being the orbital on atom $a(b)$ respectively, while the Kohn-Sham wavefunction has the form $\Phi_{gs}(\br_1,\br_2) = \phi_0(\br_1)\phi_0(\br_2)$ with $\phi_0(\br) = \sqrt{(\phi_a^2(\br) +\phi^2_b(\br)/2}$. 
 In contrast, in the closed-shell fragment case, the interacting wavefunction can be well-approximated by a single Slater determinant (i.e. can be weakly-correlated), depending on the individual fragments, but certainly strong-correlation is not introduced by the presence of the other fragment, unlike in the open-shell-fragment case. 

The static correlation in the Kohn-Sham system makes the analysis of charge-transfer excitations very different than in the previous section in several regards: the HOMO and LUMO are both delocalized over the whole long-range molecule with substantial overlap, their orbital energy difference tends to zero as the molecule stretches, and any single-excitation out of the HOMO is near-degenerate with a double-excitation where the other electron occupying the HOMO transits to the LUMO. In the following we summarize the analysis of Refs.~\cite{M05c,MT06} for this case. 

\begin{figure}[h]
\begin{center}
\includegraphics[width=0.45\textwidth,height=0.5\textwidth]{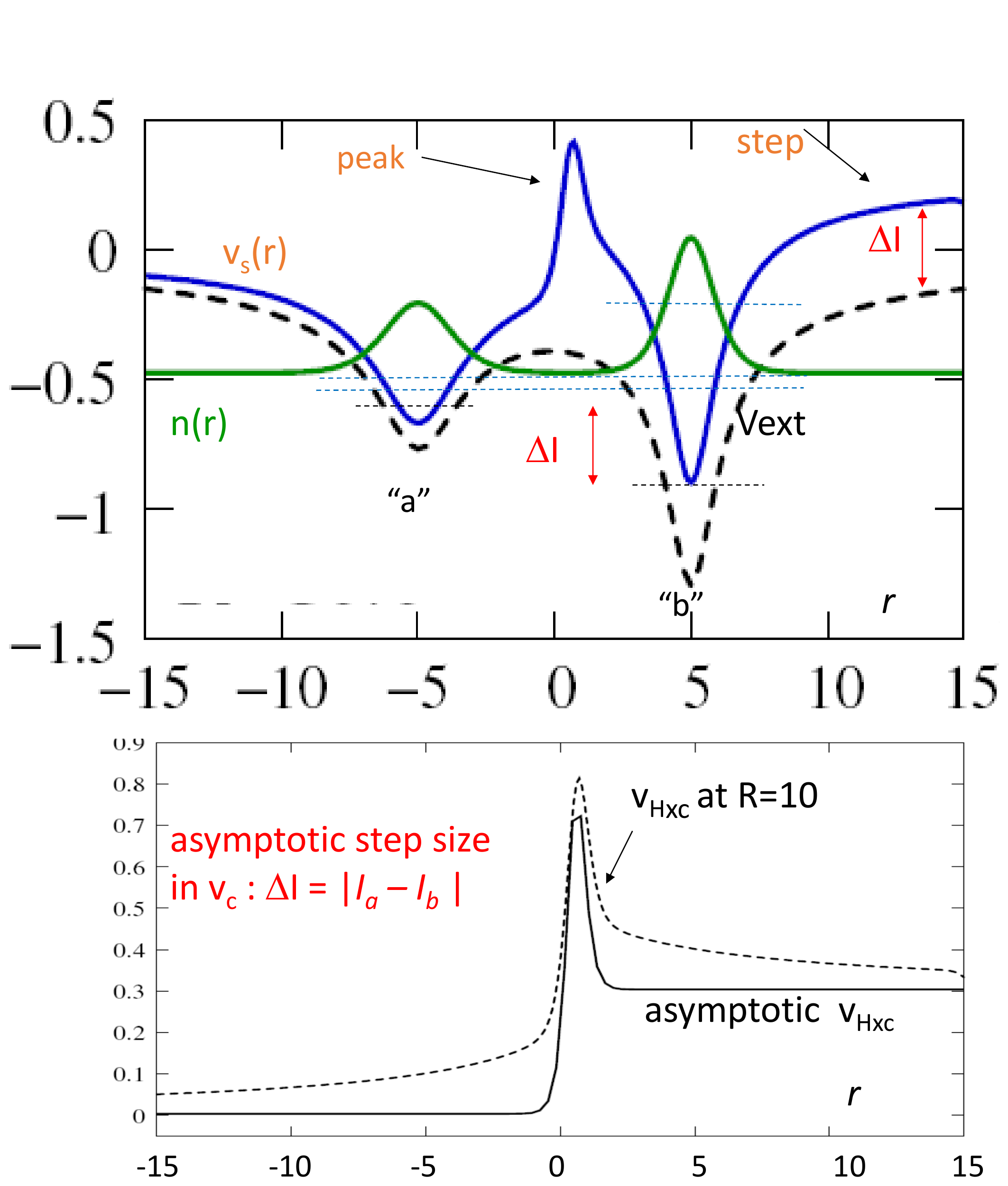}
\end{center}
\caption{Kohn-Sham potential of a model heteroatomic diatomic molecule (such as LiH).  Note that the Kohn-Sham potential (solid blue) does go back down to zero eventually on the right-hand-side (not shown); the external potential is shown as the black dashed line in the upper panel while the lower panel shows the $v\Hxc$ component at R = 10 and in the limit of infinite separation. It can be shown that the step structure appears in the correlation component $v\c$, and has a size $\Delta I$ that "re-aligns" the atomic HOMOs, whose energies are indicated as the thin horizontal black lines, in the limit of large separation. The lowest thin blue dashed horizontal line represents  the Kohn-Sham ground-state eigenvalue (of the bonding type orbital), while the one lying just above it is the lowest Kohn-Sham excitation which is to an antibonding type of orbital. The highest thin blue dashed horizontal line represents the Kohn-Sham energy of an excited state in the donor well; as discussed in the text, double-excitations are necessary for even a qualitatively correct description of both local and charge-transfer excitations in such a system.}
\label{fig:gs-step}
\end{figure}

At the heart of any TDDFT excitation energy calculation are the ground-state Kohn-Sham potential and the xc kernel. These both have unusual structure for a heteroatomic molecule consisting of open-shell fragments, arising from static correlation in the case of the exact functionals. The ground-state Kohn-Sham potential, in the limit of large separation, looks locally atomic-like near each atom up to a spatial constant: in the interatomic region a step and peak structure exists~\cite{AB85,P85b, GB96,TMM09,HTR09}, with the step size of $\Delta I = \vert I_a -I_b\vert$ in the limit of large separation; see Figure~\ref{fig:gs-step}.

One way to understand why this structure exists is as follows. First, within each atomic region, the HOMO orbital of the long-range molecule must reduce to  the respective atomic HOMO in order to produce the correct atomic density, so the potential in the Kohn-Sham equation reduces to the respective atomic Kohn-Sham potential  up to a uniform constant (plus corrections that go as $1/R$). Equating the HOMO eigenvalue of the molecule to the negative of the ionization potential, which, in the large-separation limit must be the smaller of the two ionization potentials of the individual fragment, one deduces that the uniform constant in the atomic regions must be such to pull the atom with the larger ionization potential up so that the individual atomic HOMO's then align, i.e. by the amount $\Delta I$. The step structure is in the correlation potential and prevents unphysical fractional charges in the dissociated limit; approximations like LDA/GGA that are unable to capture this feature suffer from the problem of dissociation into fractionally charged atoms instead of neutral ones (unless ensemble corrections are used with a slightly fractional electron number~\cite{KK15}). 
Ref.~\cite{TMM09} studied the interatomic-separation-dependence of the step and found that it marks the location of an avoided crossing between the ground and lowest charge-transfer excited potential energy surface, marking the point at which the molecule transitions from being a single system to two individual ones; the sharper the crossing, the more abrupt the transition from ionic to covalent character, and the sharper the step.  

What are the implications of the step structure for the xc kernel? The re-alignment of the atomic HOMO levels creates a near-degeneracy of the HOMO and LUMO orbitals of the molecule, which leads to a strong frequency-dependence of $f\xc$ throughout the frequency-range. 
The molecular HOMO and LUMO orbitals have an energy difference that falls away exponentially with interatomic separation $R$: as the limit of infinite separation is approached, $\phi_H(\br) \approx (\phi_a(\br) + \phi_b(\br))/\sqrt{2}$ has a bonding nature, while $\phi_L(\br) \approx (\phi_a(\br)  - \phi_b(\br))/\sqrt{2}$ has an antibonding nature, and their energy difference goes as the tunnel frequency through the step, $\omega^{H\to L}\s = \epsilon_L - \epsilon_L \sim e^{-cR}$. This means that in Eq.~(\ref{eq:sma}) the bare Kohn-Sham frequency goes to zero with an exponential dependence on the interatomic separation and it is up to the matrix element $f\Hxc^{qq}(\omega)$ to contain the {\it entire} charge-transfer energy, Eq.~(\ref{eq:omega_CT}).  The SPA Eq.~(\ref{eq:spa}) does not hold, since the  $f\Hxc$-correction term is much larger than the $\omega_q$; the "backward" Kohn-Sham excitation is just as important as the "forward" one since the excitation energy is so small. 

Before going further, it is worth pointing out why the HOMO-LUMO transition is actually the relevant one for the lowest charge-transfer excitation, given their delocalized nature. This can be seen by considering diagonalizing the full Hamiltonian in the minimal subspace given by the ground-state (double-occupation of $\phi_H$), a single-excitation to the LUMO, and a double-excitation to the LUMO. These three Kohn-Sham states are near degenerate, but the electron-electron interaction in the true Hamiltonian splits the degeneracy, and rotates the three to yield the Heitler-London ground-state, and two charge-transfer states $\phi_a(\br_1)\phi_a(\br_2)$, $\phi_b(\br_1)\phi_b(\br_2)$. In TDDFT, the true excitation energies are not obtained via diagonalization, but rather the effect of the mixing of the Kohn-Sham excitations is hidden in the xc kernel. Ref.~\cite{M05c} showed that by effectively solving for the xc kernel  in Eq.~(\ref{eq:dyson}) and keeping only the charge-transfer excitations of the exact system and HOMO-LUMO transition Kohn-Sham system in $\chi$ and $\chi\s$ respectively, 
\ben
f\Hxc^{qq}(\omega) = \frac{1}{\omega\s}\left(\delta^2 + \frac{\omega_1\omega_2 - \omega\s^2}{4} + \frac{\omega_1\omega_2\delta^2}{\omega^2-\omega_1\omega_2}\right)
\een
where $\omega_1 = I_b - A_a - 1/R$, $\omega_2 = I_a - A_b - 1/R$ are the charge-transfer excitation energies from fragment $b$ to fragment $a$ and vice-versa,  $\delta = (\omega_1 - \omega_2)/2$, and $\omega\s$  $\sim e^{-cR}$ is the Kohn-Sham HOMO-LUMO gap. This matrix element  would scare away any adiabatic approximation: not only is it is strongly frequency-dependent, but further it diverges exponentially with interfragment separation $R$. 
The exponential dependence on $R$ was a feature also of the transfer between two closed-shells (Sec.~\ref{sec:closed}), and also shows up in homo-atomic diatomic molecules such as H$_2$~\cite{GGGB00} where the symmetry disallows charge-transfer excitations (but one may speak of charge-transfer resonances). 

The frequency-dependent divergence with respect to separation was also seen in the exact-exchange kernel of Ref.~\cite{HG12} as discussed in the previous section, but here the effect is in the correlation potential. 
The strong frequency-dependence arises in this case because the charge-transfer excitations are a linear combination of single and double-excitations out of the Kohn-Sham ground state~\cite{MZCB04}: in fact {\it every} single excitation, be it local or charge-transfer, out of the HOMO is near-degenerate with a double-excitation where the other electron occupying the HOMO transits to the LUMO at cost exponentially small in $R$~\cite{MT06}. That this must be the case is evident from Figure~\ref{fig:gs-step} in order to avoid states which have "half" an electron excess or deficient on one atom. This endows the exact $f\xc$ with a strong frequency-dependence throughout. 

Of course approximate functionals do not typically yield Kohn-Sham potentials with the step (and as a result, yield unphysical fractional charges upon dissociation).  
Nevertheless, the HOMO and LUMO orbitals are still near-degenerate (see Fig. 13 in Ref.~\cite{TMM09} for LDA eigenvalues of LiH) so using an adiabatic kernel on top would still be problematic. A functional with explicit dependence on occupied and virtual orbitals inspired by density-matrix functional theory, was shown~\cite{GB06b,BB02} to capture the step structure.

 None of the functional approximations mentioned in Section~\ref{sec:closed} work for 
charge-transfer between open-shell fragments. Static, or strong, correlation is well known to be a difficult regime for density functional approximations in the ground-state, but its implication for excitation energies and response are clearly also very challenging. In the ground-state, spin-symmetry breaking, as occurs in a spin-unrestricted calculation beyond a critical interfragment separation, is sometimes a good resort to get good energies and was shown to also sometimes yield good charge-transfer excitations  but only in cases where the acceptor contains effectively one "active" electron~\cite{FRM11}, e.g. via pseudo potentials. 



\section{Charge-transfer dynamics}
\label{sec:CTdynamics}
As discussed in the introduction, for many applications a fully time-resolved description of the charge-transfer process is necessary, that goes beyond merely a calculation of the excitation spectrum: an electron transferring between regions in space hardly falls under a perturbation of the ground-state. The theorems of TDDFT certainly apply in the non-linear regime, and the dearth of alternative methods that are computationally feasible for systems of more than a few electrons piques the interest in TDDFT arguably more than in the linear response case. 

Of course one may make an expansion of the time-dependent electronic wavefunction in terms of its excited states,  obtaining these building blocks from linear-response theory, but ultimately to obtain the time-dependent coefficients in the expansion, properties not available from linear response theory are required.  When this process is driven solely by nuclear motion (sometimes called "charge separation"), Ehrenfest or surface-hopping methods are often used, in which the nuclei are evolved via classical equations of motion either on a mean-field surface (Ehrenfest), or on a Born-Oppenheimer surface stochastically switching between them (surface-hopping). Non-adiabatic couplings between the ground and excited states can be extracted from linear response theory, but those between excited states cannot be.  They are available from quadratic response theory but recently it was shown that the adiabatic approximation yields unphysical divergences when the difference between the energies of the two excited states equals a ground-to-excited excitation energy~\cite{LL14,LSL14,ZH15,OBFS15,PRF16}.

Ehrenfest dynamics with TDDFT was used to simulate the charge-separation dynamics in a prototypical light-harvesting molecule, the caretenoid-porphyrin-C$_{60}$ triad~\cite{RFSR13}, and in a conjugated polymer-fullerene blend~\cite{FRBM14}. The effect of triad conformation (bent versus linear) that occur in solvents, has been studied in Ref.~\cite{MBCD15}. Supramolecular assemblies of donor-acceptor dyads and triads are often considered as a model for the photosynthetic process of photoexcitation followed by charge-transfer between the components. 
The authors of Ref.~\cite{RFSR13} stressed the relevance of coherent coupled electronic and vibronic motion in driving the dynamics; the time-scales gave quite good agreement with experimental time-scales in this case, with an oscillatory electron transfer behavior  of about the same period as the dominant carbon backbone vibration, although the simulated transfer fell shy of one electron. The LDA functional was used, but somehow the typical large underestimation of the charge-transfer energy did not impact much the performance in the dynamics, so the accuracy obtained is yet to be fully understood. In general, surface-hopping tends to be preferred over Ehrenfest dynamics in modelling electron-nuclear dynamics in photo-induced dynamics~\cite{CDP05,TTR07,PDP09,CRT13}, because of its ability to describe branching of the nuclear wavepacket and non-radiative relaxation. An analysis of the subtle issues associated with developing consistent mixed quantum-classical dynamics schemes is beyond the scope of the present review, and we focus here only on aspects directly related to charge-transfer. Metal-ligand charge-transfer complexes in solution have been studied in surface-hopping, but, despite the name, these are quite short-range and the overlap between the donor and acceptor orbitals is appreciable~\cite{T15,TCR11} so GGAs gave reasonably good results. Photo-induced proton-coupled electron transfer processes in photocatalytic activity on small TiO$_2$ clusters has also successfully been studied, using global hybrids~\cite{MPBLZF17}, shedding light on the catalytic rates and mechanisms. 
In general, the underestimated energies of TDDFT  for  charge-transfer excitations over long range in large systems using standard xc approximations would create havoc in surface-hopping, as the potential energy curves will have incorrect slopes and relative gaps, yielding erroneous nuclear dynamics. Using range-separated hybrids for coupled dynamics gets rapidly too expensive; moreover, the functional must treat both local and charge-transfer excitations accurately to get the coupled dynamics correct.

\begin{figure}[h]
\begin{center}
\includegraphics[width=0.5\textwidth,height=0.4\textwidth]{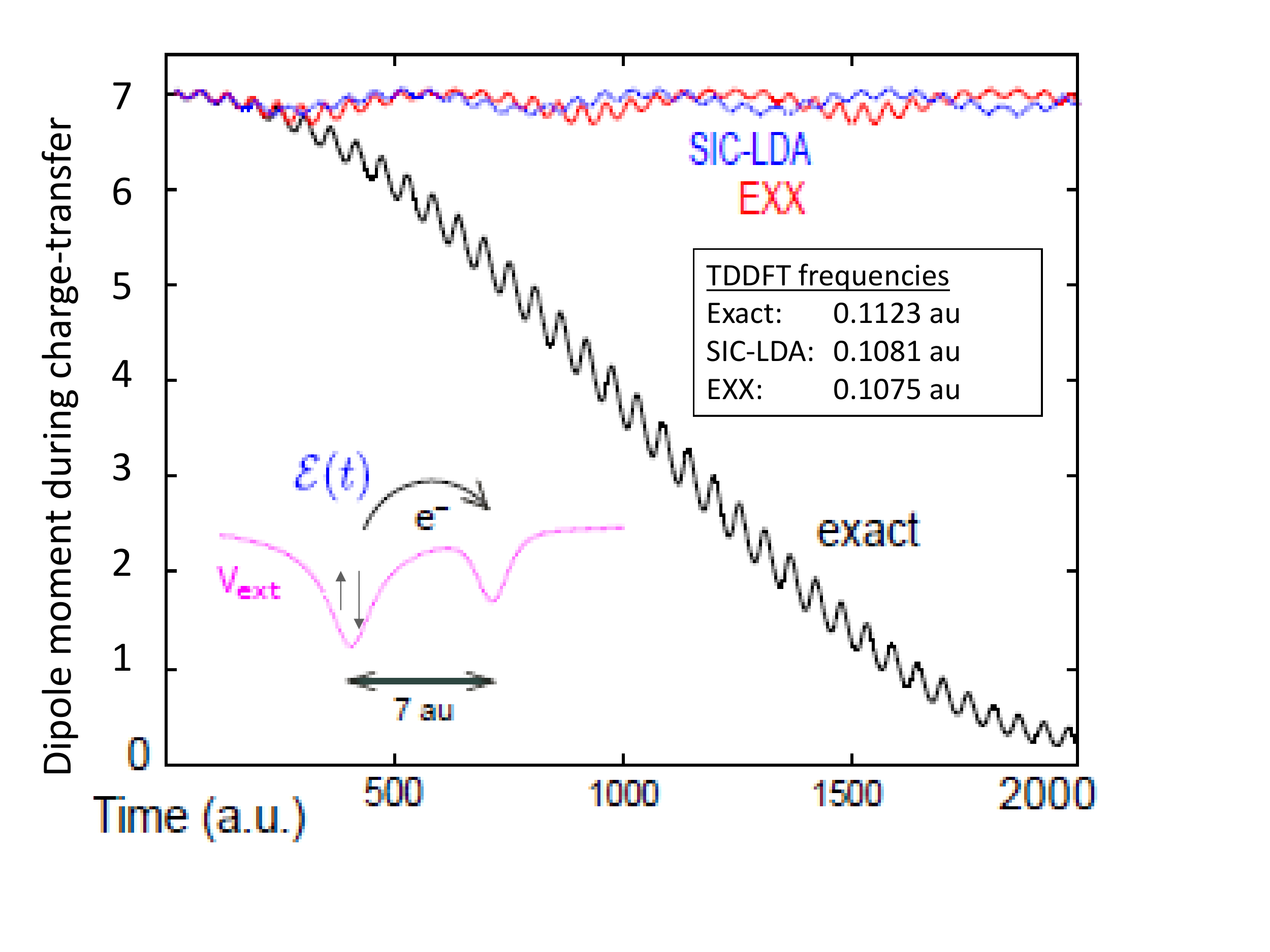}
\end{center}
\caption{Charge-transfer from a ground-state driven by a weak resonant field in a model 1D molecule whose potential is sketched on the lower left. (Details in Ref.~\cite{FERM13}). The TDDFT approximations shown yield reasonable linear response excitation frequencies compared with exact (inset) but propagating with them fails miserably. Reprinted with permission from Ref.~\cite{FERM13}, Copyright 2013, American Chemical Society.}
\label{fig:FERM}
\end{figure}

\begin{figure}[h]
\begin{center}
\includegraphics[width=0.5\textwidth,height=0.4\textwidth]{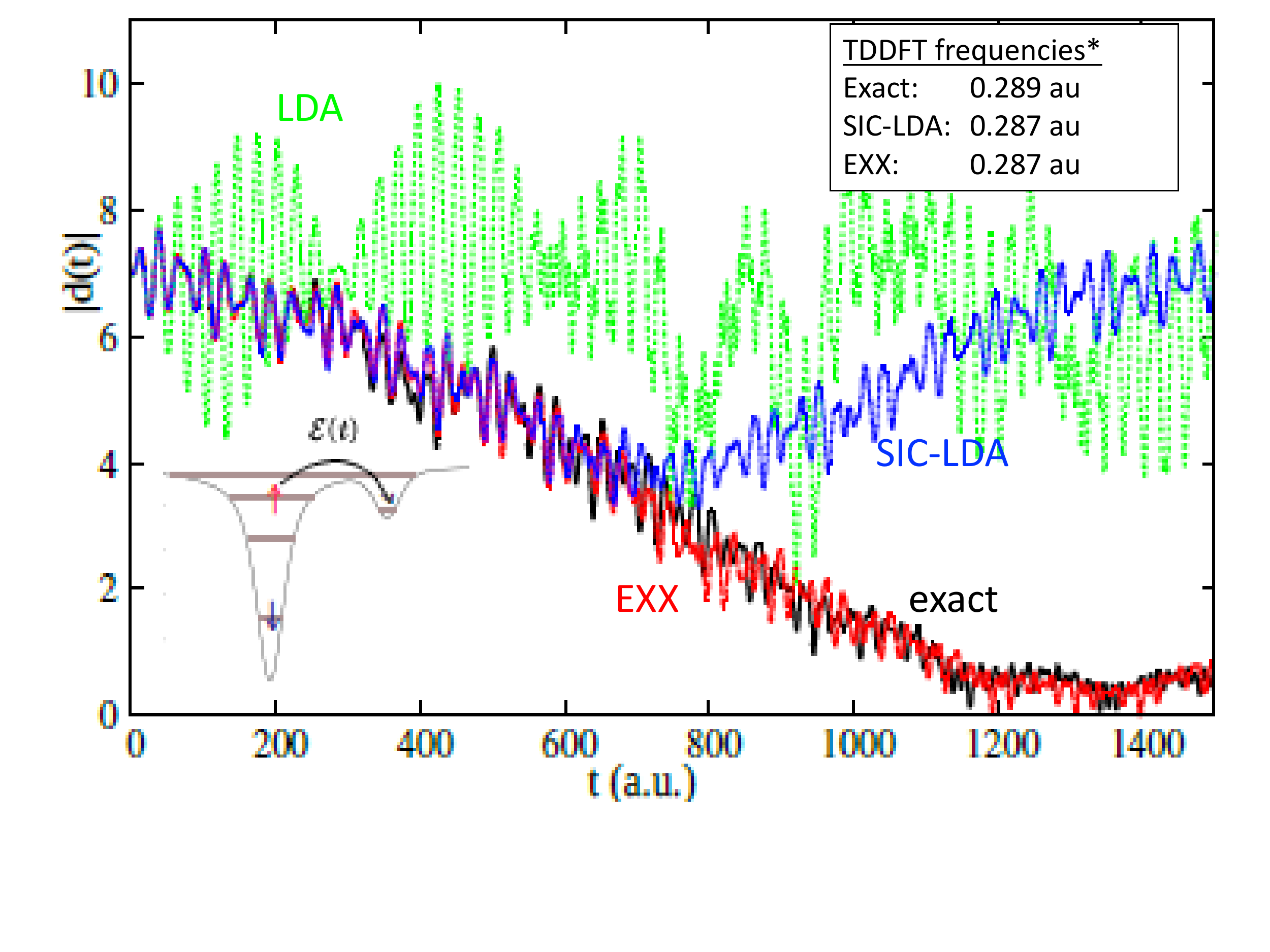}
\end{center}
\caption{Charge-transfer from an excited state driven by a weak resonant field in a model 1D molecule (Details in Ref.~\cite{FLSM15}). The TDDFT frequencies shown are those for the charge-transfer transition, as computed from the initially excited state (see Ref.~\cite{FLSM15}), and the calculations in this figure from the approximate functionals are computed at these corresponding frequencies. Despite being very accurate, only EXX is able to transfer the charge, due to the spurious changes in resonant frequencies of the other approximations during the dynamics. Reprinted figure with permission from Ref.~\cite{FLSM15}, Copyright 2015 by the American Physical Society.}
\label{fig:FLSM}
\end{figure}

Putting aside nuclei-mediated electron-transfer, we consider now electron-transfer driven by an external field. In some situations the electronic response to the field is fast enough that the nuclear motion can be neglected during the charge-transfer process. But even when it cannot be, it is still important to know whether the approximations in TDDFT are able to accurately reproduce the electronic dynamics. Will ensuring that the approximations yield good linear response properties mean that they also reproduce the time-resolved charge-transfer process well?
Studies in recent years have shown that unfortunately the answer to this question is no, and it appears that an xc potential functional with a non-adiabatic dependence on the density is needed. A dramatic example of the failure of functionals that give good charge-transfer excitation energies but poor dynamics is in Figure~\ref{fig:FERM}, while the prognosis is more subtle for charge-transfer starting in an excited state in Figure~\ref{fig:FLSM}. 



\subsection{Non-adiabatic step features}
\label{sec:nonadiab_step}
Consider first driving an electron out of the ground-state of a long-range molecule to an excited charge-transfer state.
If the molecule consists of open-shell fragments (e.g. LiH), then we expect from the outset that this is a challenge for approximate TDDFT functionals, given the problems with static correlation and the concomitant step in the initial potential and collapse in the charge-transfer excitation energy, discussed in Sec.~\ref{sec:open}. 
Instead, however, we might hope that if the molecule consists of closed-shell fragments, then the functionals that give reasonably good charge-transfer excitation energies may do a reasonable job with charge-transfer dynamics. This is unfortunately not the case, as evident in Fig.~\ref{fig:FERM}. The model system there is a simple double-well with two electrons, whose exact xc potential may be found and so allowing a detailed analysis to understand what is going on. 

 Before doing so, we point out that Ref.~\cite{RN11} showed a similar failure of approximate functionals to charge-transfer on a real molecule, LiCN. This molecule was studied in earlier calculations using time-dependent configuration interaction~\cite{KKS05} as a test system for laser-control of dipole-switching. In its ground-state the dipole moment of the molecule along its axis is relatively large, with ionic character at equilibrium geometry (Li$^+$CN$^-$).  A laser pulse is applied perpendicular to the molecular axis, with a frequency resonant with a molecular transition to an excited state  which has a relatively large perpendicular transition dipole with the ground-state, and itself has a dipole moment along the molecular axis quite different from, and opposite sign, to that of the ground state. (In fact there are two of these states, cylindrically symmetric around the axis). Hence one observes the dipole switch from one sign to the other when the resonant laser pulse is applied. Now, Ref.~\cite{RN11} asked whether the approximate functionals of TDDFT were able to reproduce this behavior, which would be very useful for single-molecule switch applications in molecular electronics, requiring larger molecules for which wavefunction methods are computationally unfeasible. The result however was very disappointing; the authors found that neither semi-local nor hybrid functionals were able to capture the dipole switch at all, and, not so dissimilar to the model system, the approximate functionals tracked the dipole for a short time, but finally simply oscillated near the initial value. 
 
\begin{figure}[h]
\begin{center}
\includegraphics[width=0.5\textwidth,height=0.6\textwidth]{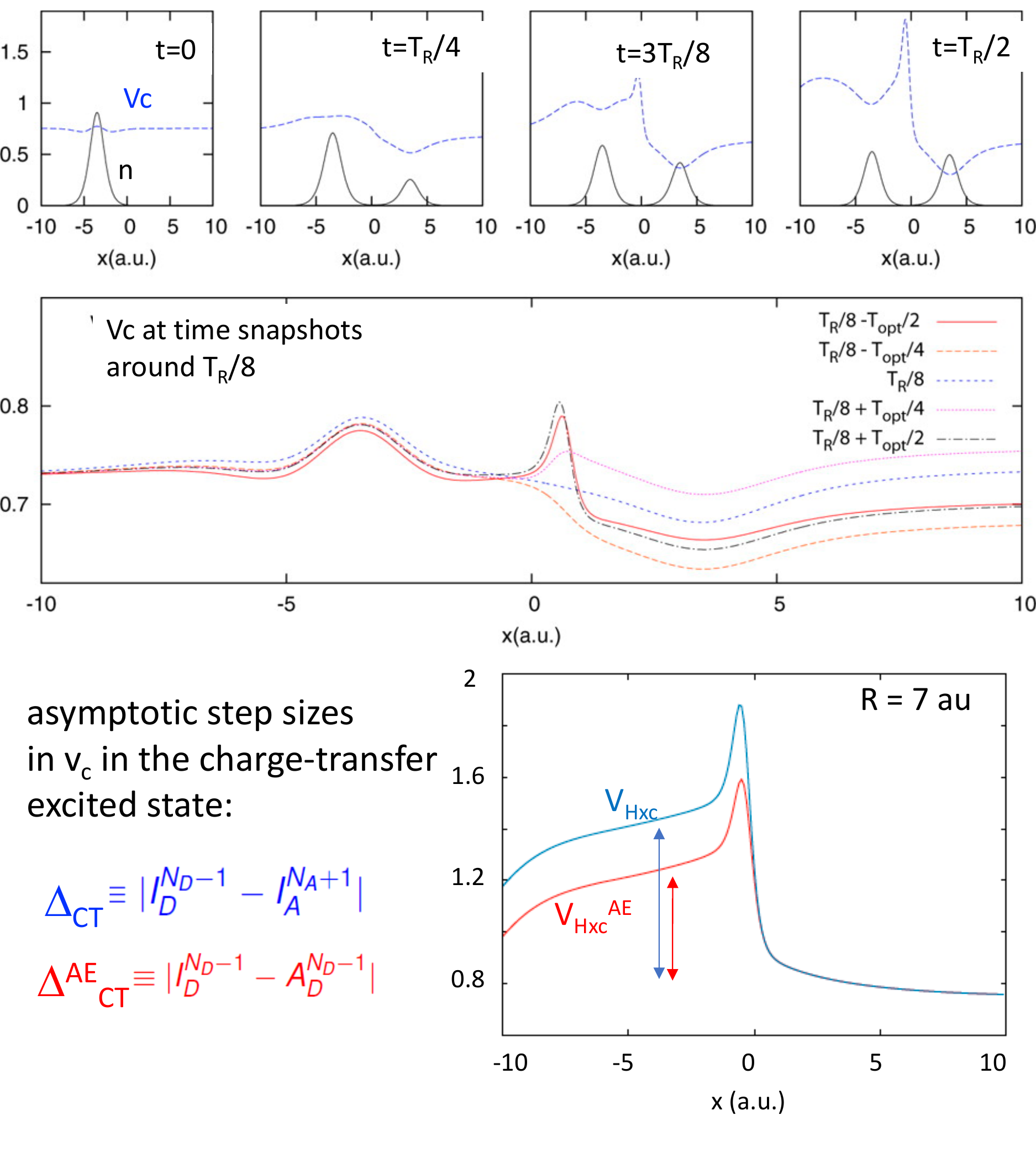}
\end{center}
\caption{Top panel: Exact correlation potential (dotted blue line) and density (black) in resonantly-driven charge-transfer in a model one-dimensional molecule. The model consists of two soft-Coulomb interacting electrons in a double well, with the centers of the wells $R = 7 au$ apart, where the ground-state places both electrons largely on the left well. A weak field resonant with the lowest charge-transfer excitation is applied, so that after half a Rabi cycle one electron has transferred to the right well, as evident in the density dynamics shown. A step associated with charge-transfer builds up in the correlation potential over time, and superimposed on this is a dynamical step that oscillates on the time-scale of the applied field (Middle plot, Reprinted with permission from Ref.~\cite{FERM13}, Copyright 2013, American Chemical Society.). The optical period, $T_{\rm opt} = 2\pi/\omega_{opt} = 56$au where $\omega_{\rm opt} =0.112$au, the frequency of the lowest charge-transfer excitation, while the Rabi period, $T_R= 4845$au. The lower plot isolates the exact Hartree-xc potential in the final charge-transfer excited state, and also the AE one, which is also able to build up a step feature but it has the wrong size; the equations on the left give the values of the two steps in the limit of infinite separation (compare with lower plot of Fig.~\ref{fig:gs-step} for ground-state step feature). }
\label{fig:ex-step}
\end{figure}

 Returning to the model system, we now ask what does the exact xc potential look like during the dynamics? As the system begins in the ground-state, it is most natural to also choose a ground-state for the Kohn-Sham initial state, which is uniquely given via ground-state DFT. The two Kohn-Sham electrons occupy the same spatial orbital, and, since the Kohn-Sham evolution is via a one-body potential, there is only ever one occupied orbital: $\Phi(x_1,x_2,t) = \phi(x_1,t)\phi(x_2,t)$. If the exact xc functional was used, then this orbital reproduces half the exact time-dependent density of the system: 
 \ben
 \vert\phi(x,t)\vert^2 = n(x,t)/2\;.
 \label{eq:phi-doublyocc}
 \een
Thinking about the true evolution, this fact immediately suggests a challenge for TDDFT in charge-transfer processes in closed-shell molecules: a single orbital must simultaneously describe an electron that gradually transfers over to another atom and one electron which stays in the donor! 
A consequence of this is that the exact correlation potential builds up an interatomic step over time, as evident in Figure~\ref{fig:ex-step}. How we obtained these exact potentials will be explained shortly. 

Although with some similarity to the step in the ground-state potential of a heteroatomic molecule discussed above (Fig.~\ref{fig:gs-step}), this step in the final charge-transfer state is, on the one hand, distinct in that it does {\it not} appear in the ground-state of the molecule (top left panel in Fig.~\ref{fig:ex-step}), and requires a non-adiabatic functional to fully capture -- the AE approximation underestimates its size, as shown in the lower panel. 
Arguments similar to those that were made for the size of the step in the ground-state potential of a molecule consisting of open-shell fragments (Section~\ref{sec:closed}) can be made here to determine the size of the
 step in the exact potential in the excited charge-transfer state in the closed-shell fragment case, yielding 
 \ben
\Delta_{\rm CT} = \vert I_D(N_D-1) - I_A(N_A+1)\vert
\label{eq:exact_CTstep}
\een
   in the large-separation limit. However, not even the exact ground-state functional can capture this step correctly as seen by the red curve in the figure. This is the AE potential, defined in Eq.~(\ref{eq:AE}). One can prove, in the limit of infinite separation, that the step in the AE potential tends to the derivative discontinuity of the $(N_D-1)$-electron donor cation, 
\ben
 \Delta^{\rm AE}_{\rm CT} = \vert I_D(N_D - 1) - A_D(N_D - 1)\vert
 \label{eq:adia_CTstep}
\een
The argument goes as follows~\cite{FERM13}. Consider the terms on the right-hand-side of Eq.~(\ref{eq:AE}) for the system in its charge-transfer excited state; assume that somehow the state is reached completely by some dynamics from the ground-state, e.g. a weak resonant laser field, which is then turned off, leaving the system in this stationary excited state. The first term, $v\s^{\rm ex. gs}$, is the potential in which two non-interacting electrons have a ground-state whose density equals that of the charge-transfer excited state, $n_{\rm CT}^*$, while the second term $v\ext^{\rm ex. gs}$ is the potential in which two interacting electrons have this density. Note that although the density is that of an excited state of the interacting system in the original potential $v\ext$, on the right-hand-side of Eq.~\ref{eq:AE} we are looking for the potentials in which it is a {\it ground-state} density (of a non-interacting and interacting system of two electrons, respectively); given that $\n_{\rm CT}^*$ has no nodes, such potentials are well-behaved~\cite{MB01}.  
From this, one determines that $v\s^{\rm ex. gs}$ must equal the exact KS potential, since the KS orbital must equal $\sqrt{n_{CT}^*(\br)/2}$, yet this is exactly what the exact orbital of Eq.~(\ref{eq:phi-doublyocc}) equals, once the charge-transfer state is fully reached. Therefore the step in the first term of Eq.~(\ref{eq:AE}) is exactly $\Delta_{\rm CT}$ of Eq.~(\ref{eq:exact_CTstep}). Although the KS state reached in the time-evolution is a ground-state of the KS potential that is reached, this is certainly not the case for the interacting system, which truly goes from a ground-state to an excited state during the charge-transfer process. The potential $v\ext^{\rm ex. gs}$ must equal the atomic potentials in the respective atomic regions, because the interacting ground-state wavefunction that has density $n_{\rm CT}^*$ has a Heitler-London form, with orbitals locally satisfying the respective atomic Schr\"odinger equation. But $v\ext^{\rm ex. gs}$ cannot be simply the sum of the atomic potentials, i.e. the exact $v\ext$, because the ground-state of that potential has two electrons in the donor well. Instead it raises the donor well up by the smallest constant such that the ground-state of the molecule has one electron in each well. One can deduce this constant has a value of $I_D(N_D) - I_A(N_A +1)$~\cite{FERM13}, and so, putting the potentials in the two terms together on the right-hand-side of Eq.~(\ref{eq:AE}), we see their steps subtract, and find the result of Eq.~(\ref{eq:adia_CTstep}). 
 
Therefore the best adiabatic approximation fails to fully capture the step in the final charge-transfer state, and most adiabatic approximations lack the spatial non-locality required to capture any kind of step. However the charge-transfer excited state is not even reached in the dynamics, as is clear from Fig.~\ref{fig:FERM} for approximate adiabatic functionals, and we will see shortly also for propagation under the AE xc potential. One finds that the time-dependent adiabatic xc potential differs significantly from the exact one, which displays not just the step associated with charge-transfer process but also a dynamical step that appears generically in dynamics~\cite{EFRM12,FERM13,RG12,M16}. First, we explain how the exact time-dependent xc potential is found. 

 Finding the exact time-dependent exchange-correlation potential that reproduces a given density evolution $n(\br,t)$ is, in the general case, a complex numerical problem, but in the case of two electrons beginning in the ground state, it is relatively simple. Requiring the only occupied orbital $\phi(x,t)$ to reproduce the same density as the true system and its first time-derivative
 \bea
 v\s(\br,t) = \frac{\nabla^2\sqrt{n}}{2\sqrt{n}}  - \frac{(\nabla\alpha(\br,t))^2}{2} - \partial_t\alpha(\br,t)\;,\\
 {\rm where}\; \nabla\cdot(n(\br,t)\nabla\alpha(\br,t)) = -\partial_t n(\br,t)\;.
 \eea
 Then to obtain $v\xc$, one would subtract the external potential, and the Hartree and exchange potential (which for two electrons occupying the same orbital, is just $-v\H/2$)
 Alternatively, an expression directly for the xc potential may be used~\cite{L99,RB09,MB01,LFSEM14}:
 \bea\nonumber
\nabla\cdot\left(n\nabla v\xc\right) &=&\nabla \cdot \Big[ \frac{1}{4}\left(\nabla' - \nabla\right)\left(\nabla^2 - \nabla'^2\right) \rho_{1c}(\br',\br,t)\vert_{\br'=\br} \\&+& n(\br,t)\int n\xc(\br',\br,t) \nabla w(\vert \br'-\br \vert) d^3r'\Big] ,
\label{eq:3Dvxc}
\eea
where 
$\rho_{1c} =\rho_1 -\rho_{1,\sss{S}}$ with 
$\rho_1(\br',\br,t) = N\sum_{\sigma_1..\sigma_N} \!\!\int d^3r_2..d^3r_N \Psi^*(\br'\sigma_1 .. \br_N\sigma_N;t) \Psi(\br\sigma_1 .. \br_N\sigma_N;t)$
is the spin-summed one-body density-matrix of the true system of electrons with two-body interaction potential $w(\vert\br - \br'\vert)$, $\rho_{1,\sss{S}}(\br',\br,t)$ is the one-body density-matrix for the KS system,  and $n\xc(\br',\br,t)$ is the xc hole, defined via the pair density, 
$P(\br',\br,t) = N(N-1)\sum_{\sigma_1..\sigma_N}\int \vert \Psi(\br'\sigma_1,\br\sigma_2,\br_3\sigma_3..\br_N\sigma_N; t) \vert^2 d^3r_3..d^3r_N 
 = n(\br,t)\left(n(\br',t) +n\xc(\br',\br,t)\right)$\;.
Eq.~(\ref{eq:3Dvxc}) is a Sturm-Liouville equation for $v\xc$ that has 
a unique solution for a given  boundary
condition~\cite{L99}. 
 
 The potentials plotted in Fig.~\ref{fig:ex-step} used either of these expressions. For cases involving more than two electrons, or with two electrons in an initial Kohn-Sham state chosen to involve more than one orbital, an iterative method must be used to find the exact xc potential; Refs.~\cite{FNRM16,RPL15,NRL13,JW16,RG12,HRCLG13} contain examples. 

 Using these equations above, the exact xc potential was found during the charge-transfer induced by a weak field resonant with the charge-transfer excitation energy. Superimposed on the gradually developing step that is associated with the transfer of charge, a "dynamical" step appears, that oscillates with the field. This step appears generically in non-perturbative dynamics, in the presence of a field, or in field-free dynamics of a non-stationary state, and represents a type of time-dependent screening that cannot be captured by any adiabatic approximation~\cite{RG12,EFRM12,LFSEM14,M16}. In a decomposition of the exact potential into kinetic terms and interaction terms, the largest contribution to the step comes from the kinetic term. 
 
 \subsection{The best that an adiabatic approximation can do}
\begin{figure}[h]
\begin{center}
\includegraphics[width=0.5\textwidth,height=0.25\textwidth]{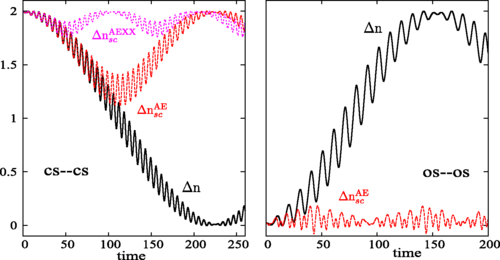}
\end{center}
\caption{Site-occupation number differences $\Delta n$ (analogous to "dipole moments") for asymmetric Hubbard dimers when driven with a weak "field" resonant with a charge-transfer excitation to the other site; the time is measured in units of 1/$U$, the Hubbard parameter. On the left, labeled "cs--cs", the charge-transfer occurs between two closed-shells, i.e. the parameters of the Hubbard dimer are such that almost 2 electrons are localized on the right-hand site in the initial ground-state, and the charge-transfer excitation has almost 1 electron on each site; the black curve shows the site-occupation number difference as the charge transfers. The magenta curve is the result when running with the adiabatic exact exchange (AEXX) approximation, not dissimilar to  the behavior in the real-space molecule of Fig.~\ref{fig:FERM}. The small Hilbert space of the Hubbard dimer allows a self-consistent calculation using the AE approximation (red) which gets a bit further, but ultimately fails also to transfer the charge. On the right panel, labeled "os--os", the Hubbard parameters are adjusted so that the ground-state has one electron on each site (hence two open-shell sites), while the charge-transfer is to a state with two electrons on the right site. In this case, the AE fails from very early on, consistent with the AE charge-transfer frequency being a significant overestimate of the actual. The details of the parameters and model are given in Ref.~\cite{FM14}. Reprinted figure with permission from Ref.~\cite{FM14}, Copyright 2015 by the American Physical Society.}
\label{fig:adia-ex}
\end{figure}

Although it completely misses the dynamical step, the AE approximation partially captures the step associated with the transfer of charge, as shown in the figures. This step is important for "holding" the charge in place once it has moved over, as is clear from the arguments in its derivation. Is this partial step enough to at least qualitatively capture the dynamics and transfer the charge? To answer this, we would need to propagate using the AE xc potential, not just to evaluate it on the exact density which was done in the previous figures and analysis. This is a very intensive process numerically, as one has to find at each time step, the ground-state potential of an interacting system whose ground-state density is equal to the instantaneous density at each time; the calculation is particularly difficult to converge when the density is very small in some regions, as in the interatomic region. Instead, one can make use of an asymmetric two-site Hubbard model, which captures the essential physics of the charge-transfer dynamics but within a small Hilbert space such that the exact ground xc potential can be computed exactly numerically as a functional of the site occupation difference~\cite{FFTAKR11,FM14,FM14b,CF12}. One can fix the bias between the two sites, together with the hopping parameter and on-site interaction,  such that in the ground-state almost 2 electrons occupy one site (the donor) and the first excited state is a charge-transfer state that has almost one electron on each site. 
Plotting the site occupation number difference as a function of time, while the system is driven by a weak field resonant with the charge-transfer excitation energy, one finds the characteristic change in the site occupation number difference, very similar to that in the dipole moment in the real-space model in Fig.~\ref{fig:FERM}, and EXX also behaves in a very similar way to that in the real-space molecule.  A self-consistent propagation using the AE xc potential begins promisingly, following the exact result for considerably longer than the adiabatic EXX approximation, but ultimately turns around, well before one electron has transferred~\cite{FM14,FM14b}, as shown in Fig.~\ref{fig:adia-ex}.  

It is worth noting that the charge-transfer {\it frequency} was very well-predicted by the AE approximation in the model of Refs~\cite{FM14,FM14b} in the cases where the initial ground-state consisted of closed-shell sites,  while the AE charge-transfer {\it dynamics} was poor. The frequencies were also were quite reasonable in the adiabatic approximations that yielded poor dynamics in Fig.~\ref{fig:FERM}.  Obtaining accurate dynamics for more than just short times needs more than just a good calculation of linear response. When the linear response predictions are poor, the dynamics will be poor too, and this is illustrated in the right panel of Fig.~\ref{fig:adia-ex}~\cite{FM14,FM14b} where the parameters of the asymmetric Hubbard dimer were adjusted to model charge-transfer  in a long-range molecule consisting of two open-shell fragments. There, the initial ground state has close to one electron on each site and driving at the exact charge-transfer frequency with the exact functional transfers one electron to the other site so the site occupation number goes from close to zero to close to 2. The AE dynamics fails from the start, consistent with the fact that the AE frequency is a drastic overestimation of the true frequency: the bare KS frequency is almost zero due to the static correlation, while the AE kernel correction overshoots the exact frequency considerably~\cite{FM14b}.
  
 \subsection{Charge-transfer dynamics from a photo-excited state}
As mentioned in Sec.~\ref{sec:nonadiab_step}, it is perhaps expected that full charge-transfer out of a ground-state is challenging in TDDFT because of the restriction to a single-Slater determinant; a single valence orbital must describe both the transferring electron as well as the one that remains. Although it sounds a bit odd, one could say that time-dependent static correlation builds up, as over the charge-transfer process the true system deviates increasingly from a single-Slater determinant.   The complete failure of the AE approximation is perhaps less expected, given that the exact ground-state xc functional is able to capture step-like features in the correlation potential. 

Instead, it would appear that charge-transfer from a locally-excited state, as reached in an initial photo-excitation, would be an easier task for TDDFT functionals. The electron that transfers now has its "own" Kohn-Sham orbital, not tied to an electron that remains, so the orbitals in principle need not become delocalized over time. There is the question of adiabatic TDDFT making errors from the very start, as this uses a ground-state xc functional to describe an initially excited state, however the error from this effect can be relatively small when the initial Kohn-Sham configuration is chosen close to that of the exact wavefunction~\cite{EM12,FNRM16}.
In fact, real simulations of photo-induced processes often begin with the Kohn-Sham system initially excited (e.g. Refs.~\cite{TRR05,RFSR13}), so investigating the exact description for such a case is very relevant. 

In Ref.~\cite{FLSM15}, charge-transfer from a locally-excited state of a long-range molecule was investigated with the help of a one-dimensional double-well to compare the exact dynamics with approximate TDDFT ones. Figure~\ref{fig:FLSM} provides an illustration of this.  An electron transfers over from the excited "donor" to the acceptor, when a weak field resonant with the energy difference between the charge-transfer state and the local excitation is applied, yielding a large change in the dipole moment. Exact exchange is able to capture this dynamics accurately, however other functionals failed, even when run at their respective resonant frequencies. The reason for this is that they violate an exact condition of TDDFT, which is a condition on the non-equilibrium density-response function, stating that the excitation frequency of a given transition is independent of the state around which the response is taken. Within the TDDFT non-equilibrium response formalism~\cite{FLSM15,LFM16,PS15,PSMS15}, this requires a subtle cancellation between time-dependences of the Kohn-Sham response function and a generalized xc kernel, which is not respected by most approximations. For this particular case of two electrons occupying different spin-orbitals (but not in the general case), exact exchange can be shown to satisfy this "resonance condition", and therefore gives good dynamics. The ability of the functional to transfer the charge is directly related to its degree of violation of this condition when the charge-transfer is resonantly induced. In fact, the failure of the approximations in the case of charge transfer out of the ground-state can also be related to the violation of the resonance condition~\cite{FLSM15,LFM16}. When a "chirped" laser is applied whose frequency is adjusted from time to time to be that of the shifting resonance of the approximation, more charge is transferred~\cite{LFM16}.

Driven dynamics is difficult, with or without charge transfer, for approximations in TDDFT, because of the spurious shifting of the resonant frequency from violation of the resonance condition. But, in many cases the charge transfer is not resonantly driven, and follows, for example, from the coupling with ionic dynamics as discussed briefly earlier. 
For example,  in photovoltaic processes, the standard picture is that after the initial photo-excitation, the system evolves in a field-free way, with the electron migrating through the system as a result of various kinds of vibrational motion, that can be quite complicated with many different aspects to consider, including relaxation, interface effects, thermal effects~\cite{PDP09}. 
 The impact of spuriously shifting resonances on the dynamics is less obvious, given that the electron transfer process does not directly depend on a resonant process, and that the ionic vibrations broaden their positions anyway. Still, this needs to be better understood, to decipher and interpret discrepancies between experiment and theory and have confidence in the theoretical predictions of future experiments.

\section{Concluding Discussion}
There has been tremendous effort in confronting the challenge of describing charge-transfer excitations and dynamics using TDDFT in recent years. It is now well understood why these excitations are severely underestimated by conventional functionals, and users are aware to apply caution when computational restrictions require them to use such functionals. Several more sophisticated functionals have been proposed and are now extensively in use for problems where charge-transfer is important, notably the long-range corrected, or range-separated hybrid functionals. However the problem is not completely solved, as several aspects are not quite resolved, from the choice of empirical parameters in some, to size-consistency violation in others, to the dearth of approximations that work for charge-transfer between open-shell fragments.  
  
Charge-transfer dynamics present yet a more challenging problem for approximate functionals in TDDFT. Although much work has been done in recent years in illuminating why functionals that may provide accurate charge-transfer excitation energies can yield terrible charge-transfer dynamics, and what are the essential features of the exact functional, there is, as of now, no good practical solution to the problem. 
 In charge-transfer out of the ground-state, non-adiabatic step-features in the xc potential that have a spatially-nonlocal- and time-non-local- dependence on the density play an important role in capturing the correct dynamics. Charge-transfer out of photo-excited states is at least conceptually easier, since the transferring electron is no longer "tied" to one that remains. One however runs into the problem of spuriously shifting resonances, which creates havoc for resonantly-driven charge-transfer, although its impact on charge-transfer driven by ionic motion is yet to be carefully gauged. Orbital-functionals may offer an avenue to build in the required non-local spatial- and time-density-dependence, but likely these must go beyond a mixing in of exact-exchange or Hartree-Fock, such as those motivated by the exact decomposition of the xc potential into kinetic and interaction components~\cite{M16}.

There are several glaring omissions in this review.  For example, we have not discussed the problem of charge-transfer in solvents~\cite{NGB06,SPN14,AJ13,LM16,DKVNF13, IMCT13,ZHHPH16}. We have also said nothing about oscillator strengths of charge-transfer excitations, which have been studied e.g. Ref.~\cite{FW12,WC08}, but generally have not garnered so much attention in the literature compared to the value of the excitation energies themselves. We also did not provide many "numbers", such as mean errors for a given functional for different classes of systems. We expect the progress in TDDFT to continue, and, together with the open issues discussed in this review, these problems will be at the forefront of density-functional development in the coming years.

\begin{acknowledgements}  NTM thanks Filipp Furche, Leeor Kronik, and Ivano Tavernelli for very helpful comments on the manuscript. 
Financial support from the National Science Foundation CHE-1566197, and Department of Energy, Office of Basic Energy Sciences, Division of Chemical Sciences, Geosciences and Biosciences under Award DE-SC0015344 is gratefully acknowledged.
\end{acknowledgements}

\addcontentsline{toc}{section}{References}
\bibliography{./ref}

\begin{thebibliography}{159}%
\makeatletter
\providecommand \@ifxundefined [1]{%
 \@ifx{#1\undefined}
}%
\providecommand \@ifnum [1]{%
 \ifnum #1\expandafter \@firstoftwo
 \else \expandafter \@secondoftwo
 \fi
}%
\providecommand \@ifx [1]{%
 \ifx #1\expandafter \@firstoftwo
 \else \expandafter \@secondoftwo
 \fi
}%
\providecommand \natexlab [1]{#1}%
\providecommand \enquote  [1]{``#1''}%
\providecommand \bibnamefont  [1]{#1}%
\providecommand \bibfnamefont [1]{#1}%
\providecommand \citenamefont [1]{#1}%
\providecommand \href@noop [0]{\@secondoftwo}%
\providecommand \href [0]{\begingroup \@sanitize@url \@href}%
\providecommand \@href[1]{\@@startlink{#1}\@@href}%
\providecommand \@@href[1]{\endgroup#1\@@endlink}%
\providecommand \@sanitize@url [0]{\catcode `\\12\catcode `\$12\catcode
  `\&12\catcode `\#12\catcode `\^12\catcode `\_12\catcode `\%12\relax}%
\providecommand \@@startlink[1]{}%
\providecommand \@@endlink[0]{}%
\providecommand \url  [0]{\begingroup\@sanitize@url \@url }%
\providecommand \@url [1]{\endgroup\@href {#1}{\urlprefix }}%
\providecommand \urlprefix  [0]{URL }%
\providecommand \Eprint [0]{\href }%
\providecommand \doibase [0]{http://dx.doi.org/}%
\providecommand \selectlanguage [0]{\@gobble}%
\providecommand \bibinfo  [0]{\@secondoftwo}%
\providecommand \bibfield  [0]{\@secondoftwo}%
\providecommand \translation [1]{[#1]}%
\providecommand \BibitemOpen [0]{}%
\providecommand \bibitemStop [0]{}%
\providecommand \bibitemNoStop [0]{.\EOS\space}%
\providecommand \EOS [0]{\spacefactor3000\relax}%
\providecommand \BibitemShut  [1]{\csname bibitem#1\endcsname}%
\let\auto@bib@innerbib\@empty
\bibitem [{\citenamefont {Runge}\ and\ \citenamefont {Gross}(1984)}]{RG84}%
  \BibitemOpen
  \bibfield  {author} {\bibinfo {author} {\bibfnamefont {E.}~\bibnamefont
  {Runge}}\ and\ \bibinfo {author} {\bibfnamefont {E.~K.~U.}\ \bibnamefont
  {Gross}},\ }\href {\doibase 10.1103/PhysRevLett.52.997} {\bibfield  {journal}
  {\bibinfo  {journal} {Phys. Rev. Lett.}\ }\textbf {\bibinfo {volume} {52}},\
  \bibinfo {pages} {997} (\bibinfo {year} {1984})}\BibitemShut {NoStop}%
\bibitem [{\citenamefont {Rappoport}\ and\ \citenamefont
  {Furche}(2004)}]{RF04}%
  \BibitemOpen
  \bibfield  {author} {\bibinfo {author} {\bibfnamefont {D.}~\bibnamefont
  {Rappoport}}\ and\ \bibinfo {author} {\bibfnamefont {F.}~\bibnamefont
  {Furche}},\ }\href@noop {} {\bibfield  {journal} {\bibinfo  {journal} {J. Am.
  Chem. Soc.}\ }\textbf {\bibinfo {volume} {126}},\ \bibinfo {pages} {1277}
  (\bibinfo {year} {2004})}\BibitemShut {NoStop}%
\bibitem [{\citenamefont {Lippert}\ \emph {et~al.}(1961)\citenamefont
  {Lippert}, \citenamefont {L\"uder}, \citenamefont {oll}, \citenamefont
  {N\"agele}, \citenamefont {Boos}, \citenamefont {Prigge},\ and\ \citenamefont
  {Seibold-Blankenstein}}]{Lippert1959}%
  \BibitemOpen
  \bibfield  {author} {\bibinfo {author} {\bibfnamefont {E.}~\bibnamefont
  {Lippert}}, \bibinfo {author} {\bibfnamefont {W.}~\bibnamefont {L\"uder}},
  \bibinfo {author} {\bibfnamefont {F.}~\bibnamefont {oll}}, \bibinfo {author}
  {\bibfnamefont {W.}~\bibnamefont {N\"agele}}, \bibinfo {author}
  {\bibfnamefont {H.}~\bibnamefont {Boos}}, \bibinfo {author} {\bibfnamefont
  {H.}~\bibnamefont {Prigge}}, \ and\ \bibinfo {author} {\bibfnamefont
  {I.}~\bibnamefont {Seibold-Blankenstein}},\ }\href@noop {} {\bibfield
  {journal} {\bibinfo  {journal} {Angew. Chem.}\ }\textbf {\bibinfo {volume}
  {73}},\ \bibinfo {pages} {695} (\bibinfo {year} {1961})}\BibitemShut
  {NoStop}%
\bibitem [{\citenamefont {Autschbach}(2009)}]{A09}%
  \BibitemOpen
  \bibfield  {author} {\bibinfo {author} {\bibfnamefont {J.}~\bibnamefont
  {Autschbach}},\ }\href@noop {} {\bibfield  {journal} {\bibinfo  {journal}
  {ChemPhysChem}\ }\textbf {\bibinfo {volume} {10}},\ \bibinfo {pages} {1757}
  (\bibinfo {year} {2009})}\BibitemShut {NoStop}%
\bibitem [{\citenamefont {Fuks}(2016)}]{F16}%
  \BibitemOpen
  \bibfield  {author} {\bibinfo {author} {\bibfnamefont {J.~I.}\ \bibnamefont
  {Fuks}},\ }\href {\doibase 10.1140/epjb/e2016-70110-y} {\bibfield  {journal}
  {\bibinfo  {journal} {The European Physical Journal B}\ }\textbf {\bibinfo
  {volume} {89}},\ \bibinfo {pages} {236} (\bibinfo {year} {2016})}\BibitemShut
  {NoStop}%
\bibitem [{\citenamefont {Maitra}(2016)}]{M16}%
  \BibitemOpen
  \bibfield  {author} {\bibinfo {author} {\bibfnamefont {N.~T.}\ \bibnamefont
  {Maitra}},\ }\href@noop {} {\bibfield  {journal} {\bibinfo  {journal} {J.
  Chem. Phys.}\ }\textbf {\bibinfo {volume} {144}},\ \bibinfo {pages} {220901}
  (\bibinfo {year} {2016})}\BibitemShut {NoStop}%
\bibitem [{\citenamefont {Gross}\ and\ \citenamefont {Maitra}()}]{GM12chap}%
  \BibitemOpen
  \bibfield  {author} {\bibinfo {author} {\bibfnamefont {E.~K.}\ \bibnamefont
  {Gross}}\ and\ \bibinfo {author} {\bibfnamefont {N.~T.}\ \bibnamefont
  {Maitra}},\ }in\ \href@noop {} {\emph {\bibinfo {booktitle} {Fundamentals of
  time-dependent density functional theory}}},\ \bibinfo {editor} {edited by\
  \bibinfo {editor} {\bibfnamefont {M.~A.}\ \bibnamefont {Marques}}, \bibinfo
  {editor} {\bibfnamefont {N.~T.}\ \bibnamefont {Maitra}}, \bibinfo {editor}
  {\bibfnamefont {F.~M.}\ \bibnamefont {Nogueira}}, \bibinfo {editor}
  {\bibfnamefont {E.~K.}\ \bibnamefont {Gross}}, \ and\ \bibinfo {editor}
  {\bibfnamefont {A.}~\bibnamefont {Rubio}}},\ pp.\ \bibinfo {pages}
  {53--97}\BibitemShut {NoStop}%
\bibitem [{\citenamefont {Marques}\ \emph {et~al.}(2012)\citenamefont
  {Marques}, \citenamefont {Maitra}, \citenamefont {Nogueira}, \citenamefont
  {Gross},\ and\ \citenamefont {Rubio}}]{TDDFTbook12}%
  \BibitemOpen
  \bibinfo {editor} {\bibfnamefont {M.~A.}\ \bibnamefont {Marques}}, \bibinfo
  {editor} {\bibfnamefont {N.~T.}\ \bibnamefont {Maitra}}, \bibinfo {editor}
  {\bibfnamefont {F.~M.}\ \bibnamefont {Nogueira}}, \bibinfo {editor}
  {\bibfnamefont {E.~K.}\ \bibnamefont {Gross}}, \ and\ \bibinfo {editor}
  {\bibfnamefont {A.}~\bibnamefont {Rubio}},\ eds.,\ \href@noop {} {\emph
  {\bibinfo {title} {Fundamentals of time-dependent density functional
  theory}}},\ Vol.\ \bibinfo {volume} {837}\ (\bibinfo  {publisher}
  {Springer},\ \bibinfo {year} {2012})\BibitemShut {NoStop}%
\bibitem [{\citenamefont {Ullrich}(2011)}]{Carstenbook}%
  \BibitemOpen
  \bibfield  {author} {\bibinfo {author} {\bibfnamefont {C.~A.}\ \bibnamefont
  {Ullrich}},\ }\href@noop {} {\emph {\bibinfo {title} {Time-dependent
  density-functional theory: concepts and applications}}}\ (\bibinfo
  {publisher} {Oxford University Press},\ \bibinfo {year} {2011})\BibitemShut
  {NoStop}%
\bibitem [{\citenamefont {van Leeuwen}(1999)}]{L99}%
  \BibitemOpen
  \bibfield  {author} {\bibinfo {author} {\bibfnamefont {R.}~\bibnamefont {van
  Leeuwen}},\ }\href {\doibase 10.1103/PhysRevLett.82.3863} {\bibfield
  {journal} {\bibinfo  {journal} {Phys. Rev. Lett.}\ }\textbf {\bibinfo
  {volume} {82}},\ \bibinfo {pages} {3863} (\bibinfo {year}
  {1999})}\BibitemShut {NoStop}%
\bibitem [{\citenamefont {Hohenberg}\ and\ \citenamefont {Kohn}(1964)}]{HK64}%
  \BibitemOpen
  \bibfield  {author} {\bibinfo {author} {\bibfnamefont {P.}~\bibnamefont
  {Hohenberg}}\ and\ \bibinfo {author} {\bibfnamefont {W.}~\bibnamefont
  {Kohn}},\ }\href {\doibase 10.1103/PhysRev.136.B864} {\bibfield  {journal}
  {\bibinfo  {journal} {Phys. Rev.}\ }\textbf {\bibinfo {volume} {136}},\
  \bibinfo {pages} {B864} (\bibinfo {year} {1964})}\BibitemShut {NoStop}%
\bibitem [{\citenamefont {Kohn}\ and\ \citenamefont {Sham}(1965)}]{KS65}%
  \BibitemOpen
  \bibfield  {author} {\bibinfo {author} {\bibfnamefont {W.}~\bibnamefont
  {Kohn}}\ and\ \bibinfo {author} {\bibfnamefont {L.~J.}\ \bibnamefont
  {Sham}},\ }\href {\doibase 10.1103/PhysRev.140.A1133} {\bibfield  {journal}
  {\bibinfo  {journal} {Phys. Rev.}\ }\textbf {\bibinfo {volume} {140}},\
  \bibinfo {pages} {A1133} (\bibinfo {year} {1965})}\BibitemShut {NoStop}%
\bibitem [{\citenamefont {Vignale}\ and\ \citenamefont {Kohn}(1996)}]{VK96}%
  \BibitemOpen
  \bibfield  {author} {\bibinfo {author} {\bibfnamefont {G.}~\bibnamefont
  {Vignale}}\ and\ \bibinfo {author} {\bibfnamefont {W.}~\bibnamefont {Kohn}},\
  }\href@noop {} {\bibfield  {journal} {\bibinfo  {journal} {Phys. Rev. Lett.}\
  }\textbf {\bibinfo {volume} {77}},\ \bibinfo {pages} {2037} (\bibinfo {year}
  {1996})}\BibitemShut {NoStop}%
\bibitem [{\citenamefont {Vignale}\ \emph {et~al.}(1997)\citenamefont
  {Vignale}, \citenamefont {Ullrich},\ and\ \citenamefont {Conti}}]{VUC97}%
  \BibitemOpen
  \bibfield  {author} {\bibinfo {author} {\bibfnamefont {G.}~\bibnamefont
  {Vignale}}, \bibinfo {author} {\bibfnamefont {C.~A.}\ \bibnamefont
  {Ullrich}}, \ and\ \bibinfo {author} {\bibfnamefont {S.}~\bibnamefont
  {Conti}},\ }\href@noop {} {\bibfield  {journal} {\bibinfo  {journal} {Phys.
  Rev. Lett.}\ }\textbf {\bibinfo {volume} {79}},\ \bibinfo {pages} {4878}
  (\bibinfo {year} {1997})}\BibitemShut {NoStop}%
\bibitem [{\citenamefont {Kurzweil}\ and\ \citenamefont {Baer}(2006)}]{KB06}%
  \BibitemOpen
  \bibfield  {author} {\bibinfo {author} {\bibfnamefont {Y.}~\bibnamefont
  {Kurzweil}}\ and\ \bibinfo {author} {\bibfnamefont {R.}~\bibnamefont
  {Baer}},\ }\href {\doibase 10.1103/PhysRevB.73.075413} {\bibfield  {journal}
  {\bibinfo  {journal} {Phys. Rev. B}\ }\textbf {\bibinfo {volume} {73}},\
  \bibinfo {pages} {075413} (\bibinfo {year} {2006})}\BibitemShut {NoStop}%
\bibitem [{\citenamefont {Fischer}\ \emph {et~al.}(2015)\citenamefont
  {Fischer}, \citenamefont {Cramer},\ and\ \citenamefont {Govind}}]{FCG15}%
  \BibitemOpen
  \bibfield  {author} {\bibinfo {author} {\bibfnamefont {S.~A.}\ \bibnamefont
  {Fischer}}, \bibinfo {author} {\bibfnamefont {C.~J.}\ \bibnamefont {Cramer}},
  \ and\ \bibinfo {author} {\bibfnamefont {N.}~\bibnamefont {Govind}},\
  }\href@noop {} {\bibfield  {journal} {\bibinfo  {journal} {J. Chem. Theory
  and Comput.}\ }\textbf {\bibinfo {volume} {11}},\ \bibinfo {pages} {4294}
  (\bibinfo {year} {2015})}\BibitemShut {NoStop}%
\bibitem [{\citenamefont {Tussupbayev}\ \emph {et~al.}(2015)\citenamefont
  {Tussupbayev}, \citenamefont {Govind}, \citenamefont {Lopata},\ and\
  \citenamefont {Cramer}}]{TGLC15}%
  \BibitemOpen
  \bibfield  {author} {\bibinfo {author} {\bibfnamefont {S.}~\bibnamefont
  {Tussupbayev}}, \bibinfo {author} {\bibfnamefont {N.}~\bibnamefont {Govind}},
  \bibinfo {author} {\bibfnamefont {K.}~\bibnamefont {Lopata}}, \ and\ \bibinfo
  {author} {\bibfnamefont {C.~J.}\ \bibnamefont {Cramer}},\ }\href {\doibase
  10.1021/ct500763y} {\bibfield  {journal} {\bibinfo  {journal} {Journal of
  Chemical Theory and Computation}\ }\textbf {\bibinfo {volume} {11}},\
  \bibinfo {pages} {1102} (\bibinfo {year} {2015})},\ \bibinfo {note} {pMID:
  26579760}\BibitemShut {NoStop}%
\bibitem [{\citenamefont {van Leeuwen}(1998)}]{L98}%
  \BibitemOpen
  \bibfield  {author} {\bibinfo {author} {\bibfnamefont {R.}~\bibnamefont {van
  Leeuwen}},\ }\href@noop {} {\bibfield  {journal} {\bibinfo  {journal} {Phys.
  Rev. Lett.}\ }\textbf {\bibinfo {volume} {80}},\ \bibinfo {pages} {1280}
  (\bibinfo {year} {1998})}\BibitemShut {NoStop}%
\bibitem [{\citenamefont {Ullrich}\ \emph {et~al.}(1995)\citenamefont
  {Ullrich}, \citenamefont {Gossmann},\ and\ \citenamefont {Gross}}]{UGG95}%
  \BibitemOpen
  \bibfield  {author} {\bibinfo {author} {\bibfnamefont {C.~A.}\ \bibnamefont
  {Ullrich}}, \bibinfo {author} {\bibfnamefont {U.~J.}\ \bibnamefont
  {Gossmann}}, \ and\ \bibinfo {author} {\bibfnamefont {E.~K.~U.}\ \bibnamefont
  {Gross}},\ }\href@noop {} {\bibfield  {journal} {\bibinfo  {journal} {Phys.
  Rev. Lett.}\ }\textbf {\bibinfo {volume} {74}},\ \bibinfo {pages} {872}
  (\bibinfo {year} {1995})}\BibitemShut {NoStop}%
\bibitem [{\citenamefont {Mundt}\ \emph {et~al.}(2007)\citenamefont {Mundt},
  \citenamefont {K\"ummel}, \citenamefont {van Leeuwen},\ and\ \citenamefont
  {Reinhard}}]{MKLR07}%
  \BibitemOpen
  \bibfield  {author} {\bibinfo {author} {\bibfnamefont {M.}~\bibnamefont
  {Mundt}}, \bibinfo {author} {\bibfnamefont {S.}~\bibnamefont {K\"ummel}},
  \bibinfo {author} {\bibfnamefont {R.}~\bibnamefont {van Leeuwen}}, \ and\
  \bibinfo {author} {\bibfnamefont {P.-G.}\ \bibnamefont {Reinhard}},\
  }\href@noop {} {\bibfield  {journal} {\bibinfo  {journal} {Phys. Rev. A}\
  }\textbf {\bibinfo {volume} {75}},\ \bibinfo {pages} {050501} (\bibinfo
  {year} {2007})}\BibitemShut {NoStop}%
\bibitem [{\citenamefont {Hofmann}\ \emph {et~al.}(2012)\citenamefont
  {Hofmann}, \citenamefont {K{\"o}rzd{\"o}rfer},\ and\ \citenamefont
  {K\"ummel}}]{HKK12}%
  \BibitemOpen
  \bibfield  {author} {\bibinfo {author} {\bibfnamefont {D.}~\bibnamefont
  {Hofmann}}, \bibinfo {author} {\bibfnamefont {T.}~\bibnamefont
  {K{\"o}rzd{\"o}rfer}}, \ and\ \bibinfo {author} {\bibfnamefont
  {S.}~\bibnamefont {K\"ummel}},\ }\href@noop {} {\bibfield  {journal}
  {\bibinfo  {journal} {Phys. Rev. Lett.}\ }\textbf {\bibinfo {volume} {108}},\
  \bibinfo {pages} {146401} (\bibinfo {year} {2012})}\BibitemShut {NoStop}%
\bibitem [{\citenamefont {Wijewardane}\ and\ \citenamefont
  {Ullrich}(2008)}]{WU08}%
  \BibitemOpen
  \bibfield  {author} {\bibinfo {author} {\bibfnamefont {H.~O.}\ \bibnamefont
  {Wijewardane}}\ and\ \bibinfo {author} {\bibfnamefont {C.~A.}\ \bibnamefont
  {Ullrich}},\ }\href@noop {} {\bibfield  {journal} {\bibinfo  {journal} {Phys.
  Rev. Lett.}\ }\textbf {\bibinfo {volume} {100}},\ \bibinfo {pages} {056404}
  (\bibinfo {year} {2008})}\BibitemShut {NoStop}%
\bibitem [{\citenamefont {Hessler}\ \emph {et~al.}(2002)\citenamefont
  {Hessler}, \citenamefont {Maitra},\ and\ \citenamefont {Burke}}]{HMB02}%
  \BibitemOpen
  \bibfield  {author} {\bibinfo {author} {\bibfnamefont {P.}~\bibnamefont
  {Hessler}}, \bibinfo {author} {\bibfnamefont {N.~T.}\ \bibnamefont {Maitra}},
  \ and\ \bibinfo {author} {\bibfnamefont {K.}~\bibnamefont {Burke}},\
  }\href@noop {} {\bibfield  {journal} {\bibinfo  {journal} {J. Chem. Phys.}\
  }\textbf {\bibinfo {volume} {117}} (\bibinfo {year} {2002})}\BibitemShut
  {NoStop}%
\bibitem [{\citenamefont {Thiele}\ \emph {et~al.}(2008)\citenamefont {Thiele},
  \citenamefont {Gross},\ and\ \citenamefont {K\"ummel}}]{TGK08}%
  \BibitemOpen
  \bibfield  {author} {\bibinfo {author} {\bibfnamefont {M.}~\bibnamefont
  {Thiele}}, \bibinfo {author} {\bibfnamefont {E.~K.~U.}\ \bibnamefont
  {Gross}}, \ and\ \bibinfo {author} {\bibfnamefont {S.}~\bibnamefont
  {K\"ummel}},\ }\href {\doibase 10.1103/PhysRevLett.100.153004} {\bibfield
  {journal} {\bibinfo  {journal} {Phys. Rev. Lett.}\ }\textbf {\bibinfo
  {volume} {100}},\ \bibinfo {pages} {153004} (\bibinfo {year}
  {2008})}\BibitemShut {NoStop}%
\bibitem [{\citenamefont {Gross}\ and\ \citenamefont {Kohn}(1985)}]{GK85}%
  \BibitemOpen
  \bibfield  {author} {\bibinfo {author} {\bibfnamefont {E.~K.~U.}\
  \bibnamefont {Gross}}\ and\ \bibinfo {author} {\bibfnamefont
  {W.}~\bibnamefont {Kohn}},\ }\href@noop {} {\bibfield  {journal} {\bibinfo
  {journal} {Phys. Rev. Lett.}\ }\textbf {\bibinfo {volume} {55}},\ \bibinfo
  {pages} {2850} (\bibinfo {year} {1985})}\BibitemShut {NoStop}%
\bibitem [{\citenamefont {Petersilka}\ \emph {et~al.}(1996)\citenamefont
  {Petersilka}, \citenamefont {Gossmann},\ and\ \citenamefont {Gross}}]{PGG96}%
  \BibitemOpen
  \bibfield  {author} {\bibinfo {author} {\bibfnamefont {M.}~\bibnamefont
  {Petersilka}}, \bibinfo {author} {\bibfnamefont {U.~J.}\ \bibnamefont
  {Gossmann}}, \ and\ \bibinfo {author} {\bibfnamefont {E.~K.~U.}\ \bibnamefont
  {Gross}},\ }\href {\doibase 10.1103/PhysRevLett.76.1212} {\bibfield
  {journal} {\bibinfo  {journal} {Phys. Rev. Lett.}\ }\textbf {\bibinfo
  {volume} {76}},\ \bibinfo {pages} {1212} (\bibinfo {year}
  {1996})}\BibitemShut {NoStop}%
\bibitem [{\citenamefont {Casida}(1995)}]{C95}%
  \BibitemOpen
  \bibfield  {author} {\bibinfo {author} {\bibfnamefont {M.}~\bibnamefont
  {Casida}},\ }in\ \href@noop {} {\emph {\bibinfo {booktitle} {Recent Advances
  in Density Functional Methods, Part I}}},\ \bibinfo {editor} {edited by\
  \bibinfo {editor} {\bibfnamefont {D.}~\bibnamefont {Chong}}}\ (\bibinfo
  {publisher} {World Scientific, Singapore},\ \bibinfo {year}
  {1995})\BibitemShut {NoStop}%
\bibitem [{\citenamefont {Casida}(1996)}]{C96}%
  \BibitemOpen
  \bibfield  {author} {\bibinfo {author} {\bibfnamefont {M.~E.}\ \bibnamefont
  {Casida}},\ }in\ \href@noop {} {\emph {\bibinfo {booktitle} {Recent
  Developments and Applications of Modern Density Functional Theory}}},\
  \bibinfo {editor} {edited by\ \bibinfo {editor} {\bibfnamefont {J.~M.}\
  \bibnamefont {Seminario}}}\ (\bibinfo  {publisher} {Elsevier},\ \bibinfo
  {address} {Amsterdam},\ \bibinfo {year} {1996})\ p.\ \bibinfo {pages}
  {391}\BibitemShut {NoStop}%
\bibitem [{\citenamefont {Bauernschmitt}\ and\ \citenamefont
  {Ahlrichs}(1996)}]{BA96a}%
  \BibitemOpen
  \bibfield  {author} {\bibinfo {author} {\bibfnamefont {R.}~\bibnamefont
  {Bauernschmitt}}\ and\ \bibinfo {author} {\bibfnamefont {R.}~\bibnamefont
  {Ahlrichs}},\ }\href@noop {} {\bibfield  {journal} {\bibinfo  {journal}
  {Chem. Phys. Lett.}\ }\textbf {\bibinfo {volume} {256}},\ \bibinfo {pages}
  {454 } (\bibinfo {year} {1996})}\BibitemShut {NoStop}%
\bibitem [{\citenamefont {Grabo}\ \emph
  {et~al.}(2000{\natexlab{a}})\citenamefont {Grabo}, \citenamefont
  {Petersilka},\ and\ \citenamefont {Gross}}]{GPG00}%
  \BibitemOpen
  \bibfield  {author} {\bibinfo {author} {\bibfnamefont {T.}~\bibnamefont
  {Grabo}}, \bibinfo {author} {\bibfnamefont {M.}~\bibnamefont {Petersilka}}, \
  and\ \bibinfo {author} {\bibfnamefont {E.}~\bibnamefont {Gross}},\
  }\href@noop {} {\bibfield  {journal} {\bibinfo  {journal} {Journal of
  Molecular Structure: THEOCHEM}\ }\textbf {\bibinfo {volume} {501}},\ \bibinfo
  {pages} {353} (\bibinfo {year} {2000}{\natexlab{a}})}\BibitemShut {NoStop}%
\bibitem [{\citenamefont {Furche}\ \emph {et~al.}(2016)\citenamefont {Furche},
  \citenamefont {Krull}, \citenamefont {Nguyen},\ and\ \citenamefont
  {Kwon}}]{FKNK16}%
  \BibitemOpen
  \bibfield  {author} {\bibinfo {author} {\bibfnamefont {F.}~\bibnamefont
  {Furche}}, \bibinfo {author} {\bibfnamefont {B.~T.}\ \bibnamefont {Krull}},
  \bibinfo {author} {\bibfnamefont {B.~D.}\ \bibnamefont {Nguyen}}, \ and\
  \bibinfo {author} {\bibfnamefont {J.}~\bibnamefont {Kwon}},\ }\href {\doibase
  10.1063/1.4947245} {\bibfield  {journal} {\bibinfo  {journal} {The Journal of
  Chemical Physics}\ }\textbf {\bibinfo {volume} {144}},\ \bibinfo {pages}
  {174105} (\bibinfo {year} {2016})},\ \Eprint
  {http://arxiv.org/abs/http://dx.doi.org/10.1063/1.4947245}
  {http://dx.doi.org/10.1063/1.4947245} \BibitemShut {NoStop}%
\bibitem [{\citenamefont {Yabana}\ \emph {et~al.}(2006)\citenamefont {Yabana},
  \citenamefont {Nakatsukasa}, \citenamefont {Iwata},\ and\ \citenamefont
  {Bertsch}}]{YNIB06}%
  \BibitemOpen
  \bibfield  {author} {\bibinfo {author} {\bibfnamefont {K.}~\bibnamefont
  {Yabana}}, \bibinfo {author} {\bibfnamefont {T.}~\bibnamefont {Nakatsukasa}},
  \bibinfo {author} {\bibfnamefont {J.-I.}\ \bibnamefont {Iwata}}, \ and\
  \bibinfo {author} {\bibfnamefont {G.}~\bibnamefont {Bertsch}},\ }\href@noop
  {} {\bibfield  {journal} {\bibinfo  {journal} {Physica Status Solidi (b)}\
  }\textbf {\bibinfo {volume} {243}},\ \bibinfo {pages} {1121} (\bibinfo {year}
  {2006})}\BibitemShut {NoStop}%
\bibitem [{\citenamefont {Andrade}\ \emph {et~al.}(2012)\citenamefont
  {Andrade}, \citenamefont {Alberdi-Rodriguez}, \citenamefont {Strubbe},
  \citenamefont {Oliveira}, \citenamefont {Nogueira}, \citenamefont {Castro},
  \citenamefont {Muguerza}, \citenamefont {Arruabarrena}, \citenamefont
  {Louie}, \citenamefont {Aspuru-Guzik} \emph {et~al.}}]{octopus}%
  \BibitemOpen
  \bibfield  {author} {\bibinfo {author} {\bibfnamefont {X.}~\bibnamefont
  {Andrade}}, \bibinfo {author} {\bibfnamefont {J.}~\bibnamefont
  {Alberdi-Rodriguez}}, \bibinfo {author} {\bibfnamefont {D.~A.}\ \bibnamefont
  {Strubbe}}, \bibinfo {author} {\bibfnamefont {M.~J.}\ \bibnamefont
  {Oliveira}}, \bibinfo {author} {\bibfnamefont {F.}~\bibnamefont {Nogueira}},
  \bibinfo {author} {\bibfnamefont {A.}~\bibnamefont {Castro}}, \bibinfo
  {author} {\bibfnamefont {J.}~\bibnamefont {Muguerza}}, \bibinfo {author}
  {\bibfnamefont {A.}~\bibnamefont {Arruabarrena}}, \bibinfo {author}
  {\bibfnamefont {S.~G.}\ \bibnamefont {Louie}}, \bibinfo {author}
  {\bibfnamefont {A.}~\bibnamefont {Aspuru-Guzik}},  \emph {et~al.},\
  }\href@noop {} {\bibfield  {journal} {\bibinfo  {journal} {J. Phys. Condens.
  Matter}\ }\textbf {\bibinfo {volume} {24}},\ \bibinfo {pages} {233202}
  (\bibinfo {year} {2012})}\BibitemShut {NoStop}%
\bibitem [{\citenamefont {Castro}\ \emph {et~al.}(2006)\citenamefont {Castro},
  \citenamefont {Appel}, \citenamefont {Oliveira}, \citenamefont {Rozzi},
  \citenamefont {Andrade}, \citenamefont {Lorenzen}, \citenamefont {Marques},
  \citenamefont {Gross},\ and\ \citenamefont {Rubio}}]{octopus2}%
  \BibitemOpen
  \bibfield  {author} {\bibinfo {author} {\bibfnamefont {A.}~\bibnamefont
  {Castro}}, \bibinfo {author} {\bibfnamefont {H.}~\bibnamefont {Appel}},
  \bibinfo {author} {\bibfnamefont {M.}~\bibnamefont {Oliveira}}, \bibinfo
  {author} {\bibfnamefont {C.~A.}\ \bibnamefont {Rozzi}}, \bibinfo {author}
  {\bibfnamefont {X.}~\bibnamefont {Andrade}}, \bibinfo {author} {\bibfnamefont
  {F.}~\bibnamefont {Lorenzen}}, \bibinfo {author} {\bibfnamefont {M.~A.~L.}\
  \bibnamefont {Marques}}, \bibinfo {author} {\bibfnamefont {E.~K.~U.}\
  \bibnamefont {Gross}}, \ and\ \bibinfo {author} {\bibfnamefont
  {A.}~\bibnamefont {Rubio}},\ }\href {\doibase 10.1002/pssb.200642067}
  {\bibfield  {journal} {\bibinfo  {journal} {physica status solidi (b)}\
  }\textbf {\bibinfo {volume} {243}},\ \bibinfo {pages} {2465} (\bibinfo {year}
  {2006})}\BibitemShut {NoStop}%
\bibitem [{\citenamefont {Valiev}\ \emph {et~al.}(2010)\citenamefont {Valiev},
  \citenamefont {Bylaska}, \citenamefont {Govind}, \citenamefont {Kowalski},
  \citenamefont {Straatsma}, \citenamefont {Dam}, \citenamefont {Wang},
  \citenamefont {Nieplocha}, \citenamefont {Apra}, \citenamefont {Windus},\
  and\ \citenamefont {de~Jong}}]{nwchem}%
  \BibitemOpen
  \bibfield  {author} {\bibinfo {author} {\bibfnamefont {M.}~\bibnamefont
  {Valiev}}, \bibinfo {author} {\bibfnamefont {E.}~\bibnamefont {Bylaska}},
  \bibinfo {author} {\bibfnamefont {N.}~\bibnamefont {Govind}}, \bibinfo
  {author} {\bibfnamefont {K.}~\bibnamefont {Kowalski}}, \bibinfo {author}
  {\bibfnamefont {T.}~\bibnamefont {Straatsma}}, \bibinfo {author}
  {\bibfnamefont {H.~V.}\ \bibnamefont {Dam}}, \bibinfo {author} {\bibfnamefont
  {D.}~\bibnamefont {Wang}}, \bibinfo {author} {\bibfnamefont {J.}~\bibnamefont
  {Nieplocha}}, \bibinfo {author} {\bibfnamefont {E.}~\bibnamefont {Apra}},
  \bibinfo {author} {\bibfnamefont {T.}~\bibnamefont {Windus}}, \ and\ \bibinfo
  {author} {\bibfnamefont {W.}~\bibnamefont {de~Jong}},\ }\href {\doibase
  http://dx.doi.org/10.1016/j.cpc.2010.04.018} {\bibfield  {journal} {\bibinfo
  {journal} {Computer Physics Communications}\ }\textbf {\bibinfo {volume}
  {181}},\ \bibinfo {pages} {1477 } (\bibinfo {year} {2010})}\BibitemShut
  {NoStop}%
\bibitem [{\citenamefont {Lopata}\ and\ \citenamefont {Govind}(2011)}]{LN11}%
  \BibitemOpen
  \bibfield  {author} {\bibinfo {author} {\bibfnamefont {K.}~\bibnamefont
  {Lopata}}\ and\ \bibinfo {author} {\bibfnamefont {N.}~\bibnamefont
  {Govind}},\ }\href {\doibase 10.1021/ct200137z} {\bibfield  {journal}
  {\bibinfo  {journal} {Journal of Chemical Theory and Computation}\ }\textbf
  {\bibinfo {volume} {7}},\ \bibinfo {pages} {1344} (\bibinfo {year} {2011})},\
  \bibinfo {note} {pMID: 26610129},\ \Eprint
  {http://arxiv.org/abs/http://dx.doi.org/10.1021/ct200137z}
  {http://dx.doi.org/10.1021/ct200137z} \BibitemShut {NoStop}%
\bibitem [{\citenamefont {Sternheimer}(1954)}]{Sternheimer54}%
  \BibitemOpen
  \bibfield  {author} {\bibinfo {author} {\bibfnamefont {R.~M.}\ \bibnamefont
  {Sternheimer}},\ }\href@noop {} {\bibfield  {journal} {\bibinfo  {journal}
  {Phys. Rev.}\ }\textbf {\bibinfo {volume} {96}},\ \bibinfo {pages} {951}
  (\bibinfo {year} {1954})}\BibitemShut {NoStop}%
\bibitem [{\citenamefont {Andrade}\ \emph {et~al.}(2007)\citenamefont
  {Andrade}, \citenamefont {Botti}, \citenamefont {Marques},\ and\
  \citenamefont {Rubio}}]{Andrade07}%
  \BibitemOpen
  \bibfield  {author} {\bibinfo {author} {\bibfnamefont {X.}~\bibnamefont
  {Andrade}}, \bibinfo {author} {\bibfnamefont {S.}~\bibnamefont {Botti}},
  \bibinfo {author} {\bibfnamefont {M.~A.~L.}\ \bibnamefont {Marques}}, \ and\
  \bibinfo {author} {\bibfnamefont {A.}~\bibnamefont {Rubio}},\ }\href@noop {}
  {\bibfield  {journal} {\bibinfo  {journal} {J. Chem. Phys.}\ }\textbf
  {\bibinfo {volume} {126}},\ \bibinfo {eid} {184106} (\bibinfo {year}
  {2007})}\BibitemShut {NoStop}%
\bibitem [{\citenamefont {Strubbe}\ \emph {et~al.}(2012)\citenamefont
  {Strubbe}, \citenamefont {Lehtovaara}, \citenamefont {Rubio}, \citenamefont
  {Marques},\ and\ \citenamefont {Louie}}]{Strubbechapter}%
  \BibitemOpen
  \bibfield  {author} {\bibinfo {author} {\bibfnamefont {D.~A.}\ \bibnamefont
  {Strubbe}}, \bibinfo {author} {\bibfnamefont {L.}~\bibnamefont {Lehtovaara}},
  \bibinfo {author} {\bibfnamefont {A.}~\bibnamefont {Rubio}}, \bibinfo
  {author} {\bibfnamefont {M.~A.}\ \bibnamefont {Marques}}, \ and\ \bibinfo
  {author} {\bibfnamefont {S.~G.}\ \bibnamefont {Louie}},\ }in\ \href@noop {}
  {\emph {\bibinfo {booktitle} {Fundamentals of Time-Dependent Density
  Functional Theory}}},\ \bibinfo {editor} {edited by\ \bibinfo {editor}
  {\bibfnamefont {M.~A.}\ \bibnamefont {Marques}}, \bibinfo {editor}
  {\bibfnamefont {N.~T.}\ \bibnamefont {Maitra}}, \bibinfo {editor}
  {\bibfnamefont {F.~M.}\ \bibnamefont {Nogueira}}, \bibinfo {editor}
  {\bibfnamefont {E.}~\bibnamefont {Gross}}, \ and\ \bibinfo {editor}
  {\bibfnamefont {A.}~\bibnamefont {Rubio}}}\ (\bibinfo  {publisher} {Springer
  Berlin Heidelberg},\ \bibinfo {year} {2012})\ pp.\ \bibinfo {pages}
  {139--166}\BibitemShut {NoStop}%
\bibitem [{\citenamefont {Tozer}\ \emph {et~al.}(1999)\citenamefont {Tozer},
  \citenamefont {Amos}, \citenamefont {Handy}, \citenamefont {Roos},\ and\
  \citenamefont {Serrano-Andres}}]{TAHR99}%
  \BibitemOpen
  \bibfield  {author} {\bibinfo {author} {\bibfnamefont {D.~J.}\ \bibnamefont
  {Tozer}}, \bibinfo {author} {\bibfnamefont {R.~D.}\ \bibnamefont {Amos}},
  \bibinfo {author} {\bibfnamefont {N.~C.}\ \bibnamefont {Handy}}, \bibinfo
  {author} {\bibfnamefont {B.~O.}\ \bibnamefont {Roos}}, \ and\ \bibinfo
  {author} {\bibfnamefont {L.}~\bibnamefont {Serrano-Andres}},\ }\href@noop {}
  {\bibfield  {journal} {\bibinfo  {journal} {Mol. Phys.}\ }\textbf {\bibinfo
  {volume} {97}},\ \bibinfo {pages} {859} (\bibinfo {year} {1999})}\BibitemShut
  {NoStop}%
\bibitem [{\citenamefont {Peach}\ \emph {et~al.}(2008)\citenamefont {Peach},
  \citenamefont {Benfield}, \citenamefont {Helgaker},\ and\ \citenamefont
  {Tozer}}]{PBHT08}%
  \BibitemOpen
  \bibfield  {author} {\bibinfo {author} {\bibfnamefont {M.~J.~G.}\
  \bibnamefont {Peach}}, \bibinfo {author} {\bibfnamefont {P.}~\bibnamefont
  {Benfield}}, \bibinfo {author} {\bibfnamefont {T.}~\bibnamefont {Helgaker}},
  \ and\ \bibinfo {author} {\bibfnamefont {D.~J.}\ \bibnamefont {Tozer}},\
  }\href@noop {} {\bibfield  {journal} {\bibinfo  {journal} {J. Chem. Phys.}\
  }\textbf {\bibinfo {volume} {128}},\ \bibinfo {eid} {044118} (\bibinfo {year}
  {2008})}\BibitemShut {NoStop}%
\bibitem [{\citenamefont {Guido}\ \emph {et~al.}(2013)\citenamefont {Guido},
  \citenamefont {Cortona}, \citenamefont {Mennucci},\ and\ \citenamefont
  {Adamo}}]{GCMA13}%
  \BibitemOpen
  \bibfield  {author} {\bibinfo {author} {\bibfnamefont {C.~A.}\ \bibnamefont
  {Guido}}, \bibinfo {author} {\bibfnamefont {P.}~\bibnamefont {Cortona}},
  \bibinfo {author} {\bibfnamefont {B.}~\bibnamefont {Mennucci}}, \ and\
  \bibinfo {author} {\bibfnamefont {C.}~\bibnamefont {Adamo}},\ }\href
  {\doibase 10.1021/ct400337e} {\bibfield  {journal} {\bibinfo  {journal}
  {Journal of Chemical Theory and Computation}\ }\textbf {\bibinfo {volume}
  {9}},\ \bibinfo {pages} {3118} (\bibinfo {year} {2013})},\ \bibinfo {note}
  {pMID: 26583991},\ \Eprint
  {http://arxiv.org/abs/http://dx.doi.org/10.1021/ct400337e}
  {http://dx.doi.org/10.1021/ct400337e} \BibitemShut {NoStop}%
\bibitem [{\citenamefont {Li}\ and\ \citenamefont {Ullrich}(2016)}]{LU16}%
  \BibitemOpen
  \bibfield  {author} {\bibinfo {author} {\bibfnamefont {Y.}~\bibnamefont
  {Li}}\ and\ \bibinfo {author} {\bibfnamefont {C.~A.}\ \bibnamefont
  {Ullrich}},\ }\href {\doibase 10.1063/1.4966036} {\bibfield  {journal}
  {\bibinfo  {journal} {The Journal of Chemical Physics}\ }\textbf {\bibinfo
  {volume} {145}},\ \bibinfo {pages} {164107} (\bibinfo {year} {2016})},\
  \Eprint {http://arxiv.org/abs/http://dx.doi.org/10.1063/1.4966036}
  {http://dx.doi.org/10.1063/1.4966036} \BibitemShut {NoStop}%
\bibitem [{\citenamefont {Mewes}\ \emph {et~al.}(2015)\citenamefont {Mewes},
  \citenamefont {Plasser},\ and\ \citenamefont {Dreuw}}]{MPD15}%
  \BibitemOpen
  \bibfield  {author} {\bibinfo {author} {\bibfnamefont {S.~A.}\ \bibnamefont
  {Mewes}}, \bibinfo {author} {\bibfnamefont {F.}~\bibnamefont {Plasser}}, \
  and\ \bibinfo {author} {\bibfnamefont {A.}~\bibnamefont {Dreuw}},\ }\href
  {\doibase 10.1063/1.4935178} {\bibfield  {journal} {\bibinfo  {journal} {The
  Journal of Chemical Physics}\ }\textbf {\bibinfo {volume} {143}},\ \bibinfo
  {pages} {171101} (\bibinfo {year} {2015})},\ \Eprint
  {http://arxiv.org/abs/http://dx.doi.org/10.1063/1.4935178}
  {http://dx.doi.org/10.1063/1.4935178} \BibitemShut {NoStop}%
\bibitem [{\citenamefont {Vasiliev}\ \emph {et~al.}(1999)\citenamefont
  {Vasiliev}, \citenamefont {\"O\ifmmode~\breve{g}\else \u{g}\fi{}\"ut},\ and\
  \citenamefont {Chelikowsky}}]{VOC99}%
  \BibitemOpen
  \bibfield  {author} {\bibinfo {author} {\bibfnamefont {I.}~\bibnamefont
  {Vasiliev}}, \bibinfo {author} {\bibfnamefont {S.}~\bibnamefont
  {\"O\ifmmode~\breve{g}\else \u{g}\fi{}\"ut}}, \ and\ \bibinfo {author}
  {\bibfnamefont {J.~R.}\ \bibnamefont {Chelikowsky}},\ }\href {\doibase
  10.1103/PhysRevLett.82.1919} {\bibfield  {journal} {\bibinfo  {journal}
  {Phys. Rev. Lett.}\ }\textbf {\bibinfo {volume} {82}},\ \bibinfo {pages}
  {1919} (\bibinfo {year} {1999})}\BibitemShut {NoStop}%
\bibitem [{\citenamefont {Appel}\ \emph {et~al.}(2003)\citenamefont {Appel},
  \citenamefont {Gross},\ and\ \citenamefont {Burke}}]{AGB03}%
  \BibitemOpen
  \bibfield  {author} {\bibinfo {author} {\bibfnamefont {H.}~\bibnamefont
  {Appel}}, \bibinfo {author} {\bibfnamefont {E.~K.~U.}\ \bibnamefont {Gross}},
  \ and\ \bibinfo {author} {\bibfnamefont {K.}~\bibnamefont {Burke}},\ }\href
  {\doibase 10.1103/PhysRevLett.90.043005} {\bibfield  {journal} {\bibinfo
  {journal} {Phys. Rev. Lett.}\ }\textbf {\bibinfo {volume} {90}},\ \bibinfo
  {pages} {043005} (\bibinfo {year} {2003})}\BibitemShut {NoStop}%
\bibitem [{\citenamefont {Grabo}\ \emph
  {et~al.}(2000{\natexlab{b}})\citenamefont {Grabo}, \citenamefont {Kreibich},
  \citenamefont {Kurth},\ and\ \citenamefont {Gross}}]{GKKG00}%
  \BibitemOpen
  \bibfield  {author} {\bibinfo {author} {\bibfnamefont {T.}~\bibnamefont
  {Grabo}}, \bibinfo {author} {\bibfnamefont {T.}~\bibnamefont {Kreibich}},
  \bibinfo {author} {\bibfnamefont {S.}~\bibnamefont {Kurth}}, \ and\ \bibinfo
  {author} {\bibfnamefont {E.}~\bibnamefont {Gross}},\ }in\ \href@noop {}
  {\emph {\bibinfo {booktitle} {Strong Coulomb Correlations in Electronic
  Structure Calculations: Beyond the Local Density Approximation}}},\ \bibinfo
  {editor} {edited by\ \bibinfo {editor} {\bibfnamefont {V.}~\bibnamefont
  {Anisimov}}}\ (\bibinfo  {publisher} {Gordon and Breach},\ \bibinfo {address}
  {Amsterdam},\ \bibinfo {year} {2000})\ p.\ \bibinfo {pages} {203}\BibitemShut
  {NoStop}%
\bibitem [{\citenamefont {Perdew}\ and\ \citenamefont {Levy}(1983)}]{PL83}%
  \BibitemOpen
  \bibfield  {author} {\bibinfo {author} {\bibfnamefont {J.~P.}\ \bibnamefont
  {Perdew}}\ and\ \bibinfo {author} {\bibfnamefont {M.}~\bibnamefont {Levy}},\
  }\href {\doibase 10.1103/PhysRevLett.51.1884} {\bibfield  {journal} {\bibinfo
   {journal} {Phys. Rev. Lett.}\ }\textbf {\bibinfo {volume} {51}},\ \bibinfo
  {pages} {1884} (\bibinfo {year} {1983})}\BibitemShut {NoStop}%
\bibitem [{\citenamefont {Sham}\ and\ \citenamefont {Schl\"uter}(1983)}]{SS83}%
  \BibitemOpen
  \bibfield  {author} {\bibinfo {author} {\bibfnamefont {L.~J.}\ \bibnamefont
  {Sham}}\ and\ \bibinfo {author} {\bibfnamefont {M.}~\bibnamefont
  {Schl\"uter}},\ }\href {\doibase 10.1103/PhysRevLett.51.1888} {\bibfield
  {journal} {\bibinfo  {journal} {Phys. Rev. Lett.}\ }\textbf {\bibinfo
  {volume} {51}},\ \bibinfo {pages} {1888} (\bibinfo {year}
  {1983})}\BibitemShut {NoStop}%
\bibitem [{\citenamefont {Perdew}(1985)}]{P85b}%
  \BibitemOpen
  \bibfield  {author} {\bibinfo {author} {\bibfnamefont {J.~P.}\ \bibnamefont
  {Perdew}},\ }in\ \href@noop {} {\emph {\bibinfo {booktitle}
  {Density-Functional Methods In Physics}}},\ \bibinfo {editor} {edited by\
  \bibinfo {editor} {\bibfnamefont {R.~M.}\ \bibnamefont {Dreizler}}\ and\
  \bibinfo {editor} {\bibfnamefont {J.}~\bibnamefont {da~Providencia}}}\
  (\bibinfo  {publisher} {Springer},\ \bibinfo {address} {New York},\ \bibinfo
  {year} {1985})\ p.\ \bibinfo {pages} {265}\BibitemShut {NoStop}%
\bibitem [{\citenamefont {Almbladh}\ and\ \citenamefont {von
  Barth}(1985)}]{AB85}%
  \BibitemOpen
  \bibfield  {author} {\bibinfo {author} {\bibfnamefont {C.-O.}\ \bibnamefont
  {Almbladh}}\ and\ \bibinfo {author} {\bibfnamefont {U.}~\bibnamefont {von
  Barth}},\ }\href@noop {} {\bibfield  {journal} {\bibinfo  {journal} {Phys.
  Rev. B}\ }\textbf {\bibinfo {volume} {31}},\ \bibinfo {pages} {3231}
  (\bibinfo {year} {1985})}\BibitemShut {NoStop}%
\bibitem [{\citenamefont {Perdew}\ \emph {et~al.}(1982)\citenamefont {Perdew},
  \citenamefont {Parr}, \citenamefont {Levy},\ and\ \citenamefont
  {Balduz}}]{PPLB82}%
  \BibitemOpen
  \bibfield  {author} {\bibinfo {author} {\bibfnamefont {J.~P.}\ \bibnamefont
  {Perdew}}, \bibinfo {author} {\bibfnamefont {R.~G.}\ \bibnamefont {Parr}},
  \bibinfo {author} {\bibfnamefont {M.}~\bibnamefont {Levy}}, \ and\ \bibinfo
  {author} {\bibfnamefont {J.~L.}\ \bibnamefont {Balduz}},\ }\href@noop {}
  {\bibfield  {journal} {\bibinfo  {journal} {Phys. Rev. Lett.}\ }\textbf
  {\bibinfo {volume} {49}},\ \bibinfo {pages} {1691} (\bibinfo {year}
  {1982})}\BibitemShut {NoStop}%
\bibitem [{\citenamefont {Seidl}\ \emph {et~al.}(1996)\citenamefont {Seidl},
  \citenamefont {G\"orling}, \citenamefont {Vogl}, \citenamefont {Majewski},\
  and\ \citenamefont {Levy}}]{SGVML96}%
  \BibitemOpen
  \bibfield  {author} {\bibinfo {author} {\bibfnamefont {A.}~\bibnamefont
  {Seidl}}, \bibinfo {author} {\bibfnamefont {A.}~\bibnamefont {G\"orling}},
  \bibinfo {author} {\bibfnamefont {P.}~\bibnamefont {Vogl}}, \bibinfo {author}
  {\bibfnamefont {J.~A.}\ \bibnamefont {Majewski}}, \ and\ \bibinfo {author}
  {\bibfnamefont {M.}~\bibnamefont {Levy}},\ }\href {\doibase
  10.1103/PhysRevB.53.3764} {\bibfield  {journal} {\bibinfo  {journal} {Phys.
  Rev. B}\ }\textbf {\bibinfo {volume} {53}},\ \bibinfo {pages} {3764}
  (\bibinfo {year} {1996})}\BibitemShut {NoStop}%
\bibitem [{\citenamefont {Cohen}\ \emph {et~al.}(2008)\citenamefont {Cohen},
  \citenamefont {Mori-S\'anchez},\ and\ \citenamefont {Yang}}]{CMY08}%
  \BibitemOpen
  \bibfield  {author} {\bibinfo {author} {\bibfnamefont {A.~J.}\ \bibnamefont
  {Cohen}}, \bibinfo {author} {\bibfnamefont {P.}~\bibnamefont
  {Mori-S\'anchez}}, \ and\ \bibinfo {author} {\bibfnamefont {W.}~\bibnamefont
  {Yang}},\ }\href {\doibase 10.1103/PhysRevB.77.115123} {\bibfield  {journal}
  {\bibinfo  {journal} {Phys. Rev. B}\ }\textbf {\bibinfo {volume} {77}},\
  \bibinfo {pages} {115123} (\bibinfo {year} {2008})}\BibitemShut {NoStop}%
\bibitem [{\citenamefont {Mori-S\'anchez}\ \emph {et~al.}(2008)\citenamefont
  {Mori-S\'anchez}, \citenamefont {Cohen},\ and\ \citenamefont {Yang}}]{MCY08}%
  \BibitemOpen
  \bibfield  {author} {\bibinfo {author} {\bibfnamefont {P.}~\bibnamefont
  {Mori-S\'anchez}}, \bibinfo {author} {\bibfnamefont {A.~J.}\ \bibnamefont
  {Cohen}}, \ and\ \bibinfo {author} {\bibfnamefont {W.}~\bibnamefont {Yang}},\
  }\href {\doibase 10.1103/PhysRevLett.100.146401} {\bibfield  {journal}
  {\bibinfo  {journal} {Phys. Rev. Lett.}\ }\textbf {\bibinfo {volume} {100}},\
  \bibinfo {pages} {146401} (\bibinfo {year} {2008})}\BibitemShut {NoStop}%
\bibitem [{\citenamefont {Yang}\ \emph {et~al.}(2012)\citenamefont {Yang},
  \citenamefont {Cohen},\ and\ \citenamefont {Mori-Sánchez}}]{YCM12}%
  \BibitemOpen
  \bibfield  {author} {\bibinfo {author} {\bibfnamefont {W.}~\bibnamefont
  {Yang}}, \bibinfo {author} {\bibfnamefont {A.~J.}\ \bibnamefont {Cohen}}, \
  and\ \bibinfo {author} {\bibfnamefont {P.}~\bibnamefont {Mori-Sánchez}},\
  }\href {\doibase 10.1063/1.3702391} {\bibfield  {journal} {\bibinfo
  {journal} {The Journal of Chemical Physics}\ }\textbf {\bibinfo {volume}
  {136}},\ \bibinfo {pages} {204111} (\bibinfo {year} {2012})},\ \Eprint
  {http://arxiv.org/abs/http://dx.doi.org/10.1063/1.3702391}
  {http://dx.doi.org/10.1063/1.3702391} \BibitemShut {NoStop}%
\bibitem [{\citenamefont {Mundt}\ and\ \citenamefont {K\"ummel}(2005)}]{MK05}%
  \BibitemOpen
  \bibfield  {author} {\bibinfo {author} {\bibfnamefont {M.}~\bibnamefont
  {Mundt}}\ and\ \bibinfo {author} {\bibfnamefont {S.}~\bibnamefont
  {K\"ummel}},\ }\href {\doibase 10.1103/PhysRevLett.95.203004} {\bibfield
  {journal} {\bibinfo  {journal} {Phys. Rev. Lett.}\ }\textbf {\bibinfo
  {volume} {95}},\ \bibinfo {pages} {203004} (\bibinfo {year}
  {2005})}\BibitemShut {NoStop}%
\bibitem [{\citenamefont {Hellgren}\ and\ \citenamefont {Gross}(2012)}]{HG12}%
  \BibitemOpen
  \bibfield  {author} {\bibinfo {author} {\bibfnamefont {M.}~\bibnamefont
  {Hellgren}}\ and\ \bibinfo {author} {\bibfnamefont {E.~K.~U.}\ \bibnamefont
  {Gross}},\ }\href@noop {} {\bibfield  {journal} {\bibinfo  {journal} {Phys.
  Rev. A}\ }\textbf {\bibinfo {volume} {85}},\ \bibinfo {pages} {022514}
  (\bibinfo {year} {2012})}\BibitemShut {NoStop}%
\bibitem [{\citenamefont {Kraisler}\ and\ \citenamefont {Kronik}(2013)}]{KK13}%
  \BibitemOpen
  \bibfield  {author} {\bibinfo {author} {\bibfnamefont {E.}~\bibnamefont
  {Kraisler}}\ and\ \bibinfo {author} {\bibfnamefont {L.}~\bibnamefont
  {Kronik}},\ }\href@noop {} {\bibfield  {journal} {\bibinfo  {journal} {Phys.
  Rev. Lett.}\ }\textbf {\bibinfo {volume} {110}},\ \bibinfo {pages} {126403}
  (\bibinfo {year} {2013})}\BibitemShut {NoStop}%
\bibitem [{\citenamefont {G\"orling}(2015)}]{G15}%
  \BibitemOpen
  \bibfield  {author} {\bibinfo {author} {\bibfnamefont {A.}~\bibnamefont
  {G\"orling}},\ }\href@noop {} {\bibfield  {journal} {\bibinfo  {journal}
  {Phys. Rev. B}\ }\textbf {\bibinfo {volume} {91}},\ \bibinfo {pages} {245120}
  (\bibinfo {year} {2015})}\BibitemShut {NoStop}%
\bibitem [{\citenamefont {Dreuw}\ \emph {et~al.}(2003)\citenamefont {Dreuw},
  \citenamefont {Weisman},\ and\ \citenamefont {Head-Gordon}}]{DWH03}%
  \BibitemOpen
  \bibfield  {author} {\bibinfo {author} {\bibfnamefont {A.}~\bibnamefont
  {Dreuw}}, \bibinfo {author} {\bibfnamefont {J.~L.}\ \bibnamefont {Weisman}},
  \ and\ \bibinfo {author} {\bibfnamefont {M.}~\bibnamefont {Head-Gordon}},\
  }\href@noop {} {\bibfield  {journal} {\bibinfo  {journal} {J. Chem. Phys.}\
  }\textbf {\bibinfo {volume} {119}},\ \bibinfo {pages} {2943} (\bibinfo {year}
  {2003})}\BibitemShut {NoStop}%
\bibitem [{\citenamefont {Tozer}(2003)}]{T03}%
  \BibitemOpen
  \bibfield  {author} {\bibinfo {author} {\bibfnamefont {D.~J.}\ \bibnamefont
  {Tozer}},\ }\href@noop {} {\bibfield  {journal} {\bibinfo  {journal} {J.
  Chem. Phys.}\ }\textbf {\bibinfo {volume} {119}},\ \bibinfo {pages} {12697}
  (\bibinfo {year} {2003})}\BibitemShut {NoStop}%
\bibitem [{\citenamefont {Dreuw}\ and\ \citenamefont
  {Head-Gordon}(2004)}]{DH04}%
  \BibitemOpen
  \bibfield  {author} {\bibinfo {author} {\bibfnamefont {A.}~\bibnamefont
  {Dreuw}}\ and\ \bibinfo {author} {\bibfnamefont {M.}~\bibnamefont
  {Head-Gordon}},\ }\href@noop {} {\bibfield  {journal} {\bibinfo  {journal}
  {J. Am. Chem. Soc.}\ }\textbf {\bibinfo {volume} {126}},\ \bibinfo {pages}
  {4007} (\bibinfo {year} {2004})}\BibitemShut {NoStop}%
\bibitem [{\citenamefont {Gritsenko}\ and\ \citenamefont
  {Baerends}(2004)}]{GB04c}%
  \BibitemOpen
  \bibfield  {author} {\bibinfo {author} {\bibfnamefont {O.}~\bibnamefont
  {Gritsenko}}\ and\ \bibinfo {author} {\bibfnamefont {E.~J.}\ \bibnamefont
  {Baerends}},\ }\href@noop {} {\bibfield  {journal} {\bibinfo  {journal} {J.
  Chem. Phys.}\ }\textbf {\bibinfo {volume} {121}},\ \bibinfo {pages} {655}
  (\bibinfo {year} {2004})}\BibitemShut {NoStop}%
\bibitem [{\citenamefont {Subotnik}(2011)}]{S11}%
  \BibitemOpen
  \bibfield  {author} {\bibinfo {author} {\bibfnamefont {J.~E.}\ \bibnamefont
  {Subotnik}},\ }\href {\doibase 10.1063/1.3627152} {\bibfield  {journal}
  {\bibinfo  {journal} {The Journal of Chemical Physics}\ }\textbf {\bibinfo
  {volume} {135}},\ \bibinfo {pages} {071104} (\bibinfo {year} {2011})},\
  \Eprint {http://arxiv.org/abs/http://dx.doi.org/10.1063/1.3627152}
  {http://dx.doi.org/10.1063/1.3627152} \BibitemShut {NoStop}%
\bibitem [{\citenamefont {He\ss{}elmann}\ \emph {et~al.}(2009)\citenamefont
  {He\ss{}elmann}, \citenamefont {Ipatov},\ and\ \citenamefont
  {G{\"o}rling}}]{HIG09}%
  \BibitemOpen
  \bibfield  {author} {\bibinfo {author} {\bibfnamefont {A.}~\bibnamefont
  {He\ss{}elmann}}, \bibinfo {author} {\bibfnamefont {A.}~\bibnamefont
  {Ipatov}}, \ and\ \bibinfo {author} {\bibfnamefont {A.}~\bibnamefont
  {G{\"o}rling}},\ }\href@noop {} {\bibfield  {journal} {\bibinfo  {journal}
  {Phys. Rev. A}\ }\textbf {\bibinfo {volume} {80}},\ \bibinfo {pages} {012507}
  (\bibinfo {year} {2009})}\BibitemShut {NoStop}%
\bibitem [{\citenamefont {Gimon}\ \emph {et~al.}(2009)\citenamefont {Gimon},
  \citenamefont {Ipatov}, \citenamefont {He{\ss}elmann},\ and\ \citenamefont
  {G\"orling}}]{GIHG09}%
  \BibitemOpen
  \bibfield  {author} {\bibinfo {author} {\bibfnamefont {T.}~\bibnamefont
  {Gimon}}, \bibinfo {author} {\bibfnamefont {A.}~\bibnamefont {Ipatov}},
  \bibinfo {author} {\bibfnamefont {A.}~\bibnamefont {He{\ss}elmann}}, \ and\
  \bibinfo {author} {\bibfnamefont {A.}~\bibnamefont {G\"orling}},\ }\href
  {\doibase 10.1021/ct800539a} {\bibfield  {journal} {\bibinfo  {journal}
  {Journal of Chemical Theory and Computation}\ }\textbf {\bibinfo {volume}
  {5}},\ \bibinfo {pages} {781} (\bibinfo {year} {2009})},\ \bibinfo {note}
  {pMID: 26609583},\ \Eprint
  {http://arxiv.org/abs/http://dx.doi.org/10.1021/ct800539a}
  {http://dx.doi.org/10.1021/ct800539a} \BibitemShut {NoStop}%
\bibitem [{\citenamefont {G\"orling}(1998)}]{G98}%
  \BibitemOpen
  \bibfield  {author} {\bibinfo {author} {\bibfnamefont {A.}~\bibnamefont
  {G\"orling}},\ }\href {\doibase 10.1103/PhysRevA.57.3433} {\bibfield
  {journal} {\bibinfo  {journal} {Phys. Rev. A}\ }\textbf {\bibinfo {volume}
  {57}},\ \bibinfo {pages} {3433} (\bibinfo {year} {1998})}\BibitemShut
  {NoStop}%
\bibitem [{\citenamefont {Hellgren}\ and\ \citenamefont {Gross}(2013)}]{HG13}%
  \BibitemOpen
  \bibfield  {author} {\bibinfo {author} {\bibfnamefont {M.}~\bibnamefont
  {Hellgren}}\ and\ \bibinfo {author} {\bibfnamefont {E.~K.~U.}\ \bibnamefont
  {Gross}},\ }\href {\doibase 10.1103/PhysRevA.88.052507} {\bibfield  {journal}
  {\bibinfo  {journal} {Phys. Rev. A}\ }\textbf {\bibinfo {volume} {88}},\
  \bibinfo {pages} {052507} (\bibinfo {year} {2013})}\BibitemShut {NoStop}%
\bibitem [{\citenamefont {G\"orling}\ and\ \citenamefont {Levy}(1993)}]{GL93}%
  \BibitemOpen
  \bibfield  {author} {\bibinfo {author} {\bibfnamefont {A.}~\bibnamefont
  {G\"orling}}\ and\ \bibinfo {author} {\bibfnamefont {M.}~\bibnamefont
  {Levy}},\ }\href {\doibase 10.1103/PhysRevB.47.13105} {\bibfield  {journal}
  {\bibinfo  {journal} {Phys. Rev. B}\ }\textbf {\bibinfo {volume} {47}},\
  \bibinfo {pages} {13105} (\bibinfo {year} {1993})}\BibitemShut {NoStop}%
\bibitem [{\citenamefont {Gonze}\ and\ \citenamefont {Scheffler}(1999)}]{GS99}%
  \BibitemOpen
  \bibfield  {author} {\bibinfo {author} {\bibfnamefont {X.}~\bibnamefont
  {Gonze}}\ and\ \bibinfo {author} {\bibfnamefont {M.}~\bibnamefont
  {Scheffler}},\ }\href {\doibase 10.1103/PhysRevLett.82.4416} {\bibfield
  {journal} {\bibinfo  {journal} {Phys. Rev. Lett.}\ }\textbf {\bibinfo
  {volume} {82}},\ \bibinfo {pages} {4416} (\bibinfo {year}
  {1999})}\BibitemShut {NoStop}%
\bibitem [{\citenamefont {Perdew}(1990)}]{P90}%
  \BibitemOpen
  \bibfield  {author} {\bibinfo {author} {\bibfnamefont {J.~P.}\ \bibnamefont
  {Perdew}},\ }\href {\doibase http://dx.doi.org/10.1016/S0065-3276(08)60594-8}
  {\bibfield  {journal} {\bibinfo  {journal} {Advances in Quantum Chemistry}\
  }\textbf {\bibinfo {volume} {21}},\ \bibinfo {pages} {113 } (\bibinfo {year}
  {1990})},\ \bibinfo {note} {density Functional Theory of Many-Fermion
  Systems}\BibitemShut {NoStop}%
\bibitem [{\citenamefont {K\"orzd\"orfer}\ \emph {et~al.}(2008)\citenamefont
  {K\"orzd\"orfer}, \citenamefont {K\"ummel},\ and\ \citenamefont
  {Mundt}}]{KKM08}%
  \BibitemOpen
  \bibfield  {author} {\bibinfo {author} {\bibfnamefont {T.}~\bibnamefont
  {K\"orzd\"orfer}}, \bibinfo {author} {\bibfnamefont {S.}~\bibnamefont
  {K\"ummel}}, \ and\ \bibinfo {author} {\bibfnamefont {M.}~\bibnamefont
  {Mundt}},\ }\href {\doibase 10.1063/1.2944272} {\bibfield  {journal}
  {\bibinfo  {journal} {The Journal of Chemical Physics}\ }\textbf {\bibinfo
  {volume} {129}},\ \bibinfo {pages} {014110} (\bibinfo {year} {2008})},\
  \Eprint {http://arxiv.org/abs/http://dx.doi.org/10.1063/1.2944272}
  {http://dx.doi.org/10.1063/1.2944272} \BibitemShut {NoStop}%
\bibitem [{\citenamefont {Becke}(1993)}]{B93}%
  \BibitemOpen
  \bibfield  {author} {\bibinfo {author} {\bibfnamefont {A.}~\bibnamefont
  {Becke}},\ }\href@noop {} {\bibfield  {journal} {\bibinfo  {journal} {J.
  Chem. Phys.}\ }\textbf {\bibinfo {volume} {98}},\ \bibinfo {pages} {5648}
  (\bibinfo {year} {1993})}\BibitemShut {NoStop}%
\bibitem [{\citenamefont {Lee}\ \emph {et~al.}(1988)\citenamefont {Lee},
  \citenamefont {Yang},\ and\ \citenamefont {Parr}}]{LYP88}%
  \BibitemOpen
  \bibfield  {author} {\bibinfo {author} {\bibfnamefont {C.}~\bibnamefont
  {Lee}}, \bibinfo {author} {\bibfnamefont {W.}~\bibnamefont {Yang}}, \ and\
  \bibinfo {author} {\bibfnamefont {R.~G.}\ \bibnamefont {Parr}},\ }\href
  {\doibase 10.1103/PhysRevB.37.785} {\bibfield  {journal} {\bibinfo  {journal}
  {Phys. Rev. B}\ }\textbf {\bibinfo {volume} {37}},\ \bibinfo {pages} {785}
  (\bibinfo {year} {1988})}\BibitemShut {NoStop}%
\bibitem [{\citenamefont {Leininger}\ \emph {et~al.}(1997)\citenamefont
  {Leininger}, \citenamefont {Stoll}, \citenamefont {Werner},\ and\
  \citenamefont {Savin}}]{LSWS97}%
  \BibitemOpen
  \bibfield  {author} {\bibinfo {author} {\bibfnamefont {T.}~\bibnamefont
  {Leininger}}, \bibinfo {author} {\bibfnamefont {H.}~\bibnamefont {Stoll}},
  \bibinfo {author} {\bibfnamefont {H.-J.}\ \bibnamefont {Werner}}, \ and\
  \bibinfo {author} {\bibfnamefont {A.}~\bibnamefont {Savin}},\ }\href@noop {}
  {\bibfield  {journal} {\bibinfo  {journal} {Chem. Phys. Lett.}\ }\textbf
  {\bibinfo {volume} {275}},\ \bibinfo {pages} {151 } (\bibinfo {year}
  {1997})}\BibitemShut {NoStop}%
\bibitem [{\citenamefont {Savin}(1996)}]{Savin96}%
  \BibitemOpen
  \bibfield  {author} {\bibinfo {author} {\bibfnamefont {A.}~\bibnamefont
  {Savin}},\ }in\ \href@noop {} {\emph {\bibinfo {booktitle} {Recent
  Developments and Applications of Modern Density Functional Theory}}},\
  \bibinfo {editor} {edited by\ \bibinfo {editor} {\bibfnamefont {J.~M.}\
  \bibnamefont {Seminario}}}\ (\bibinfo  {publisher} {Elsevier},\ \bibinfo
  {year} {1996})\ pp.\ \bibinfo {pages} {327--354}\BibitemShut {NoStop}%
\bibitem [{\citenamefont {G\"orling}\ and\ \citenamefont {Levy}(1997)}]{GL97}%
  \BibitemOpen
  \bibfield  {author} {\bibinfo {author} {\bibfnamefont {A.}~\bibnamefont
  {G\"orling}}\ and\ \bibinfo {author} {\bibfnamefont {M.}~\bibnamefont
  {Levy}},\ }\href {\doibase 10.1063/1.473369} {\bibfield  {journal} {\bibinfo
  {journal} {The Journal of Chemical Physics}\ }\textbf {\bibinfo {volume}
  {106}},\ \bibinfo {pages} {2675} (\bibinfo {year} {1997})},\ \Eprint
  {http://arxiv.org/abs/http://dx.doi.org/10.1063/1.473369}
  {http://dx.doi.org/10.1063/1.473369} \BibitemShut {NoStop}%
\bibitem [{\citenamefont {Tretiak}\ and\ \citenamefont
  {Chernyak}(2003)}]{TC03}%
  \BibitemOpen
  \bibfield  {author} {\bibinfo {author} {\bibfnamefont {S.}~\bibnamefont
  {Tretiak}}\ and\ \bibinfo {author} {\bibfnamefont {V.}~\bibnamefont
  {Chernyak}},\ }\href {\doibase 10.1063/1.1614240} {\bibfield  {journal}
  {\bibinfo  {journal} {The Journal of Chemical Physics}\ }\textbf {\bibinfo
  {volume} {119}},\ \bibinfo {pages} {8809} (\bibinfo {year} {2003})},\ \Eprint
  {http://arxiv.org/abs/http://dx.doi.org/10.1063/1.1614240}
  {http://dx.doi.org/10.1063/1.1614240} \BibitemShut {NoStop}%
\bibitem [{\citenamefont {Grimme}\ and\ \citenamefont {Neese}(2007)}]{GN07}%
  \BibitemOpen
  \bibfield  {author} {\bibinfo {author} {\bibfnamefont {S.}~\bibnamefont
  {Grimme}}\ and\ \bibinfo {author} {\bibfnamefont {F.}~\bibnamefont {Neese}},\
  }\href {\doibase 10.1063/1.2772854} {\bibfield  {journal} {\bibinfo
  {journal} {The Journal of Chemical Physics}\ }\textbf {\bibinfo {volume}
  {127}},\ \bibinfo {pages} {154116} (\bibinfo {year} {2007})},\ \Eprint
  {http://arxiv.org/abs/http://dx.doi.org/10.1063/1.2772854}
  {http://dx.doi.org/10.1063/1.2772854} \BibitemShut {NoStop}%
\bibitem [{\citenamefont {Zhao}\ \emph {et~al.}(2006)\citenamefont {Zhao}, ,\
  and\ \citenamefont {Truhlar}}]{ZT06}%
  \BibitemOpen
  \bibfield  {author} {\bibinfo {author} {\bibfnamefont {Y.}~\bibnamefont
  {Zhao}}, , \ and\ \bibinfo {author} {\bibfnamefont {D.~G.}\ \bibnamefont
  {Truhlar}},\ }\href@noop {} {\bibfield  {journal} {\bibinfo  {journal} {J.
  Phys. Chem. A}\ }\textbf {\bibinfo {volume} {110}},\ \bibinfo {pages} {13126}
  (\bibinfo {year} {2006})}\BibitemShut {NoStop}%
\bibitem [{\citenamefont {Yu}\ \emph {et~al.}(2016)\citenamefont {Yu},
  \citenamefont {He}, \citenamefont {Li},\ and\ \citenamefont
  {Truhlar}}]{YHLT15}%
  \BibitemOpen
  \bibfield  {author} {\bibinfo {author} {\bibfnamefont {H.~S.}\ \bibnamefont
  {Yu}}, \bibinfo {author} {\bibfnamefont {X.}~\bibnamefont {He}}, \bibinfo
  {author} {\bibfnamefont {S.~L.}\ \bibnamefont {Li}}, \ and\ \bibinfo {author}
  {\bibfnamefont {D.~G.}\ \bibnamefont {Truhlar}},\ }\href {\doibase
  10.1039/C6SC00705H} {\bibfield  {journal} {\bibinfo  {journal} {Chem. Sci.}\
  }\textbf {\bibinfo {volume} {7}},\ \bibinfo {pages} {5032} (\bibinfo {year}
  {2016})}\BibitemShut {NoStop}%
\bibitem [{\citenamefont {Stoll}\ and\ \citenamefont {Savin}(1985)}]{SS85}%
  \BibitemOpen
  \bibfield  {author} {\bibinfo {author} {\bibfnamefont {H.}~\bibnamefont
  {Stoll}}\ and\ \bibinfo {author} {\bibfnamefont {A.}~\bibnamefont {Savin}},\
  }in\ \href@noop {} {\emph {\bibinfo {booktitle} {Density Functional Methods
  in Physics}}},\ \bibinfo {editor} {edited by\ \bibinfo {editor}
  {\bibfnamefont {R.}~\bibnamefont {Dreizler}}\ and\ \bibinfo {editor}
  {\bibfnamefont {J.}~\bibnamefont {da~Providencia}}}\ (\bibinfo  {publisher}
  {Plenum, New York},\ \bibinfo {year} {1985})\ p.\ \bibinfo {pages}
  {177}\BibitemShut {NoStop}%
\bibitem [{\citenamefont {Henderson}\ \emph
  {et~al.}(2008{\natexlab{a}})\citenamefont {Henderson}, \citenamefont
  {Janesko},\ and\ \citenamefont {Scuseria}}]{HJS08}%
  \BibitemOpen
  \bibfield  {author} {\bibinfo {author} {\bibfnamefont {T.~M.}\ \bibnamefont
  {Henderson}}, \bibinfo {author} {\bibfnamefont {B.~G.}\ \bibnamefont
  {Janesko}}, \ and\ \bibinfo {author} {\bibfnamefont {G.~E.}\ \bibnamefont
  {Scuseria}},\ }\href {\doibase 10.1021/jp806573k} {\bibfield  {journal}
  {\bibinfo  {journal} {The Journal of Physical Chemistry A}\ }\textbf
  {\bibinfo {volume} {112}},\ \bibinfo {pages} {12530} (\bibinfo {year}
  {2008}{\natexlab{a}})},\ \bibinfo {note} {pMID: 19006280},\ \Eprint
  {http://arxiv.org/abs/http://dx.doi.org/10.1021/jp806573k}
  {http://dx.doi.org/10.1021/jp806573k} \BibitemShut {NoStop}%
\bibitem [{\citenamefont {Kronik}\ \emph {et~al.}(2012)\citenamefont {Kronik},
  \citenamefont {Stein}, \citenamefont {Refaely-Abramson},\ and\ \citenamefont
  {Baer}}]{KSRB12}%
  \BibitemOpen
  \bibfield  {author} {\bibinfo {author} {\bibfnamefont {L.}~\bibnamefont
  {Kronik}}, \bibinfo {author} {\bibfnamefont {T.}~\bibnamefont {Stein}},
  \bibinfo {author} {\bibfnamefont {S.}~\bibnamefont {Refaely-Abramson}}, \
  and\ \bibinfo {author} {\bibfnamefont {R.}~\bibnamefont {Baer}},\ }\href@noop
  {} {\bibfield  {journal} {\bibinfo  {journal} {J. Chem. Theory and Comput.}\
  }\textbf {\bibinfo {volume} {8}},\ \bibinfo {pages} {1515} (\bibinfo {year}
  {2012})}\BibitemShut {NoStop}%
\bibitem [{\citenamefont {Tawada}\ \emph {et~al.}(2004)\citenamefont {Tawada},
  \citenamefont {Tsuneda}, \citenamefont {Yanagisawa}, \citenamefont {Yanai},\
  and\ \citenamefont {Hirao}}]{TTYY04}%
  \BibitemOpen
  \bibfield  {author} {\bibinfo {author} {\bibfnamefont {Y.}~\bibnamefont
  {Tawada}}, \bibinfo {author} {\bibfnamefont {T.}~\bibnamefont {Tsuneda}},
  \bibinfo {author} {\bibfnamefont {S.}~\bibnamefont {Yanagisawa}}, \bibinfo
  {author} {\bibfnamefont {T.}~\bibnamefont {Yanai}}, \ and\ \bibinfo {author}
  {\bibfnamefont {K.}~\bibnamefont {Hirao}},\ }\href@noop {} {\bibfield
  {journal} {\bibinfo  {journal} {J. Chem. Phys.}\ }\textbf {\bibinfo {volume}
  {120}},\ \bibinfo {pages} {8425} (\bibinfo {year} {2004})}\BibitemShut
  {NoStop}%
\bibitem [{\citenamefont {Chiba}\ \emph {et~al.}(2007)\citenamefont {Chiba},
  \citenamefont {Tsuneda},\ and\ \citenamefont {Hirao}}]{CTH07}%
  \BibitemOpen
  \bibfield  {author} {\bibinfo {author} {\bibfnamefont {M.}~\bibnamefont
  {Chiba}}, \bibinfo {author} {\bibfnamefont {T.}~\bibnamefont {Tsuneda}}, \
  and\ \bibinfo {author} {\bibfnamefont {K.}~\bibnamefont {Hirao}},\ }\href
  {\doibase 10.1063/1.2426335} {\bibfield  {journal} {\bibinfo  {journal} {The
  Journal of Chemical Physics}\ }\textbf {\bibinfo {volume} {126}},\ \bibinfo
  {pages} {034504} (\bibinfo {year} {2007})},\ \Eprint
  {http://arxiv.org/abs/http://dx.doi.org/10.1063/1.2426335}
  {http://dx.doi.org/10.1063/1.2426335} \BibitemShut {NoStop}%
\bibitem [{\citenamefont {Iikura}\ \emph {et~al.}(2001)\citenamefont {Iikura},
  \citenamefont {Tsuneda}, \citenamefont {Yanai},\ and\ \citenamefont
  {Hirao}}]{ITYH01}%
  \BibitemOpen
  \bibfield  {author} {\bibinfo {author} {\bibfnamefont {H.}~\bibnamefont
  {Iikura}}, \bibinfo {author} {\bibfnamefont {T.}~\bibnamefont {Tsuneda}},
  \bibinfo {author} {\bibfnamefont {T.}~\bibnamefont {Yanai}}, \ and\ \bibinfo
  {author} {\bibfnamefont {K.}~\bibnamefont {Hirao}},\ }\href {\doibase
  10.1063/1.1383587} {\bibfield  {journal} {\bibinfo  {journal} {The Journal of
  Chemical Physics}\ }\textbf {\bibinfo {volume} {115}},\ \bibinfo {pages}
  {3540} (\bibinfo {year} {2001})},\ \Eprint
  {http://arxiv.org/abs/http://dx.doi.org/10.1063/1.1383587}
  {http://dx.doi.org/10.1063/1.1383587} \BibitemShut {NoStop}%
\bibitem [{\citenamefont {Yanai}\ \emph {et~al.}(2004)\citenamefont {Yanai},
  \citenamefont {Tew},\ and\ \citenamefont {Handy}}]{YTH04}%
  \BibitemOpen
  \bibfield  {author} {\bibinfo {author} {\bibfnamefont {T.}~\bibnamefont
  {Yanai}}, \bibinfo {author} {\bibfnamefont {D.~P.}\ \bibnamefont {Tew}}, \
  and\ \bibinfo {author} {\bibfnamefont {N.~C.}\ \bibnamefont {Handy}},\
  }\href@noop {} {\bibfield  {journal} {\bibinfo  {journal} {Chem. Phys.
  Lett.}\ }\textbf {\bibinfo {volume} {393}},\ \bibinfo {pages} {51 } (\bibinfo
  {year} {2004})}\BibitemShut {NoStop}%
\bibitem [{\citenamefont {Rohrdanz}\ \emph {et~al.}(2009)\citenamefont
  {Rohrdanz}, \citenamefont {Martins},\ and\ \citenamefont {Herbert}}]{RMH09}%
  \BibitemOpen
  \bibfield  {author} {\bibinfo {author} {\bibfnamefont {M.~A.}\ \bibnamefont
  {Rohrdanz}}, \bibinfo {author} {\bibfnamefont {K.~M.}\ \bibnamefont
  {Martins}}, \ and\ \bibinfo {author} {\bibfnamefont {J.~M.}\ \bibnamefont
  {Herbert}},\ }\href@noop {} {\bibfield  {journal} {\bibinfo  {journal} {J.
  Chem. Phys.}\ }\textbf {\bibinfo {volume} {130}},\ \bibinfo {eid} {054112}
  (\bibinfo {year} {2009})}\BibitemShut {NoStop}%
\bibitem [{\citenamefont {Henderson}\ \emph
  {et~al.}(2008{\natexlab{b}})\citenamefont {Henderson}, \citenamefont
  {Janesko},\ and\ \citenamefont {Scuseria}}]{HJS08b}%
  \BibitemOpen
  \bibfield  {author} {\bibinfo {author} {\bibfnamefont {T.~M.}\ \bibnamefont
  {Henderson}}, \bibinfo {author} {\bibfnamefont {B.~G.}\ \bibnamefont
  {Janesko}}, \ and\ \bibinfo {author} {\bibfnamefont {G.~E.}\ \bibnamefont
  {Scuseria}},\ }\href {\doibase 10.1063/1.2921797} {\bibfield  {journal}
  {\bibinfo  {journal} {The Journal of Chemical Physics}\ }\textbf {\bibinfo
  {volume} {128}},\ \bibinfo {pages} {194105} (\bibinfo {year}
  {2008}{\natexlab{b}})},\ \Eprint
  {http://arxiv.org/abs/http://dx.doi.org/10.1063/1.2921797}
  {http://dx.doi.org/10.1063/1.2921797} \BibitemShut {NoStop}%
\bibitem [{\citenamefont {Baer}\ and\ \citenamefont {Neuhauser}(2005)}]{BN05}%
  \BibitemOpen
  \bibfield  {author} {\bibinfo {author} {\bibfnamefont {R.}~\bibnamefont
  {Baer}}\ and\ \bibinfo {author} {\bibfnamefont {D.}~\bibnamefont
  {Neuhauser}},\ }\href {\doibase 10.1103/PhysRevLett.94.043002} {\bibfield
  {journal} {\bibinfo  {journal} {Phys. Rev. Lett.}\ }\textbf {\bibinfo
  {volume} {94}},\ \bibinfo {pages} {043002} (\bibinfo {year}
  {2005})}\BibitemShut {NoStop}%
\bibitem [{\citenamefont {Baer}\ \emph {et~al.}(2006)\citenamefont {Baer},
  \citenamefont {Livshits},\ and\ \citenamefont {Neuhauser}}]{BLN06}%
  \BibitemOpen
  \bibfield  {author} {\bibinfo {author} {\bibfnamefont {R.}~\bibnamefont
  {Baer}}, \bibinfo {author} {\bibfnamefont {E.}~\bibnamefont {Livshits}}, \
  and\ \bibinfo {author} {\bibfnamefont {D.}~\bibnamefont {Neuhauser}},\ }\href
  {\doibase http://dx.doi.org/10.1016/j.chemphys.2006.06.041} {\bibfield
  {journal} {\bibinfo  {journal} {Chemical Physics}\ }\textbf {\bibinfo
  {volume} {329}},\ \bibinfo {pages} {266 } (\bibinfo {year} {2006})},\
  \bibinfo {note} {electron Correlation and Multimode Dynamics in
  Molecules}\BibitemShut {NoStop}%
\bibitem [{\citenamefont {Baer}\ \emph {et~al.}(2010)\citenamefont {Baer},
  \citenamefont {Livshits},\ and\ \citenamefont {Salzner}}]{BLS10}%
  \BibitemOpen
  \bibfield  {author} {\bibinfo {author} {\bibfnamefont {R.}~\bibnamefont
  {Baer}}, \bibinfo {author} {\bibfnamefont {E.}~\bibnamefont {Livshits}}, \
  and\ \bibinfo {author} {\bibfnamefont {U.}~\bibnamefont {Salzner}},\
  }\href@noop {} {\bibfield  {journal} {\bibinfo  {journal} {Ann. Rev. Phys.
  Chem.}\ }\textbf {\bibinfo {volume} {61}},\ \bibinfo {pages} {85} (\bibinfo
  {year} {2010})}\BibitemShut {NoStop}%
\bibitem [{\citenamefont {Stein}\ \emph {et~al.}(2009)\citenamefont {Stein},
  \citenamefont {Kronik},\ and\ \citenamefont {Baer}}]{SKB09}%
  \BibitemOpen
  \bibfield  {author} {\bibinfo {author} {\bibfnamefont {T.}~\bibnamefont
  {Stein}}, \bibinfo {author} {\bibfnamefont {L.}~\bibnamefont {Kronik}}, \
  and\ \bibinfo {author} {\bibfnamefont {R.}~\bibnamefont {Baer}},\ }\href@noop
  {} {\bibfield  {journal} {\bibinfo  {journal} {J. Am. Chem. Soc.}\ }\textbf
  {\bibinfo {volume} {131}},\ \bibinfo {pages} {2818} (\bibinfo {year}
  {2009})}\BibitemShut {NoStop}%
\bibitem [{\citenamefont {K\"orzd\"orfer}\ and\ \citenamefont
  {Br\'edas}(2014)}]{KB14}%
  \BibitemOpen
  \bibfield  {author} {\bibinfo {author} {\bibfnamefont {T.}~\bibnamefont
  {K\"orzd\"orfer}}\ and\ \bibinfo {author} {\bibfnamefont {J.-L.}\
  \bibnamefont {Br\'edas}},\ }\href {\doibase 10.1021/ar500021t} {\bibfield
  {journal} {\bibinfo  {journal} {Accounts of Chemical Research}\ }\textbf
  {\bibinfo {volume} {47}},\ \bibinfo {pages} {3284} (\bibinfo {year}
  {2014})},\ \bibinfo {note} {pMID: 24784485},\ \Eprint
  {http://arxiv.org/abs/http://dx.doi.org/10.1021/ar500021t}
  {http://dx.doi.org/10.1021/ar500021t} \BibitemShut {NoStop}%
\bibitem [{\citenamefont {Egger}\ \emph {et~al.}(2014)\citenamefont {Egger},
  \citenamefont {Weissman}, \citenamefont {Refaely-Abramson}, \citenamefont
  {Sharifzadeh}, \citenamefont {Dauth}, \citenamefont {Baer}, \citenamefont
  {Kümmel}, \citenamefont {Neaton}, \citenamefont {Zojer},\ and\ \citenamefont
  {Kronik}}]{EWRS14}%
  \BibitemOpen
  \bibfield  {author} {\bibinfo {author} {\bibfnamefont {D.~A.}\ \bibnamefont
  {Egger}}, \bibinfo {author} {\bibfnamefont {S.}~\bibnamefont {Weissman}},
  \bibinfo {author} {\bibfnamefont {S.}~\bibnamefont {Refaely-Abramson}},
  \bibinfo {author} {\bibfnamefont {S.}~\bibnamefont {Sharifzadeh}}, \bibinfo
  {author} {\bibfnamefont {M.}~\bibnamefont {Dauth}}, \bibinfo {author}
  {\bibfnamefont {R.}~\bibnamefont {Baer}}, \bibinfo {author} {\bibfnamefont
  {S.}~\bibnamefont {Kümmel}}, \bibinfo {author} {\bibfnamefont {J.~B.}\
  \bibnamefont {Neaton}}, \bibinfo {author} {\bibfnamefont {E.}~\bibnamefont
  {Zojer}}, \ and\ \bibinfo {author} {\bibfnamefont {L.}~\bibnamefont
  {Kronik}},\ }\href {\doibase 10.1021/ct400956h} {\bibfield  {journal}
  {\bibinfo  {journal} {Journal of Chemical Theory and Computation}\ }\textbf
  {\bibinfo {volume} {10}},\ \bibinfo {pages} {1934} (\bibinfo {year}
  {2014})},\ \bibinfo {note} {pMID: 24839410},\ \Eprint
  {http://arxiv.org/abs/http://dx.doi.org/10.1021/ct400956h}
  {http://dx.doi.org/10.1021/ct400956h} \BibitemShut {NoStop}%
\bibitem [{\citenamefont {Manna}\ \emph
  {et~al.}(2015{\natexlab{a}})\citenamefont {Manna}, \citenamefont {Lee},
  \citenamefont {McMahon},\ and\ \citenamefont {Dunietz}}]{MLMD15}%
  \BibitemOpen
  \bibfield  {author} {\bibinfo {author} {\bibfnamefont {A.~K.}\ \bibnamefont
  {Manna}}, \bibinfo {author} {\bibfnamefont {M.~H.}\ \bibnamefont {Lee}},
  \bibinfo {author} {\bibfnamefont {K.~L.}\ \bibnamefont {McMahon}}, \ and\
  \bibinfo {author} {\bibfnamefont {B.~D.}\ \bibnamefont {Dunietz}},\ }\href
  {\doibase 10.1021/ct501018n} {\bibfield  {journal} {\bibinfo  {journal}
  {Journal of Chemical Theory and Computation}\ }\textbf {\bibinfo {volume}
  {11}},\ \bibinfo {pages} {1110} (\bibinfo {year} {2015}{\natexlab{a}})},\
  \bibinfo {note} {pMID: 26579761},\ \Eprint
  {http://arxiv.org/abs/http://dx.doi.org/10.1021/ct501018n}
  {http://dx.doi.org/10.1021/ct501018n} \BibitemShut {NoStop}%
\bibitem [{\citenamefont {Sears}\ \emph {et~al.}(2011)\citenamefont {Sears},
  \citenamefont {K\"orzdoerfer}, \citenamefont {Zhang},\ and\ \citenamefont
  {Br\'edas}}]{SKZB11}%
  \BibitemOpen
  \bibfield  {author} {\bibinfo {author} {\bibfnamefont {J.~S.}\ \bibnamefont
  {Sears}}, \bibinfo {author} {\bibfnamefont {T.}~\bibnamefont
  {K\"orzdoerfer}}, \bibinfo {author} {\bibfnamefont {C.-R.}\ \bibnamefont
  {Zhang}}, \ and\ \bibinfo {author} {\bibfnamefont {J.-L.}\ \bibnamefont
  {Br\'edas}},\ }\href {\doibase 10.1063/1.3656734} {\bibfield  {journal}
  {\bibinfo  {journal} {The Journal of Chemical Physics}\ }\textbf {\bibinfo
  {volume} {135}},\ \bibinfo {pages} {151103} (\bibinfo {year} {2011})},\
  \Eprint {http://arxiv.org/abs/http://dx.doi.org/10.1063/1.3656734}
  {http://dx.doi.org/10.1063/1.3656734} \BibitemShut {NoStop}%
\bibitem [{\citenamefont {Peach}\ \emph {et~al.}(2011)\citenamefont {Peach},
  \citenamefont {Williamson},\ and\ \citenamefont {Tozer}}]{PWT11}%
  \BibitemOpen
  \bibfield  {author} {\bibinfo {author} {\bibfnamefont {M.~J.~G.}\
  \bibnamefont {Peach}}, \bibinfo {author} {\bibfnamefont {M.~J.}\ \bibnamefont
  {Williamson}}, \ and\ \bibinfo {author} {\bibfnamefont {D.~J.}\ \bibnamefont
  {Tozer}},\ }\href {\doibase 10.1021/ct200651r} {\bibfield  {journal}
  {\bibinfo  {journal} {Journal of Chemical Theory and Computation}\ }\textbf
  {\bibinfo {volume} {7}},\ \bibinfo {pages} {3578} (\bibinfo {year} {2011})},\
  \bibinfo {note} {pMID: 26598256},\ \Eprint
  {http://arxiv.org/abs/http://dx.doi.org/10.1021/ct200651r}
  {http://dx.doi.org/10.1021/ct200651r} \BibitemShut {NoStop}%
\bibitem [{\citenamefont {Karolewski}\ \emph {et~al.}(2013)\citenamefont
  {Karolewski}, \citenamefont {Kronik},\ and\ \citenamefont
  {K{\"u}mmel}}]{KKK13}%
  \BibitemOpen
  \bibfield  {author} {\bibinfo {author} {\bibfnamefont {A.}~\bibnamefont
  {Karolewski}}, \bibinfo {author} {\bibfnamefont {L.}~\bibnamefont {Kronik}},
  \ and\ \bibinfo {author} {\bibfnamefont {S.}~\bibnamefont {K{\"u}mmel}},\
  }\href@noop {} {\bibfield  {journal} {\bibinfo  {journal} {J. Chem. Phys.}\
  }\textbf {\bibinfo {volume} {138}},\ \bibinfo {eid} {204115} (\bibinfo {year}
  {2013})}\BibitemShut {NoStop}%
\bibitem [{\citenamefont {Maitra}(2005)}]{M05c}%
  \BibitemOpen
  \bibfield  {author} {\bibinfo {author} {\bibfnamefont {N.~T.}\ \bibnamefont
  {Maitra}},\ }\href@noop {} {\bibfield  {journal} {\bibinfo  {journal} {J.
  Chem. Phys.}\ }\textbf {\bibinfo {volume} {122}},\ \bibinfo {eid} {234104}
  (\bibinfo {year} {2005})}\BibitemShut {NoStop}%
\bibitem [{\citenamefont {Maitra}\ and\ \citenamefont {Tempel}(2006)}]{MT06}%
  \BibitemOpen
  \bibfield  {author} {\bibinfo {author} {\bibfnamefont {N.~T.}\ \bibnamefont
  {Maitra}}\ and\ \bibinfo {author} {\bibfnamefont {D.~G.}\ \bibnamefont
  {Tempel}},\ }\href@noop {} {\bibfield  {journal} {\bibinfo  {journal} {J.
  Chem. Phys.}\ }\textbf {\bibinfo {volume} {125}},\ \bibinfo {eid} {184111}
  (\bibinfo {year} {2006})}\BibitemShut {NoStop}%
\bibitem [{\citenamefont {Gritsenko}\ and\ \citenamefont
  {Baerends}(1996)}]{GB96}%
  \BibitemOpen
  \bibfield  {author} {\bibinfo {author} {\bibfnamefont {O.~V.}\ \bibnamefont
  {Gritsenko}}\ and\ \bibinfo {author} {\bibfnamefont {E.~J.}\ \bibnamefont
  {Baerends}},\ }\href {\doibase 10.1103/PhysRevA.54.1957} {\bibfield
  {journal} {\bibinfo  {journal} {Phys. Rev. A}\ }\textbf {\bibinfo {volume}
  {54}},\ \bibinfo {pages} {1957} (\bibinfo {year} {1996})}\BibitemShut
  {NoStop}%
\bibitem [{\citenamefont {Tempel}\ \emph {et~al.}(2009)\citenamefont {Tempel},
  \citenamefont {Martinez},\ and\ \citenamefont {Maitra}}]{TMM09}%
  \BibitemOpen
  \bibfield  {author} {\bibinfo {author} {\bibfnamefont {D.~G.}\ \bibnamefont
  {Tempel}}, \bibinfo {author} {\bibfnamefont {T.~J.}\ \bibnamefont
  {Martinez}}, \ and\ \bibinfo {author} {\bibfnamefont {N.~T.}\ \bibnamefont
  {Maitra}},\ }\href {\doibase 10.1021/ct800535c} {\bibfield  {journal}
  {\bibinfo  {journal} {Journal of Chemical Theory and Computation}\ }\textbf
  {\bibinfo {volume} {5}},\ \bibinfo {pages} {770} (\bibinfo {year} {2009})},\
  \bibinfo {note} {pMID: 26609582},\ \Eprint
  {http://arxiv.org/abs/http://dx.doi.org/10.1021/ct800535c}
  {http://dx.doi.org/10.1021/ct800535c} \BibitemShut {NoStop}%
\bibitem [{\citenamefont {Helbig}\ \emph {et~al.}(2009)\citenamefont {Helbig},
  \citenamefont {Tokatly},\ and\ \citenamefont {Rubio}}]{HTR09}%
  \BibitemOpen
  \bibfield  {author} {\bibinfo {author} {\bibfnamefont {N.}~\bibnamefont
  {Helbig}}, \bibinfo {author} {\bibfnamefont {I.~V.}\ \bibnamefont {Tokatly}},
  \ and\ \bibinfo {author} {\bibfnamefont {A.}~\bibnamefont {Rubio}},\ }\href
  {\doibase 10.1063/1.3271392} {\bibfield  {journal} {\bibinfo  {journal} {The
  Journal of Chemical Physics}\ }\textbf {\bibinfo {volume} {131}},\ \bibinfo
  {pages} {224105} (\bibinfo {year} {2009})},\ \Eprint
  {http://arxiv.org/abs/http://dx.doi.org/10.1063/1.3271392}
  {http://dx.doi.org/10.1063/1.3271392} \BibitemShut {NoStop}%
\bibitem [{\citenamefont {Kraisler}\ and\ \citenamefont {Kronik}(2015)}]{KK15}%
  \BibitemOpen
  \bibfield  {author} {\bibinfo {author} {\bibfnamefont {E.}~\bibnamefont
  {Kraisler}}\ and\ \bibinfo {author} {\bibfnamefont {L.}~\bibnamefont
  {Kronik}},\ }\href {\doibase 10.1103/PhysRevA.91.032504} {\bibfield
  {journal} {\bibinfo  {journal} {Phys. Rev. A}\ }\textbf {\bibinfo {volume}
  {91}},\ \bibinfo {pages} {032504} (\bibinfo {year} {2015})}\BibitemShut
  {NoStop}%
\bibitem [{\citenamefont {Gritsenko}\ \emph {et~al.}(2000)\citenamefont
  {Gritsenko}, \citenamefont {van Gisbergen}, \citenamefont {G\"orling},\ and\
  \citenamefont {Baerends}}]{GGGB00}%
  \BibitemOpen
  \bibfield  {author} {\bibinfo {author} {\bibfnamefont {O.~V.}\ \bibnamefont
  {Gritsenko}}, \bibinfo {author} {\bibfnamefont {S.~J.~A.}\ \bibnamefont {van
  Gisbergen}}, \bibinfo {author} {\bibfnamefont {A.}~\bibnamefont {G\"orling}},
  \ and\ \bibinfo {author} {\bibfnamefont {E.~J.}\ \bibnamefont {Baerends}},\
  }\href {\doibase 10.1063/1.1318750} {\bibfield  {journal} {\bibinfo
  {journal} {The Journal of Chemical Physics}\ }\textbf {\bibinfo {volume}
  {113}},\ \bibinfo {pages} {8478} (\bibinfo {year} {2000})},\ \Eprint
  {http://arxiv.org/abs/http://dx.doi.org/10.1063/1.1318750}
  {http://dx.doi.org/10.1063/1.1318750} \BibitemShut {NoStop}%
\bibitem [{\citenamefont {Maitra}\ \emph {et~al.}(2004)\citenamefont {Maitra},
  \citenamefont {Zhang}, \citenamefont {Cave},\ and\ \citenamefont
  {Burke}}]{MZCB04}%
  \BibitemOpen
  \bibfield  {author} {\bibinfo {author} {\bibfnamefont {N.~T.}\ \bibnamefont
  {Maitra}}, \bibinfo {author} {\bibfnamefont {F.}~\bibnamefont {Zhang}},
  \bibinfo {author} {\bibfnamefont {R.~J.}\ \bibnamefont {Cave}}, \ and\
  \bibinfo {author} {\bibfnamefont {K.}~\bibnamefont {Burke}},\ }\href@noop {}
  {\bibfield  {journal} {\bibinfo  {journal} {J. Chem. Phys.}\ }\textbf
  {\bibinfo {volume} {120}} (\bibinfo {year} {2004})}\BibitemShut {NoStop}%
\bibitem [{\citenamefont {Gritsenko}\ and\ \citenamefont
  {Baerends}(2006)}]{GB06b}%
  \BibitemOpen
  \bibfield  {author} {\bibinfo {author} {\bibfnamefont {O.}~\bibnamefont
  {Gritsenko}}\ and\ \bibinfo {author} {\bibfnamefont {E.~J.}\ \bibnamefont
  {Baerends}},\ }\href {\doibase 10.1002/qua.21100} {\bibfield  {journal}
  {\bibinfo  {journal} {International Journal of Quantum Chemistry}\ }\textbf
  {\bibinfo {volume} {106}},\ \bibinfo {pages} {3167} (\bibinfo {year}
  {2006})}\BibitemShut {NoStop}%
\bibitem [{\citenamefont {BUIJSE}\ and\ \citenamefont {BAERENDS}(2002)}]{BB02}%
  \BibitemOpen
  \bibfield  {author} {\bibinfo {author} {\bibfnamefont {M.~A.}\ \bibnamefont
  {BUIJSE}}\ and\ \bibinfo {author} {\bibfnamefont {E.~J.}\ \bibnamefont
  {BAERENDS}},\ }\href {\doibase 10.1080/00268970110070243} {\bibfield
  {journal} {\bibinfo  {journal} {Molecular Physics}\ }\textbf {\bibinfo
  {volume} {100}},\ \bibinfo {pages} {401} (\bibinfo {year} {2002})},\ \Eprint
  {http://arxiv.org/abs/http://dx.doi.org/10.1080/00268970110070243}
  {http://dx.doi.org/10.1080/00268970110070243} \BibitemShut {NoStop}%
\bibitem [{\citenamefont {Fuks}\ \emph {et~al.}(2011)\citenamefont {Fuks},
  \citenamefont {Rubio},\ and\ \citenamefont {Maitra}}]{FRM11}%
  \BibitemOpen
  \bibfield  {author} {\bibinfo {author} {\bibfnamefont {J.~I.}\ \bibnamefont
  {Fuks}}, \bibinfo {author} {\bibfnamefont {A.}~\bibnamefont {Rubio}}, \ and\
  \bibinfo {author} {\bibfnamefont {N.~T.}\ \bibnamefont {Maitra}},\ }\href
  {\doibase 10.1103/PhysRevA.83.042501} {\bibfield  {journal} {\bibinfo
  {journal} {Phys. Rev. A}\ }\textbf {\bibinfo {volume} {83}},\ \bibinfo
  {pages} {042501} (\bibinfo {year} {2011})}\BibitemShut {NoStop}%
\bibitem [{\citenamefont {Li}\ and\ \citenamefont {Liu}(2014)}]{LL14}%
  \BibitemOpen
  \bibfield  {author} {\bibinfo {author} {\bibfnamefont {Z.}~\bibnamefont
  {Li}}\ and\ \bibinfo {author} {\bibfnamefont {W.}~\bibnamefont {Liu}},\
  }\href@noop {} {\bibfield  {journal} {\bibinfo  {journal} {J. Chem. Phys.}\
  }\textbf {\bibinfo {volume} {141}},\ \bibinfo {eid} {014110} (\bibinfo {year}
  {2014})}\BibitemShut {NoStop}%
\bibitem [{\citenamefont {Li}\ \emph {et~al.}(2014)\citenamefont {Li},
  \citenamefont {Suo},\ and\ \citenamefont {Liu}}]{LSL14}%
  \BibitemOpen
  \bibfield  {author} {\bibinfo {author} {\bibfnamefont {Z.}~\bibnamefont
  {Li}}, \bibinfo {author} {\bibfnamefont {B.}~\bibnamefont {Suo}}, \ and\
  \bibinfo {author} {\bibfnamefont {W.}~\bibnamefont {Liu}},\ }\href@noop {}
  {\bibfield  {journal} {\bibinfo  {journal} {J. Chem. Phys.}\ }\textbf
  {\bibinfo {volume} {141}},\ \bibinfo {eid} {244105} (\bibinfo {year}
  {2014})}\BibitemShut {NoStop}%
\bibitem [{\citenamefont {Zhang}\ and\ \citenamefont {Herbert}(2015)}]{ZH15}%
  \BibitemOpen
  \bibfield  {author} {\bibinfo {author} {\bibfnamefont {X.}~\bibnamefont
  {Zhang}}\ and\ \bibinfo {author} {\bibfnamefont {J.~M.}\ \bibnamefont
  {Herbert}},\ }\href
  {http://scitation.aip.org/content/aip/journal/jcp/142/6/10.1063/1.4907376}
  {\bibfield  {journal} {\bibinfo  {journal} {J. Chem. Phys.}\ }\textbf
  {\bibinfo {volume} {142}},\ \bibinfo {eid} {064109} (\bibinfo {year}
  {2015})}\BibitemShut {NoStop}%
\bibitem [{\citenamefont {Ou}\ \emph {et~al.}(2015)\citenamefont {Ou},
  \citenamefont {Bellchambers}, \citenamefont {Furche},\ and\ \citenamefont
  {Subotnik}}]{OBFS15}%
  \BibitemOpen
  \bibfield  {author} {\bibinfo {author} {\bibfnamefont {Q.}~\bibnamefont
  {Ou}}, \bibinfo {author} {\bibfnamefont {G.~D.}\ \bibnamefont
  {Bellchambers}}, \bibinfo {author} {\bibfnamefont {F.}~\bibnamefont
  {Furche}}, \ and\ \bibinfo {author} {\bibfnamefont {J.~E.}\ \bibnamefont
  {Subotnik}},\ }\href@noop {} {\bibfield  {journal} {\bibinfo  {journal} {J.
  Chem. Phys.}\ }\textbf {\bibinfo {volume} {142}},\ \bibinfo {eid} {064114}
  (\bibinfo {year} {2015})}\BibitemShut {NoStop}%
\bibitem [{\citenamefont {Parker}\ \emph {et~al.}(2016)\citenamefont {Parker},
  \citenamefont {Roy},\ and\ \citenamefont {Furche}}]{PRF16}%
  \BibitemOpen
  \bibfield  {author} {\bibinfo {author} {\bibfnamefont {S.~M.}\ \bibnamefont
  {Parker}}, \bibinfo {author} {\bibfnamefont {S.}~\bibnamefont {Roy}}, \ and\
  \bibinfo {author} {\bibfnamefont {F.}~\bibnamefont {Furche}},\ }\href
  {\doibase 10.1063/1.4963749} {\bibfield  {journal} {\bibinfo  {journal} {The
  Journal of Chemical Physics}\ }\textbf {\bibinfo {volume} {145}},\ \bibinfo
  {pages} {134105} (\bibinfo {year} {2016})},\ \Eprint
  {http://arxiv.org/abs/http://dx.doi.org/10.1063/1.4963749}
  {http://dx.doi.org/10.1063/1.4963749} \BibitemShut {NoStop}%
\bibitem [{\citenamefont {Rozzi}\ \emph {et~al.}(2013)\citenamefont {Rozzi},
  \citenamefont {Falke}, \citenamefont {Spallanzani}, \citenamefont {Rubio},
  \citenamefont {Molinari}, \citenamefont {Brida}, \citenamefont {Maiuri},
  \citenamefont {Cerullo}, \citenamefont {Schramm}, \citenamefont
  {Christoffers} \emph {et~al.}}]{RFSR13}%
  \BibitemOpen
  \bibfield  {author} {\bibinfo {author} {\bibfnamefont {C.~A.}\ \bibnamefont
  {Rozzi}}, \bibinfo {author} {\bibfnamefont {S.~M.}\ \bibnamefont {Falke}},
  \bibinfo {author} {\bibfnamefont {N.}~\bibnamefont {Spallanzani}}, \bibinfo
  {author} {\bibfnamefont {A.}~\bibnamefont {Rubio}}, \bibinfo {author}
  {\bibfnamefont {E.}~\bibnamefont {Molinari}}, \bibinfo {author}
  {\bibfnamefont {D.}~\bibnamefont {Brida}}, \bibinfo {author} {\bibfnamefont
  {M.}~\bibnamefont {Maiuri}}, \bibinfo {author} {\bibfnamefont
  {G.}~\bibnamefont {Cerullo}}, \bibinfo {author} {\bibfnamefont
  {H.}~\bibnamefont {Schramm}}, \bibinfo {author} {\bibfnamefont
  {J.}~\bibnamefont {Christoffers}},  \emph {et~al.},\ }\href@noop {}
  {\bibfield  {journal} {\bibinfo  {journal} {Nature Communications}\ }\textbf
  {\bibinfo {volume} {4}},\ \bibinfo {pages} {1602} (\bibinfo {year}
  {2013})}\BibitemShut {NoStop}%
\bibitem [{\citenamefont {Falke}\ \emph {et~al.}(2014)\citenamefont {Falke},
  \citenamefont {Rozzi}, \citenamefont {Brida}, \citenamefont {Maiuri},
  \citenamefont {Amato}, \citenamefont {Sommer}, \citenamefont {De~Sio},
  \citenamefont {Rubio}, \citenamefont {Cerullo}, \citenamefont {Molinari},\
  and\ \citenamefont {Lienau}}]{FRBM14}%
  \BibitemOpen
  \bibfield  {author} {\bibinfo {author} {\bibfnamefont {S.~M.}\ \bibnamefont
  {Falke}}, \bibinfo {author} {\bibfnamefont {C.~A.}\ \bibnamefont {Rozzi}},
  \bibinfo {author} {\bibfnamefont {D.}~\bibnamefont {Brida}}, \bibinfo
  {author} {\bibfnamefont {M.}~\bibnamefont {Maiuri}}, \bibinfo {author}
  {\bibfnamefont {M.}~\bibnamefont {Amato}}, \bibinfo {author} {\bibfnamefont
  {E.}~\bibnamefont {Sommer}}, \bibinfo {author} {\bibfnamefont
  {A.}~\bibnamefont {De~Sio}}, \bibinfo {author} {\bibfnamefont
  {A.}~\bibnamefont {Rubio}}, \bibinfo {author} {\bibfnamefont
  {G.}~\bibnamefont {Cerullo}}, \bibinfo {author} {\bibfnamefont
  {E.}~\bibnamefont {Molinari}}, \ and\ \bibinfo {author} {\bibfnamefont
  {C.}~\bibnamefont {Lienau}},\ }\href@noop {} {\bibfield  {journal} {\bibinfo
  {journal} {Science}\ }\textbf {\bibinfo {volume} {344}},\ \bibinfo {pages}
  {1001} (\bibinfo {year} {2014})}\BibitemShut {NoStop}%
\bibitem [{\citenamefont {Manna}\ \emph
  {et~al.}(2015{\natexlab{b}})\citenamefont {Manna}, \citenamefont
  {Balamurugan}, \citenamefont {Cheung},\ and\ \citenamefont
  {Dunietz}}]{MBCD15}%
  \BibitemOpen
  \bibfield  {author} {\bibinfo {author} {\bibfnamefont {A.~K.}\ \bibnamefont
  {Manna}}, \bibinfo {author} {\bibfnamefont {D.}~\bibnamefont {Balamurugan}},
  \bibinfo {author} {\bibfnamefont {M.~S.}\ \bibnamefont {Cheung}}, \ and\
  \bibinfo {author} {\bibfnamefont {B.~D.}\ \bibnamefont {Dunietz}},\ }\href
  {\doibase 10.1021/acs.jpclett.5b00074} {\bibfield  {journal} {\bibinfo
  {journal} {The Journal of Physical Chemistry Letters}\ }\textbf {\bibinfo
  {volume} {6}},\ \bibinfo {pages} {1231} (\bibinfo {year}
  {2015}{\natexlab{b}})},\ \bibinfo {note} {pMID: 26262978},\ \Eprint
  {http://arxiv.org/abs/http://dx.doi.org/10.1021/acs.jpclett.5b00074}
  {http://dx.doi.org/10.1021/acs.jpclett.5b00074} \BibitemShut {NoStop}%
\bibitem [{\citenamefont {Craig}\ \emph {et~al.}(2005)\citenamefont {Craig},
  \citenamefont {Duncan},\ and\ \citenamefont {Prezhdo}}]{CDP05}%
  \BibitemOpen
  \bibfield  {author} {\bibinfo {author} {\bibfnamefont {C.~F.}\ \bibnamefont
  {Craig}}, \bibinfo {author} {\bibfnamefont {W.~R.}\ \bibnamefont {Duncan}}, \
  and\ \bibinfo {author} {\bibfnamefont {O.~V.}\ \bibnamefont {Prezhdo}},\
  }\href@noop {} {\bibfield  {journal} {\bibinfo  {journal} {Phys. Rev. Lett.}\
  }\textbf {\bibinfo {volume} {95}},\ \bibinfo {pages} {163001} (\bibinfo
  {year} {2005})}\BibitemShut {NoStop}%
\bibitem [{\citenamefont {Tapavicza}\ \emph {et~al.}(2007)\citenamefont
  {Tapavicza}, \citenamefont {Tavernelli},\ and\ \citenamefont
  {Rothlisberger}}]{TTR07}%
  \BibitemOpen
  \bibfield  {author} {\bibinfo {author} {\bibfnamefont {E.}~\bibnamefont
  {Tapavicza}}, \bibinfo {author} {\bibfnamefont {I.}~\bibnamefont
  {Tavernelli}}, \ and\ \bibinfo {author} {\bibfnamefont {U.}~\bibnamefont
  {Rothlisberger}},\ }\href@noop {} {\bibfield  {journal} {\bibinfo  {journal}
  {Phys. Rev. Lett.}\ }\textbf {\bibinfo {volume} {98}},\ \bibinfo {pages}
  {023001} (\bibinfo {year} {2007})}\BibitemShut {NoStop}%
\bibitem [{\citenamefont {Prezhdo}\ \emph {et~al.}(2009)\citenamefont
  {Prezhdo}, \citenamefont {Duncan},\ and\ \citenamefont {Prezhdo}}]{PDP09}%
  \BibitemOpen
  \bibfield  {author} {\bibinfo {author} {\bibfnamefont {O.~V.}\ \bibnamefont
  {Prezhdo}}, \bibinfo {author} {\bibfnamefont {W.~R.}\ \bibnamefont {Duncan}},
  \ and\ \bibinfo {author} {\bibfnamefont {V.~V.}\ \bibnamefont {Prezhdo}},\
  }\href@noop {} {\bibfield  {journal} {\bibinfo  {journal} {Progress in
  Surface Science}\ }\textbf {\bibinfo {volume} {84}},\ \bibinfo {pages} {30 }
  (\bibinfo {year} {2009})}\BibitemShut {NoStop}%
\bibitem [{\citenamefont {Curchod}\ \emph {et~al.}(2013)\citenamefont
  {Curchod}, \citenamefont {Rothlisberger},\ and\ \citenamefont
  {Tavernelli}}]{CRT13}%
  \BibitemOpen
  \bibfield  {author} {\bibinfo {author} {\bibfnamefont {B.~F.}\ \bibnamefont
  {Curchod}}, \bibinfo {author} {\bibfnamefont {U.}~\bibnamefont
  {Rothlisberger}}, \ and\ \bibinfo {author} {\bibfnamefont {I.}~\bibnamefont
  {Tavernelli}},\ }\href@noop {} {\bibfield  {journal} {\bibinfo  {journal}
  {ChemPhysChem}\ }\textbf {\bibinfo {volume} {14}},\ \bibinfo {pages} {1314}
  (\bibinfo {year} {2013})}\BibitemShut {NoStop}%
\bibitem [{\citenamefont {Tavernelli}(2015)}]{T15}%
  \BibitemOpen
  \bibfield  {author} {\bibinfo {author} {\bibfnamefont {I.}~\bibnamefont
  {Tavernelli}},\ }\href {\doibase 10.1021/ar500357y} {\bibfield  {journal}
  {\bibinfo  {journal} {Accounts of Chemical Research}\ }\textbf {\bibinfo
  {volume} {48}},\ \bibinfo {pages} {792} (\bibinfo {year} {2015})},\ \bibinfo
  {note} {pMID: 25647401},\ \Eprint
  {http://arxiv.org/abs/http://dx.doi.org/10.1021/ar500357y}
  {http://dx.doi.org/10.1021/ar500357y} \BibitemShut {NoStop}%
\bibitem [{\citenamefont {Tavernelli}\ \emph {et~al.}(2011)\citenamefont
  {Tavernelli}, \citenamefont {Curchod},\ and\ \citenamefont
  {Rothlisberger}}]{TCR11}%
  \BibitemOpen
  \bibfield  {author} {\bibinfo {author} {\bibfnamefont {I.}~\bibnamefont
  {Tavernelli}}, \bibinfo {author} {\bibfnamefont {B.~F.}\ \bibnamefont
  {Curchod}}, \ and\ \bibinfo {author} {\bibfnamefont {U.}~\bibnamefont
  {Rothlisberger}},\ }\href {\doibase
  https://doi.org/10.1016/j.chemphys.2011.03.021} {\bibfield  {journal}
  {\bibinfo  {journal} {Chemical Physics}\ }\textbf {\bibinfo {volume} {391}},\
  \bibinfo {pages} {101 } (\bibinfo {year} {2011})},\ \bibinfo {note} {open
  problems and new solutions in time dependent density functional
  theory}\BibitemShut {NoStop}%
\bibitem [{\citenamefont {Muuronen}\ \emph {et~al.}(2017)\citenamefont
  {Muuronen}, \citenamefont {Parker}, \citenamefont {Berardo}, \citenamefont
  {Le}, \citenamefont {Zwijnenburg},\ and\ \citenamefont {Furche}}]{MPBLZF17}%
  \BibitemOpen
  \bibfield  {author} {\bibinfo {author} {\bibfnamefont {M.}~\bibnamefont
  {Muuronen}}, \bibinfo {author} {\bibfnamefont {S.~M.}\ \bibnamefont
  {Parker}}, \bibinfo {author} {\bibfnamefont {E.}~\bibnamefont {Berardo}},
  \bibinfo {author} {\bibfnamefont {A.}~\bibnamefont {Le}}, \bibinfo {author}
  {\bibfnamefont {M.~A.}\ \bibnamefont {Zwijnenburg}}, \ and\ \bibinfo {author}
  {\bibfnamefont {F.}~\bibnamefont {Furche}},\ }\href {\doibase
  10.1039/C6SC04378J} {\bibfield  {journal} {\bibinfo  {journal} {Chem. Sci.}\
  }\textbf {\bibinfo {volume} {8}},\ \bibinfo {pages} {2179} (\bibinfo {year}
  {2017})}\BibitemShut {NoStop}%
\bibitem [{\citenamefont {Fuks}\ \emph
  {et~al.}(2013{\natexlab{a}})\citenamefont {Fuks}, \citenamefont {Elliott},
  \citenamefont {Rubio},\ and\ \citenamefont {Maitra}}]{FERM13}%
  \BibitemOpen
  \bibfield  {author} {\bibinfo {author} {\bibfnamefont {J.~I.}\ \bibnamefont
  {Fuks}}, \bibinfo {author} {\bibfnamefont {P.}~\bibnamefont {Elliott}},
  \bibinfo {author} {\bibfnamefont {A.}~\bibnamefont {Rubio}}, \ and\ \bibinfo
  {author} {\bibfnamefont {N.~T.}\ \bibnamefont {Maitra}},\ }\href@noop {}
  {\bibfield  {journal} {\bibinfo  {journal} {J. Phys. Chem. Lett.}\ }\textbf
  {\bibinfo {volume} {4}},\ \bibinfo {pages} {735} (\bibinfo {year}
  {2013}{\natexlab{a}})}\BibitemShut {NoStop}%
\bibitem [{\citenamefont {Fuks}\ \emph {et~al.}(2015)\citenamefont {Fuks},
  \citenamefont {Luo}, \citenamefont {Sandoval},\ and\ \citenamefont
  {Maitra}}]{FLSM15}%
  \BibitemOpen
  \bibfield  {author} {\bibinfo {author} {\bibfnamefont {J.~I.}\ \bibnamefont
  {Fuks}}, \bibinfo {author} {\bibfnamefont {K.}~\bibnamefont {Luo}}, \bibinfo
  {author} {\bibfnamefont {E.~D.}\ \bibnamefont {Sandoval}}, \ and\ \bibinfo
  {author} {\bibfnamefont {N.~T.}\ \bibnamefont {Maitra}},\ }\href@noop {}
  {\bibfield  {journal} {\bibinfo  {journal} {Phys. Rev. Lett.}\ }\textbf
  {\bibinfo {volume} {114}},\ \bibinfo {pages} {183002} (\bibinfo {year}
  {2015})}\BibitemShut {NoStop}%
\bibitem [{\citenamefont {Raghunathan}\ and\ \citenamefont
  {Nest}(2011)}]{RN11}%
  \BibitemOpen
  \bibfield  {author} {\bibinfo {author} {\bibfnamefont {S.}~\bibnamefont
  {Raghunathan}}\ and\ \bibinfo {author} {\bibfnamefont {M.}~\bibnamefont
  {Nest}},\ }\href {\doibase 10.1021/ct200270t} {\bibfield  {journal} {\bibinfo
   {journal} {J. Chem. Theory and Comput.}\ }\textbf {\bibinfo {volume} {7}},\
  \bibinfo {pages} {2492} (\bibinfo {year} {2011})}\BibitemShut {NoStop}%
\bibitem [{\citenamefont {Krause}\ \emph {et~al.}(2005)\citenamefont {Krause},
  \citenamefont {Klamroth},\ and\ \citenamefont {Saalfrank}}]{KKS05}%
  \BibitemOpen
  \bibfield  {author} {\bibinfo {author} {\bibfnamefont {P.}~\bibnamefont
  {Krause}}, \bibinfo {author} {\bibfnamefont {T.}~\bibnamefont {Klamroth}}, \
  and\ \bibinfo {author} {\bibfnamefont {P.}~\bibnamefont {Saalfrank}},\ }\href
  {\doibase 10.1063/1.1999636} {\bibfield  {journal} {\bibinfo  {journal} {The
  Journal of Chemical Physics}\ }\textbf {\bibinfo {volume} {123}},\ \bibinfo
  {pages} {074105} (\bibinfo {year} {2005})},\ \Eprint
  {http://arxiv.org/abs/http://dx.doi.org/10.1063/1.1999636}
  {http://dx.doi.org/10.1063/1.1999636} \BibitemShut {NoStop}%
\bibitem [{\citenamefont {Maitra}\ and\ \citenamefont {Burke}(2001)}]{MB01}%
  \BibitemOpen
  \bibfield  {author} {\bibinfo {author} {\bibfnamefont {N.~T.}\ \bibnamefont
  {Maitra}}\ and\ \bibinfo {author} {\bibfnamefont {K.}~\bibnamefont {Burke}},\
  }\href@noop {} {\bibfield  {journal} {\bibinfo  {journal} {Phys. Rev. A}\
  }\textbf {\bibinfo {volume} {63}},\ \bibinfo {pages} {042501} (\bibinfo
  {year} {2001})}\BibitemShut {NoStop}%
\bibitem [{\citenamefont {Elliott}\ \emph {et~al.}(2012)\citenamefont
  {Elliott}, \citenamefont {Fuks}, \citenamefont {Rubio},\ and\ \citenamefont
  {Maitra}}]{EFRM12}%
  \BibitemOpen
  \bibfield  {author} {\bibinfo {author} {\bibfnamefont {P.}~\bibnamefont
  {Elliott}}, \bibinfo {author} {\bibfnamefont {J.~I.}\ \bibnamefont {Fuks}},
  \bibinfo {author} {\bibfnamefont {A.}~\bibnamefont {Rubio}}, \ and\ \bibinfo
  {author} {\bibfnamefont {N.~T.}\ \bibnamefont {Maitra}},\ }\href {\doibase
  10.1103/PhysRevLett.109.266404} {\bibfield  {journal} {\bibinfo  {journal}
  {Phys. Rev. Lett.}\ }\textbf {\bibinfo {volume} {109}},\ \bibinfo {pages}
  {266404} (\bibinfo {year} {2012})}\BibitemShut {NoStop}%
\bibitem [{\citenamefont {Ramsden}\ and\ \citenamefont {Godby}(2012)}]{RG12}%
  \BibitemOpen
  \bibfield  {author} {\bibinfo {author} {\bibfnamefont {J.}~\bibnamefont
  {Ramsden}}\ and\ \bibinfo {author} {\bibfnamefont {R.}~\bibnamefont
  {Godby}},\ }\href@noop {} {\bibfield  {journal} {\bibinfo  {journal} {Phys.
  Rev. Lett.}\ }\textbf {\bibinfo {volume} {109}},\ \bibinfo {pages} {036402}
  (\bibinfo {year} {2012})}\BibitemShut {NoStop}%
\bibitem [{\citenamefont {Ruggenthaler}\ and\ \citenamefont
  {Bauer}(2009)}]{RB09}%
  \BibitemOpen
  \bibfield  {author} {\bibinfo {author} {\bibfnamefont {M.}~\bibnamefont
  {Ruggenthaler}}\ and\ \bibinfo {author} {\bibfnamefont {D.}~\bibnamefont
  {Bauer}},\ }\href {\doibase 10.1103/PhysRevLett.102.233001} {\bibfield
  {journal} {\bibinfo  {journal} {Phys. Rev. Lett.}\ }\textbf {\bibinfo
  {volume} {102}},\ \bibinfo {pages} {233001} (\bibinfo {year}
  {2009})}\BibitemShut {NoStop}%
\bibitem [{\citenamefont {Luo}\ \emph {et~al.}(2014)\citenamefont {Luo},
  \citenamefont {Fuks}, \citenamefont {Sandoval}, \citenamefont {Elliott},\
  and\ \citenamefont {Maitra}}]{LFSEM14}%
  \BibitemOpen
  \bibfield  {author} {\bibinfo {author} {\bibfnamefont {K.}~\bibnamefont
  {Luo}}, \bibinfo {author} {\bibfnamefont {J.~I.}\ \bibnamefont {Fuks}},
  \bibinfo {author} {\bibfnamefont {E.~D.}\ \bibnamefont {Sandoval}}, \bibinfo
  {author} {\bibfnamefont {P.}~\bibnamefont {Elliott}}, \ and\ \bibinfo
  {author} {\bibfnamefont {N.~T.}\ \bibnamefont {Maitra}},\ }\href@noop {}
  {\bibfield  {journal} {\bibinfo  {journal} {J. Chem. Phys}\ }\textbf
  {\bibinfo {volume} {140}},\ \bibinfo {pages} {18A515} (\bibinfo {year}
  {2014})}\BibitemShut {NoStop}%
\bibitem [{\citenamefont {Fuks}\ \emph {et~al.}(2016)\citenamefont {Fuks},
  \citenamefont {Nielsen}, \citenamefont {Ruggenthaler},\ and\ \citenamefont
  {Maitra}}]{FNRM16}%
  \BibitemOpen
  \bibfield  {author} {\bibinfo {author} {\bibfnamefont {J.~I.}\ \bibnamefont
  {Fuks}}, \bibinfo {author} {\bibfnamefont {S.}~\bibnamefont {Nielsen}},
  \bibinfo {author} {\bibfnamefont {M.}~\bibnamefont {Ruggenthaler}}, \ and\
  \bibinfo {author} {\bibfnamefont {N.}~\bibnamefont {Maitra}},\ }\href
  {\doibase 10.1039/C6CP00722H} {\bibfield  {journal} {\bibinfo  {journal}
  {Phys. Chem. Chem. Phys.}\ }\textbf {\bibinfo {volume} {18}},\ \bibinfo
  {pages} {20976} (\bibinfo {year} {2016})}\BibitemShut {NoStop}%
\bibitem [{\citenamefont {Ruggenthaler}\ \emph {et~al.}(2015)\citenamefont
  {Ruggenthaler}, \citenamefont {Penz},\ and\ \citenamefont {van
  Leeuwen}}]{RPL15}%
  \BibitemOpen
  \bibfield  {author} {\bibinfo {author} {\bibfnamefont {M.}~\bibnamefont
  {Ruggenthaler}}, \bibinfo {author} {\bibfnamefont {M.}~\bibnamefont {Penz}},
  \ and\ \bibinfo {author} {\bibfnamefont {R.}~\bibnamefont {van Leeuwen}},\
  }\href@noop {} {\bibfield  {journal} {\bibinfo  {journal} {J. Phys. Condens.
  Matter}\ }\textbf {\bibinfo {volume} {27}},\ \bibinfo {pages} {203202}
  (\bibinfo {year} {2015})}\BibitemShut {NoStop}%
\bibitem [{\citenamefont {Nielsen}\ \emph {et~al.}(2013)\citenamefont
  {Nielsen}, \citenamefont {Ruggenthaler},\ and\ \citenamefont {van
  Leeuwen}}]{NRL13}%
  \BibitemOpen
  \bibfield  {author} {\bibinfo {author} {\bibfnamefont {S.}~\bibnamefont
  {Nielsen}}, \bibinfo {author} {\bibfnamefont {M.}~\bibnamefont
  {Ruggenthaler}}, \ and\ \bibinfo {author} {\bibnamefont {van Leeuwen}},\
  }\href@noop {} {\bibfield  {journal} {\bibinfo  {journal} {Europhys. Lett.}\
  }\textbf {\bibinfo {volume} {101}},\ \bibinfo {pages} {33001} (\bibinfo
  {year} {2013})}\BibitemShut {NoStop}%
\bibitem [{\citenamefont {Jensen}\ and\ \citenamefont
  {Wasserman}(2016)}]{JW16}%
  \BibitemOpen
  \bibfield  {author} {\bibinfo {author} {\bibfnamefont {D.~S.}\ \bibnamefont
  {Jensen}}\ and\ \bibinfo {author} {\bibfnamefont {A.}~\bibnamefont
  {Wasserman}},\ }\href {\doibase 10.1039/C6CP00312E} {\bibfield  {journal}
  {\bibinfo  {journal} {Phys. Chem. Chem. Phys.}\ }\textbf {\bibinfo {volume}
  {18}},\ \bibinfo {pages} {21079} (\bibinfo {year} {2016})}\BibitemShut
  {NoStop}%
\bibitem [{\citenamefont {Hodgson}\ \emph {et~al.}(2013)\citenamefont
  {Hodgson}, \citenamefont {Ramsden}, \citenamefont {Chapman}, \citenamefont
  {Lillystone},\ and\ \citenamefont {Godby}}]{HRCLG13}%
  \BibitemOpen
  \bibfield  {author} {\bibinfo {author} {\bibfnamefont {M.~J.~P.}\
  \bibnamefont {Hodgson}}, \bibinfo {author} {\bibfnamefont {J.~D.}\
  \bibnamefont {Ramsden}}, \bibinfo {author} {\bibfnamefont {J.~B.~J.}\
  \bibnamefont {Chapman}}, \bibinfo {author} {\bibfnamefont {P.}~\bibnamefont
  {Lillystone}}, \ and\ \bibinfo {author} {\bibfnamefont {R.~W.}\ \bibnamefont
  {Godby}},\ }\href@noop {} {\bibfield  {journal} {\bibinfo  {journal} {Phys.
  Rev. B}\ }\textbf {\bibinfo {volume} {88}},\ \bibinfo {pages} {241102}
  (\bibinfo {year} {2013})}\BibitemShut {NoStop}%
\bibitem [{\citenamefont {Fuks}\ \emph
  {et~al.}(2013{\natexlab{b}})\citenamefont {Fuks}, \citenamefont
  {Farzanehpour}, \citenamefont {Tokatly}, \citenamefont {Appel}, \citenamefont
  {Kurth},\ and\ \citenamefont {Rubio}}]{FFTAKR11}%
  \BibitemOpen
  \bibfield  {author} {\bibinfo {author} {\bibfnamefont {J.~I.}\ \bibnamefont
  {Fuks}}, \bibinfo {author} {\bibfnamefont {M.}~\bibnamefont {Farzanehpour}},
  \bibinfo {author} {\bibfnamefont {I.~V.}\ \bibnamefont {Tokatly}}, \bibinfo
  {author} {\bibfnamefont {H.}~\bibnamefont {Appel}}, \bibinfo {author}
  {\bibfnamefont {S.}~\bibnamefont {Kurth}}, \ and\ \bibinfo {author}
  {\bibfnamefont {A.}~\bibnamefont {Rubio}},\ }\href {\doibase
  10.1103/PhysRevA.88.062512} {\bibfield  {journal} {\bibinfo  {journal} {Phys.
  Rev. A}\ }\textbf {\bibinfo {volume} {88}},\ \bibinfo {pages} {062512}
  (\bibinfo {year} {2013}{\natexlab{b}})}\BibitemShut {NoStop}%
\bibitem [{\citenamefont {Fuks}\ and\ \citenamefont
  {Maitra}(2014{\natexlab{a}})}]{FM14}%
  \BibitemOpen
  \bibfield  {author} {\bibinfo {author} {\bibfnamefont {J.~I.}\ \bibnamefont
  {Fuks}}\ and\ \bibinfo {author} {\bibfnamefont {N.~T.}\ \bibnamefont
  {Maitra}},\ }\href {\doibase 10.1039/C4CP00118D} {\bibfield  {journal}
  {\bibinfo  {journal} {Phys. Chem. Chem. Phys.}\ }\textbf {\bibinfo {volume}
  {16}},\ \bibinfo {pages} {14504} (\bibinfo {year}
  {2014}{\natexlab{a}})}\BibitemShut {NoStop}%
\bibitem [{\citenamefont {Fuks}\ and\ \citenamefont
  {Maitra}(2014{\natexlab{b}})}]{FM14b}%
  \BibitemOpen
  \bibfield  {author} {\bibinfo {author} {\bibfnamefont {J.~I.}\ \bibnamefont
  {Fuks}}\ and\ \bibinfo {author} {\bibfnamefont {N.~T.}\ \bibnamefont
  {Maitra}},\ }\href {\doibase 10.1103/PhysRevA.89.062502} {\bibfield
  {journal} {\bibinfo  {journal} {Phys. Rev. A}\ }\textbf {\bibinfo {volume}
  {89}},\ \bibinfo {pages} {062502} (\bibinfo {year}
  {2014}{\natexlab{b}})}\BibitemShut {NoStop}%
\bibitem [{\citenamefont {Carrascal}\ and\ \citenamefont
  {Ferrer}(2012)}]{CF12}%
  \BibitemOpen
  \bibfield  {author} {\bibinfo {author} {\bibfnamefont {D.~J.}\ \bibnamefont
  {Carrascal}}\ and\ \bibinfo {author} {\bibfnamefont {J.}~\bibnamefont
  {Ferrer}},\ }\href {\doibase 10.1103/PhysRevB.85.045110} {\bibfield
  {journal} {\bibinfo  {journal} {Phys. Rev. B}\ }\textbf {\bibinfo {volume}
  {85}},\ \bibinfo {pages} {045110} (\bibinfo {year} {2012})}\BibitemShut
  {NoStop}%
\bibitem [{\citenamefont {Elliott}\ and\ \citenamefont {Maitra}(2012)}]{EM12}%
  \BibitemOpen
  \bibfield  {author} {\bibinfo {author} {\bibfnamefont {P.}~\bibnamefont
  {Elliott}}\ and\ \bibinfo {author} {\bibfnamefont {N.~T.}\ \bibnamefont
  {Maitra}},\ }\href@noop {} {\bibfield  {journal} {\bibinfo  {journal} {Phys.
  Rev. A}\ }\textbf {\bibinfo {volume} {85}},\ \bibinfo {pages} {052510}
  (\bibinfo {year} {2012})}\BibitemShut {NoStop}%
\bibitem [{\citenamefont {Tavernelli}\ \emph {et~al.}(2005)\citenamefont
  {Tavernelli}, \citenamefont {R\"ohrig},\ and\ \citenamefont
  {Rothlisberger}}]{TRR05}%
  \BibitemOpen
  \bibfield  {author} {\bibinfo {author} {\bibfnamefont {I.}~\bibnamefont
  {Tavernelli}}, \bibinfo {author} {\bibfnamefont {U.~F.}\ \bibnamefont
  {R\"ohrig}}, \ and\ \bibinfo {author} {\bibfnamefont {U.}~\bibnamefont
  {Rothlisberger}},\ }\href {\doibase 10.1080/00268970512331339378} {\bibfield
  {journal} {\bibinfo  {journal} {Molecular Physics}\ }\textbf {\bibinfo
  {volume} {103}},\ \bibinfo {pages} {963} (\bibinfo {year} {2005})},\ \Eprint
  {http://arxiv.org/abs/http://dx.doi.org/10.1080/00268970512331339378}
  {http://dx.doi.org/10.1080/00268970512331339378} \BibitemShut {NoStop}%
\bibitem [{\citenamefont {Luo}\ \emph {et~al.}(2016)\citenamefont {Luo},
  \citenamefont {Fuks},\ and\ \citenamefont {Maitra}}]{LFM16}%
  \BibitemOpen
  \bibfield  {author} {\bibinfo {author} {\bibfnamefont {K.}~\bibnamefont
  {Luo}}, \bibinfo {author} {\bibfnamefont {J.~I.}\ \bibnamefont {Fuks}}, \
  and\ \bibinfo {author} {\bibfnamefont {N.~T.}\ \bibnamefont {Maitra}},\
  }\href {\doibase 10.1063/1.4955447} {\bibfield  {journal} {\bibinfo
  {journal} {The Journal of Chemical Physics}\ }\textbf {\bibinfo {volume}
  {145}},\ \bibinfo {pages} {044101} (\bibinfo {year} {2016})},\ \Eprint
  {http://arxiv.org/abs/http://dx.doi.org/10.1063/1.4955447}
  {http://dx.doi.org/10.1063/1.4955447} \BibitemShut {NoStop}%
\bibitem [{\citenamefont {Perfetto}\ and\ \citenamefont
  {Stefanucci}(2015)}]{PS15}%
  \BibitemOpen
  \bibfield  {author} {\bibinfo {author} {\bibfnamefont {E.}~\bibnamefont
  {Perfetto}}\ and\ \bibinfo {author} {\bibfnamefont {G.}~\bibnamefont
  {Stefanucci}},\ }\href {\doibase 10.1103/PhysRevA.91.033416} {\bibfield
  {journal} {\bibinfo  {journal} {Phys. Rev. A}\ }\textbf {\bibinfo {volume}
  {91}},\ \bibinfo {pages} {033416} (\bibinfo {year} {2015})}\BibitemShut
  {NoStop}%
\bibitem [{\citenamefont {Perfetto}\ \emph {et~al.}(2015)\citenamefont
  {Perfetto}, \citenamefont {Sangalli}, \citenamefont {Marini},\ and\
  \citenamefont {Stefanucci}}]{PSMS15}%
  \BibitemOpen
  \bibfield  {author} {\bibinfo {author} {\bibfnamefont {E.}~\bibnamefont
  {Perfetto}}, \bibinfo {author} {\bibfnamefont {D.}~\bibnamefont {Sangalli}},
  \bibinfo {author} {\bibfnamefont {A.}~\bibnamefont {Marini}}, \ and\ \bibinfo
  {author} {\bibfnamefont {G.}~\bibnamefont {Stefanucci}},\ }\href {\doibase
  10.1103/PhysRevB.92.205304} {\bibfield  {journal} {\bibinfo  {journal} {Phys.
  Rev. B}\ }\textbf {\bibinfo {volume} {92}},\ \bibinfo {pages} {205304}
  (\bibinfo {year} {2015})}\BibitemShut {NoStop}%
\bibitem [{\citenamefont {Neugebauer}\ \emph {et~al.}(2006)\citenamefont
  {Neugebauer}, \citenamefont {Gritsenko},\ and\ \citenamefont
  {Baerends}}]{NGB06}%
  \BibitemOpen
  \bibfield  {author} {\bibinfo {author} {\bibfnamefont {J.}~\bibnamefont
  {Neugebauer}}, \bibinfo {author} {\bibfnamefont {O.}~\bibnamefont
  {Gritsenko}}, \ and\ \bibinfo {author} {\bibfnamefont {E.~J.}\ \bibnamefont
  {Baerends}},\ }\href {\doibase 10.1063/1.2197829} {\bibfield  {journal}
  {\bibinfo  {journal} {The Journal of Chemical Physics}\ }\textbf {\bibinfo
  {volume} {124}},\ \bibinfo {pages} {214102} (\bibinfo {year} {2006})},\
  \Eprint {http://arxiv.org/abs/http://dx.doi.org/10.1063/1.2197829}
  {http://dx.doi.org/10.1063/1.2197829} \BibitemShut {NoStop}%
\bibitem [{\citenamefont {Solovyeva}\ \emph {et~al.}(2014)\citenamefont
  {Solovyeva}, \citenamefont {Pavanello},\ and\ \citenamefont
  {Neugebauer}}]{SPN14}%
  \BibitemOpen
  \bibfield  {author} {\bibinfo {author} {\bibfnamefont {A.}~\bibnamefont
  {Solovyeva}}, \bibinfo {author} {\bibfnamefont {M.}~\bibnamefont
  {Pavanello}}, \ and\ \bibinfo {author} {\bibfnamefont {J.}~\bibnamefont
  {Neugebauer}},\ }\href {\doibase 10.1063/1.4871301} {\bibfield  {journal}
  {\bibinfo  {journal} {The Journal of Chemical Physics}\ }\textbf {\bibinfo
  {volume} {140}},\ \bibinfo {pages} {164103} (\bibinfo {year} {2014})},\
  \Eprint {http://arxiv.org/abs/http://dx.doi.org/10.1063/1.4871301}
  {http://dx.doi.org/10.1063/1.4871301} \BibitemShut {NoStop}%
\bibitem [{\citenamefont {Adamo}\ and\ \citenamefont {Jacquemin}(2013)}]{AJ13}%
  \BibitemOpen
  \bibfield  {author} {\bibinfo {author} {\bibfnamefont {C.}~\bibnamefont
  {Adamo}}\ and\ \bibinfo {author} {\bibfnamefont {D.}~\bibnamefont
  {Jacquemin}},\ }\href@noop {} {\bibfield  {journal} {\bibinfo  {journal}
  {Chem. Soc. Rev.}\ }\textbf {\bibinfo {volume} {42}},\ \bibinfo {pages} {845}
  (\bibinfo {year} {2013})}\BibitemShut {NoStop}%
\bibitem [{\citenamefont {Lipparini}\ and\ \citenamefont
  {Mennucci}(2016)}]{LM16}%
  \BibitemOpen
  \bibfield  {author} {\bibinfo {author} {\bibfnamefont {F.}~\bibnamefont
  {Lipparini}}\ and\ \bibinfo {author} {\bibfnamefont {B.}~\bibnamefont
  {Mennucci}},\ }\href {\doibase 10.1063/1.4947236} {\bibfield  {journal}
  {\bibinfo  {journal} {The Journal of Chemical Physics}\ }\textbf {\bibinfo
  {volume} {144}},\ \bibinfo {pages} {160901} (\bibinfo {year} {2016})},\
  \Eprint {http://arxiv.org/abs/http://dx.doi.org/10.1063/1.4947236}
  {http://dx.doi.org/10.1063/1.4947236} \BibitemShut {NoStop}%
\bibitem [{\citenamefont {Daday}\ \emph {et~al.}(2013)\citenamefont {Daday},
  \citenamefont {K\"onig}, \citenamefont {Valsson}, \citenamefont
  {Neugebauer},\ and\ \citenamefont {Filippi}}]{DKVNF13}%
  \BibitemOpen
  \bibfield  {author} {\bibinfo {author} {\bibfnamefont {C.}~\bibnamefont
  {Daday}}, \bibinfo {author} {\bibfnamefont {C.}~\bibnamefont {K\"onig}},
  \bibinfo {author} {\bibfnamefont {O.}~\bibnamefont {Valsson}}, \bibinfo
  {author} {\bibfnamefont {J.}~\bibnamefont {Neugebauer}}, \ and\ \bibinfo
  {author} {\bibfnamefont {C.}~\bibnamefont {Filippi}},\ }\href {\doibase
  10.1021/ct400086a} {\bibfield  {journal} {\bibinfo  {journal} {Journal of
  Chemical Theory and Computation}\ }\textbf {\bibinfo {volume} {9}},\ \bibinfo
  {pages} {2355} (\bibinfo {year} {2013})},\ \bibinfo {note} {pMID: 26583726},\
  \Eprint {http://arxiv.org/abs/http://dx.doi.org/10.1021/ct400086a}
  {http://dx.doi.org/10.1021/ct400086a} \BibitemShut {NoStop}%
\bibitem [{\citenamefont {Isborn}\ \emph {et~al.}(2013)\citenamefont {Isborn},
  \citenamefont {Mar}, \citenamefont {Curchod}, \citenamefont {Tavernelli},\
  and\ \citenamefont {Martínez}}]{IMCT13}%
  \BibitemOpen
  \bibfield  {author} {\bibinfo {author} {\bibfnamefont {C.~M.}\ \bibnamefont
  {Isborn}}, \bibinfo {author} {\bibfnamefont {B.~D.}\ \bibnamefont {Mar}},
  \bibinfo {author} {\bibfnamefont {B.~F.~E.}\ \bibnamefont {Curchod}},
  \bibinfo {author} {\bibfnamefont {I.}~\bibnamefont {Tavernelli}}, \ and\
  \bibinfo {author} {\bibfnamefont {T.~J.}\ \bibnamefont {Martínez}},\ }\href
  {\doibase 10.1021/jp4058274} {\bibfield  {journal} {\bibinfo  {journal} {The
  Journal of Physical Chemistry B}\ }\textbf {\bibinfo {volume} {117}},\
  \bibinfo {pages} {12189} (\bibinfo {year} {2013})},\ \bibinfo {note} {pMID:
  23964865},\ \Eprint
  {http://arxiv.org/abs/http://dx.doi.org/10.1021/jp4058274}
  {http://dx.doi.org/10.1021/jp4058274} \BibitemShut {NoStop}%
\bibitem [{\citenamefont {Zuehlsdorff}\ \emph {et~al.}(2016)\citenamefont
  {Zuehlsdorff}, \citenamefont {Haynes}, \citenamefont {Hanke}, \citenamefont
  {Payne},\ and\ \citenamefont {Hine}}]{ZHHPH16}%
  \BibitemOpen
  \bibfield  {author} {\bibinfo {author} {\bibfnamefont {T.~J.}\ \bibnamefont
  {Zuehlsdorff}}, \bibinfo {author} {\bibfnamefont {P.~D.}\ \bibnamefont
  {Haynes}}, \bibinfo {author} {\bibfnamefont {F.}~\bibnamefont {Hanke}},
  \bibinfo {author} {\bibfnamefont {M.~C.}\ \bibnamefont {Payne}}, \ and\
  \bibinfo {author} {\bibfnamefont {N.~D.~M.}\ \bibnamefont {Hine}},\ }\href
  {\doibase 10.1021/acs.jctc.5b01014} {\bibfield  {journal} {\bibinfo
  {journal} {Journal of Chemical Theory and Computation}\ }\textbf {\bibinfo
  {volume} {12}},\ \bibinfo {pages} {1853} (\bibinfo {year} {2016})},\ \bibinfo
  {note} {pMID: 26967019},\ \Eprint
  {http://arxiv.org/abs/http://dx.doi.org/10.1021/acs.jctc.5b01014}
  {http://dx.doi.org/10.1021/acs.jctc.5b01014} \BibitemShut {NoStop}%
\bibitem [{\citenamefont {Foster}\ and\ \citenamefont {Wong}(2012)}]{FW12}%
  \BibitemOpen
  \bibfield  {author} {\bibinfo {author} {\bibfnamefont {M.~E.}\ \bibnamefont
  {Foster}}\ and\ \bibinfo {author} {\bibfnamefont {B.~M.}\ \bibnamefont
  {Wong}},\ }\href {\doibase 10.1021/ct300420f} {\bibfield  {journal} {\bibinfo
   {journal} {Journal of Chemical Theory and Computation}\ }\textbf {\bibinfo
  {volume} {8}},\ \bibinfo {pages} {2682} (\bibinfo {year} {2012})},\ \bibinfo
  {note} {pMID: 22904693},\ \Eprint
  {http://arxiv.org/abs/http://dx.doi.org/10.1021/ct300420f}
  {http://dx.doi.org/10.1021/ct300420f} \BibitemShut {NoStop}%
\bibitem [{\citenamefont {Wong}\ and\ \citenamefont {Cordaro}(2008)}]{WC08}%
  \BibitemOpen
  \bibfield  {author} {\bibinfo {author} {\bibfnamefont {B.~M.}\ \bibnamefont
  {Wong}}\ and\ \bibinfo {author} {\bibfnamefont {J.~G.}\ \bibnamefont
  {Cordaro}},\ }\href {\doibase 10.1063/1.3025924} {\bibfield  {journal}
  {\bibinfo  {journal} {The Journal of Chemical Physics}\ }\textbf {\bibinfo
  {volume} {129}},\ \bibinfo {pages} {214703} (\bibinfo {year} {2008})},\
  \Eprint {http://arxiv.org/abs/http://dx.doi.org/10.1063/1.3025924}
  {http://dx.doi.org/10.1063/1.3025924} \BibitemShut {NoStop}%
\end{thebibliography}%
  
\end{document}